\documentclass[a4paper,11pt]{article}
\usepackage[english]{babel}
\usepackage{jheppub}
\pdfoutput=1
\usepackage[T1]{fontenc}
\usepackage{bbold}
\usepackage{amsmath}
\usepackage{empheq}
\usepackage{booktabs}
\usepackage{color}
\usepackage[dvipsnames]{xcolor}
\usepackage[utf8]{inputenc}
\usepackage{xspace}
\usepackage{scalerel}
\usepackage[most]{tcolorbox}
\usepackage{multirow}
\usepackage{scalefnt}
\usepackage{bold-extra}
\usepackage{subcaption}
\usepackage{afterpage}
\usepackage[normalem]{ulem}
\usepackage{makecell}

\providecommand{\href}[2]{#2}

\newcommand\PhiB{\Phi_{\scriptscriptstyle \rm F}}

\newcommand{\CF}{C_{\mathrm{F}}}
\newcommand{\CA}{C_{\mathrm{A}}}

\newcommand{\pt}{p_{\text{\scalefont{0.77}T}}}

\newcommand{\ptarg}[1]{{p_{\text{\scalefont{0.77}T,$#1$}}}}
\newcommand{\marg}[1]{{m_{\text{\scalefont{0.77}$#1$}}}}

\newcommand{\ptrad}{{p_{\text{\scalefont{0.77}T,rad}}}}
\newcommand{\pth}{{p_{\text{\scalefont{0.77}T,$H$}}}}

\newcommand{\yh}{{y_{\text{\scalefont{0.77}H}}}}

\newcommand{\etah}{{\eta_{\text{\scalefont{0.77}H}}}}

\newcommand{\muF}{{\mu_{\text{\scalefont{0.77}F}}}}
\newcommand{\muR}{{\mu_{\text{\scalefont{0.77}R}}}}

\newcommand{\muRy}{{\mu_{\text{\scalefont{0.77}R}}^{(0),y}}}
\newcommand{\muRb}{{\mu_{\text{\scalefont{0.77}R}}^{(0),\alpha}}}
\newcommand{\KF}{K_{\text{\scalefont{0.77}F}}}
\newcommand{\KR}{K_{\text{\scalefont{0.77}R}}}

\newcommand{\KQ}{{K_{\text{\scalefont{0.77}Q}}}}

\newcommand{\noun}[1]{{\scshape #1}}

\newcommand{\POWHEG}{\noun{Powheg}}

\newcommand{\POWHEGBOXRES}{\noun{Powheg-Box-Res}}
\newcommand{\POWHEGBOXVTWO}{\noun{Powheg-Box-V2}}

\newcommand{\minlo}{{\noun{MiNLO$^{\prime}$}}}
\newcommand{\minnlo}{{\noun{MiNNLO$_{\rm PS}$}}}

\newcommand{\OpenLoops}{{\noun{OpenLoops}}}
\newcommand{\PYTHIA}[1]{\noun{Pythia{#1}}}

\newcommand{\mQQF}{m_{Q\bar Q{\rm F}}}
\newcommand{\muQ}{\mu_{Q}}
\newcommand{\Mdiv}{\mathcal{M}^\textrm{IR-div}}

\usepackage{xcolor}
\newcommand{\mathd}{\mathrm{d}}
\newcommand{\tmop}[1]{\ensuremath{\operatorname{#1}}}

\def\to{\rightarrow}

\def\mbbggs{m_{2b2\gamma}^{\star}}

\def\GeV{\mathrm{GeV}}

\newcommand{\eqn}[1]{eq.\,(\ref{#1})}
\newcommand{\neqn}[1]{eqs.\,(\ref{#1})}
\newcommand{\fig}[1]{figure\,\ref{#1}}

\newcommand{\tab}[1]{table\,\ref{#1}}
\newcommand{\sct}[1]{section\,\ref{#1}}

\def\refeq#1{\mbox{eq.\,\eqref{#1}}}

\def\citere#1{\mbox{ref.\,\cite{#1}}}
\def\citeres#1{\mbox{refs.\,\cite{#1}}}

\usepackage{etoolbox}
\makeatletter
\patchcmd{\@sect}{#8}{\boldmath #8}{}{}
\let\ori@chapter\@chapter
\def\@chapter[#1]#2{\ori@chapter[\boldmath#1]{\boldmath#2}}
\makeatother

\newcommand{\dd}{\mathop{}\!\mathrm{d}}

\newcommand{\bbH}{\ensuremath{b\bar{b}H}}
\newcommand{\ttH}{\ensuremath{t\bar{t}H}}

\newcommand{\ytsq}{\ensuremath{y_t^2}}

\newcommand{\ybsq}{\ensuremath{y_b^2}}

\definecolor{mypink}{RGB}{219, 48, 122}
\definecolor{mygreen}{rgb}{0,0.7,0}
\definecolor{raspberry}{rgb}{0.53,0.15,0.34}

 \usepackage{tikz-feynman}
\usetikzlibrary{positioning,arrows}
\usetikzlibrary{decorations.pathmorphing}
\usetikzlibrary{decorations.markings}
\usetikzlibrary{shapes.geometric}

\DeclareRobustCommand{\ensuremathrm}[1]{\ensuremath{\mathrm{#1}}\xspace}

\DeclareRobustCommand{\GeV}{\ensuremathrm{GeV}\xspace}

\newcommand{\as}{\alpha_s}

\newcommand{\atpt}[1]{{#1}(\pt)}

\newcommand{\ccbar}{{c\bar{c}}}

\usepackage{xcolor}

\DeclareMathOperator{\Tr}{Tr}

\newtcolorbox{empheqboxed}{colback=white!35, 
 colframe=black,
 width=\textwidth,
 sharpish corners,
 top=-2mm,
 bottom=0pt
}

\title{Higgs boson production in association with massive bottom quarks at NNLO+PS}

\author[a]{Christian Biello,}
\author[b]{Javier Mazzitelli,}
\author[a,c]{Aparna Sankar,}
\author[a]{Marius Wiesemann,}
\author[a,c]{\hspace{1cm}Giulia Zanderighi}
\affiliation[a]{
Max-Planck-Institut f\"ur Physik, Boltzmannstrasse 8, 85748 Garching, Germany
}
\affiliation[b]{
PSI Center for Neutron and Muon Sciences, 5232 Villigen PSI, Switzerland
}
\affiliation[c]{
Physik Department T31, James-Franck-Straße 1,\\
Technische Universität München, D-85748 Garching, Germany
}
\emailAdd{
biello@mpp.mpg.de, javier.mazzitelli@psi.ch, aparna@mpp.mpg.de, marius.wiesemann@mpp.mpg.de, zanderi@mpp.mpg.de}

\abstract{
    We study the production of a Higgs boson in association with a bottom-quark 
    pair (\bbH{}) at hadron colliders. Our calculation is performed in the four-flavour scheme
     with massive bottom quarks. This work presents the first computation of next-to-next-to-leading-order (NNLO) QCD corrections to this process, and we combine them with all-order radiative corrections from a parton shower simulation (NNLO+PS). 
     The calculation is exact, except for the two-loop amplitude, which is evaluated in the 
     small quark mass expansion, which is an excellent approximation for bottom quarks at LHC energies.
     For the NNLO+PS matching, we employ the \minnlo{} method for heavy-quark plus colour-singlet production within the \POWHEG{} framework. 
      We present an extensive phenomenological analysis both at the inclusive level and considering bottom jets using flavour-tagging algorithms. 
       By comparing four-flavour and five-flavour scheme predictions at NNLO+PS, we find
       that the NNLO corrections in the four-flavour scheme resolve the long-standing tension between the two schemes. Finally, we show that our  NNLO+PS predictions 
       also have important implications on modelling the $b\bar b H$ background in Higgs-pair measurements.
}
  

\preprint{
\vspace{-24pt}
  \begin{flushright}
  MPP-2024-239\\
  PSI-PR-24-28
  \end{flushright}
}

\begin{document}

\maketitle

\section{Introduction}
\label{sec:intro}
The discovery of the Higgs boson in 2012 by the ATLAS~\cite{ATLAS} and CMS~\cite{CMS:2008xjf} collaborations marked a significant milestone in our understanding of the Standard Model (SM) of particle physics~\cite{H125ATLAS,H125CMS,H125CMS2}. Over the past decade, extensive efforts have been dedicated to investigating the properties of this particle~\cite{ATLAS:Hig102022,CMS:Hig102022}. Measurements of its couplings to top ($t$) and bottom ($b$) quarks, $W$ and $Z$ bosons, and tau ($\tau$) leptons are consistent with SM predictions so far~\cite{ATLAS:2022vkf,CMS:2022dwd}. However, since the experiments at the Large Hadron Collider (LHC) continue to collect data at increasing rates, the higher precision of future measurements improves the sensitivity to potential deviations from the SM. Furthermore, other Higgs couplings, such as the self-interaction of the Higgs boson, are expected to become accessible when statistical (and potentially systematic) uncertainties decrease in future analyses. 

The accurate simulation of all relevant Higgs-boson production and decay modes at the LHC is essential for extracting Higgs properties in high-precision measurements and for identifying any deviation from SM predictions.
Higgs-boson production at the LHC proceeds through several mechanisms in the SM~\cite{LHCHiggsCrossSectionWorkingGroup:2016ypw}. The most prominent ones, ranked by their cross section size, are gluon-gluon fusion ($ggF$), vector-boson fusion (VBF), Higgsstrahlung ($VH$), and the associated production with top quarks (\ttH{}) and with bottom quarks (\bbH). Except for the \bbH{}, all these production mechanisms have been experimentally observed~\cite{ATLAS:Hig102022,CMS:Hig102022}. 

Among these processes, \bbH{} production is particularly interesting despite its experimental challenges. With a predicted cross section of $0.48_{-0.11}^{+0.10}$ pb at  13\,TeV in proton--proton collisions \cite{LHCHiggsCrossSectionWorkingGroup:2016ypw}, it occurs at a rate comparable to \ttH{} production. However, the experimental signature of \bbH{} is less distinct due to the absence of clear decay products, such as those from top quarks, which facilitated the observation of \ttH{} production by ATLAS and CMS in 2018~\cite{ATLAS:tth2018,CMS:tth2018}. Furthermore, \bbH{} production is complicated by its overlap and interference 
with the $ggF$ process~\cite{Pagani:2020rsg}, making it challenging to use \bbH{} to constrain the bottom-Yukawa coupling directly. Instead, measurements of Higgs decays to bottom quarks offer more precise constraints to the Yukawa coupling. At the same time, \bbH{} production yields a contribution (of about 1\%)
to the total inclusive Higgs-boson rate that is relevant for the precision goal of LHC measurements, and it
remains valuable for exploring the interplay between the Higgs-boson couplings to bottom and top quarks. 
A precise simulation of the \bbH{} production also plays an important role in constraining the light-quark Yukawa couplings, such as the charm quark,
through the Higgs transverse-momentum spectrum \cite{Bishara:2016jga}. 

Apart from that, \bbH{} production holds significant importance in two key contexts. One notable aspect arises in beyond-the-Standard-Model (BSM) scenarios, where an enhanced bottom-Yukawa coupling renders the \bbH{} process the dominant mechanism for producing (typically heavy) Higgs bosons. This is especially evident in models like the Two-Higgs-Doublet Model (2HDM) or its supersymmetric extension, the MSSM, when $\tan \beta$ is large. Another critical role of \bbH{} production is as the primary irreducible background in SM searches for Higgs-pair ($HH$) production \cite{DiMicco:2019ngk,Manzoni:2023qaf} 
in decay channels involving bottom quarks. Accurate modelling of this background is essential for improving the sensitivity of $HH$ studies, especially at the High-Luminosity LHC (HL-LHC). In this phase, the $HH$ cross section in the SM is projected to be measured with a significance of $3.4\sigma$, potentially rising to $4.9\sigma$ assuming minimal systematic uncertainties~\cite{ATLAS:2022faz}. Consequently, the precise simulation of \bbH{} production will be indispensable for achieving the level of precision required for LHC measurements, especially in the  HL-LHC phase.
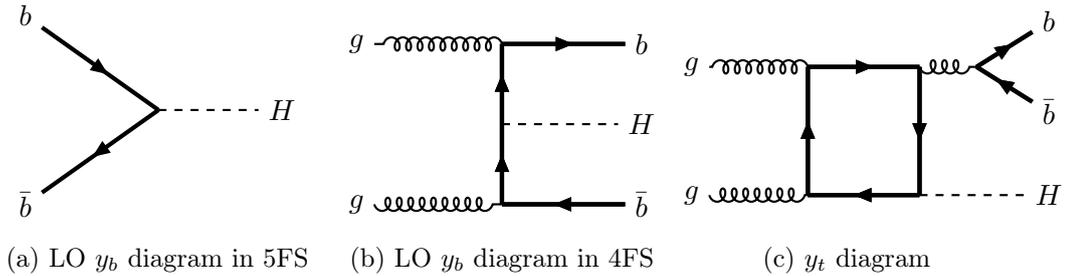
\begin{figure*}[t]
  \begin{center}
    \begin{subfigure}[b]{.3\linewidth}
      \centering
\begin{tikzpicture}[scale=1.25]
\begin{feynman}
	\vertex (a1) at (0,0) {\( b\)};
	\vertex (a2) at (0,-2) {\(\bar b\)};
	\vertex (a3) at (1.4,-1);
	\vertex (a4) at (2.7,-1){\( H\)};
        \diagram* {
          {[edges=fermion]
            (a1)--[ultra thick](a3)--[ultra thick](a2),
          },
          (a3) -- [scalar,thick] (a4),
        };
      \end{feynman}
\end{tikzpicture}
\caption{LO $y_b$ diagram in 5FS}
        \label{subfig:qq}
\end{subfigure}%
\begin{subfigure}[b]{.3\linewidth}
  \centering
\begin{tikzpicture}[scale=1.25]
  \begin{feynman}
	\vertex (a1) at (0,0) {\( g\)};
	\vertex (a2) at (0,-1.7) {\( g\)};
	\vertex (a3) at (1.53,0);
	\vertex (a4) at (1.53,-1.7);
	\vertex (a42) at (1.53,-0.85);
	\vertex (a43) at (3,-0.85) {\( H\)};
	\vertex (a5) at (3,0){\( b\)};
	\vertex (a6) at (3,-1.7){\(\bar b\)};
        \diagram* {
          {[edges=fermion]
            (a6)--[fermion, ultra thick](a4)--[fermion, ultra thick](a42)--[fermion, ultra thick](a3)--[fermion, ultra thick](a5),
          },
          (a2)--[gluon,thick](a4),
          (a3)--[gluon,thick](a1),
          (a42)--[scalar,thick](a43),
        };
  \end{feynman}
\end{tikzpicture}
\caption{LO $y_b$ diagram in 4FS}
        \label{subfig:gg}
\end{subfigure}%
\begin{subfigure}[b]{.3\linewidth}
  \centering
\begin{tikzpicture}
  \begin{feynman}
	\vertex (a1) at (0,0) {\( g\)};
	\vertex (a2) at (0,-1.7) {\( g\)};
	\vertex (a3) at (1.53,0);
	\vertex (a4) at (1.53,-1.7);
	\vertex (a5) at (3,0);
	\vertex (a6) at (3,-1.7);
	\vertex (a62) at (3.75,0);
	\vertex (a63) at (4.7,0.6){\( b\)};
	\vertex (a64) at (4.7,-0.6){\( \bar b\)};
	\vertex (a7) at (4.7,-1.7) {\( H\)};
        \diagram* {
          {[edges=fermion]
            (a6)--[fermion, ultra thick](a4)--[fermion, ultra thick](a3)--[fermion, ultra thick](a5)--[fermion, ultra thick](a6),
            (a64)--[fermion, ultra thick](a62)--[fermion, ultra thick](a63)
          },
          (a5)--[gluon,thick](a62),
          (a2)--[gluon,thick](a4),
          (a3)--[gluon,thick](a1), 
          (a6)--[scalar,thick](a7),
        };
  \end{feynman}
\end{tikzpicture}\vspace{0.15cm}
\caption{$y_t$ diagram}
        \label{subfig:gg}
\end{subfigure}
\end{center}
\caption{\label{fig:bbh} Sample Feynman diagrams for 
   Higgs production in association with bottom quarks.}
\end{figure*}

In addition to its experimental relevance, \bbH{} production is theoretically very interesting and challenging, as discussed in the following.
The dominant contributions to the \bbH{} process arise from two production mechanisms. The first one involves terms proportional to the bottom-Yukawa coupling ($y_b$), where the Higgs boson directly couples to a bottom-quark line, as illustrated in \fig{fig:bbh}\,(a) and (b). The second one stems from terms proportional to the top Yukawa coupling ($y_t$), where the Higgs couples to a closed top-quark loop, as shown in \fig{fig:bbh}\,(c). Interestingly, the latter mechanism, which corresponds to $ggF$ process with a $b\bar{b}$ pair produced through a QCD splitting, yields a slightly larger cross section. Its relative contribution becomes even larger when tagging one or more bottom-quark jets. Both mechanisms are relevant within the SM, and their accurate description requires higher-order QCD calculations due to the significant size of perturbative corrections. Subleading contributions to \bbH{} production include associated $VH$ production, where the vector boson decays to bottom quarks, and bottom-associated vector-boson fusion. While these channels contribute only a few percent to the total cross section, or even less depending on the selection criteria, dedicated simulations for these processes are available~\cite{Pagani:2020rsg}.

Theoretical predictions for \bbH{} production at the LHC rely on two main approaches for treating the bottom-quark mass: the five-flavor scheme (5FS) and the four-flavor scheme (4FS). In the 5FS, the bottom quark is treated as a massless parton, with logarithmic
contributions of collinear origin resummed into the parton distribution functions (PDFs). This assumption simplifies calculations in the 5FS, c.f.\,the leading-order (LO) diagram in \fig{fig:bbh}\,(a), allowing higher-order corrections in the strong coupling constant to be more readily computed.
Substantial progress has been made in recent years on contributions to the cross section 
proportional to $\ybsq{}$ in the 5FS \cite{Das:2023rif,Dicus:1998hs,Balazs:1998sb,Harlander:2003ai,Campbell:2002zm,Harlander:2010cz,Ozeren:2010qp,Harlander:2011fx,Buehler:2012cu,Belyaev:2005bs,Harlander:2014hya,Ahmed:2014pka,Gehrmann:2014vha,Duhr:2019kwi,Mondini:2021nck,Wiesemann:2014ioa,Krauss:2016orf,Ajjath:2019ixh,Ajjath:2019neu,Forte:2019hjc,Badger:2021ega,Das:2024pac,Cal:2023mib}. Notably, the computation of the third-order QCD cross section \cite{Duhr:2019kwi} represents a significant milestone. Pure QED and mixed QCD-QED corrections in this framework are minimal, typically below $0.03\%$ of the LO cross section \cite{AH:2019xds}, while mixed QCD-electroweak corrections contribute around $1\%$~\cite{Pagani:2020rsg}. More recently, the matching of NNLO QCD with parton showers (NNLO+PS) has been performed in the 5FS in \citere{Biello:2024vdh} by some of us.

By contrast, the 4FS treats the bottom quark as a massive particle, which increases the complexity of the calculations, c.f.\,the LO diagram in \fig{fig:bbh}\,(b), but provides a more accurate description of observables involving bottom quarks. In this scheme, the \bbH{} cross section is known up to next-to-leading order (NLO) in QCD~\cite{Dittmaier:2003ej,Dawson:2003kb,Wiesemann:2014ioa,Deutschmann:2018avk}. Combined studies of contributions from $\ybsq{}$, $\ytsq{}$, and their interference ($y_by_t$) at NLO and NLO+PS have been performed exclusively in the 4FS, see \citere{Deutschmann:2018avk} and \citere{Manzoni:2023qaf}, respectively.
It has been a long-standing issue that predictions in the 4FS and the 5FS are 
not compatible. Therefore, many studies have examined these differences 
and for the total inclusive cross section a consistent 
combination of the two schemes has been achieved. 
For works on these topics see for instance \citeres{Aivazis:1993pi,Cacciari:1998it,Forte:2010ta,Harlander:2011aa,Maltoni:2012pa,Bonvini:2015pxa,Forte:2015hba,Bonvini:2016fgf,Forte:2016sja,Lim:2016wjo,Duhr:2020kzd}.
However, without their combination differences between 4FS and the 5FS remain 
beyond their theoretical uncertainties and at the differential level no combined
4FS and 5FS predictions are available.

In this paper, we focus on the \ybsq{} contribution to the \bbH{} process and present the first NNLO QCD calculation in the 4FS. 
Additionally, we match our NNLO results to a parton shower simulation to obtain a fully exclusive event generation at NNLO+PS accuracy. This is achieved using the \minnlo{} method for the production of a heavy-quark pair in association with colour-singlet particles  ($Q\bar{Q}$F), as presented in \citere{Mazzitelli:2024ura}. We have adapted
this approach to account for a scale-dependent Yukawa coupling renormalised 
in the $\overline{\rm MS}$ scheme.
We keep the bottom-mass dependence exact throughout the calculation, except for the two-loop contribution, where we apply a small-mass expansion \cite{Mitov:2006xs,Wang:2023qbf}, which is expected to be an excellent approximation for bottom quarks at the LHC. Indeed, assessing the uncertainties associated with this approximation at NLO QCD, we find them negligible compared to the scale uncertainties.
We provide an extensive phenomenological study of our novel 4FS predictions and compare them to the 5FS results from \citere{Biello:2024vdh}. Additionally, we examine the \bbH{} process as a background for $HH$ searches.

\section{Outline of the calculation}
\label{sec:outline}

We consider the process of Higgs-boson production associated with bottom quarks
\begin{align}
pp \to b\bar{b} H +X\,,
\end{align}
where the final state is inclusive over the radiation of additional particles $X$. Contributions from the loop-induced $ggF$ process
proportional to $y_t$, as illustrated in \fig{fig:bbh}\,(c), are excluded throughout this paper. Instead, our focus is on the terms 
proportional to $y_b^2$, which form a gauge-invariant subset of the cross section, which we shall refer to as $b\bar bH$ production
for the remainder of this manuscript. For convenience, we leave the inclusion of the $y_t$-dependent contributions to future work, as these terms can be treated separately in perturbation theory. However, the leading-order $y_t^2$ contributions are already accounted for in NNLO calculations of the $ggF$ process. In our calculation of the $b\bar bH$ process we consider the bottom quarks 
to be massive, i.e.\ we employ the 4FS, see \fig{fig:bbh}\,(b) for a representative LO diagram.

The corresponding calculation in the 5FS with massless bottom quarks, namely Higgs-boson production in bottom-quark 
annihilation ($b\bar{b}\to H$), see \fig{fig:bbh}\,(a) for the LO diagram, has already been completed in \citere{Biello:2024vdh} 
by some of us. That calculation was based on the \minnlo{} method for the colour-singlet production \cite{Monni:2019whf,Monni:2020nks}.
We will make use of these results in \sct{sec:4FSvs5FS} to compare 4FS and 5FS predictions.

We implement a fully differential computation of Higgs-boson production associated with a bottom-quark pair in the 4FS up to NNLO in QCD perturbation theory and consistently match it
to a parton shower. To this end, we have adapted the \minnlo{} method for  heavy-quarks plus colour-singlet 
production ($Q\bar{Q}$F) presented in \citeres{Mazzitelli:2020jio,Mazzitelli:2021mmm,Mazzitelli:2023znt,Mazzitelli:2024ura} to account for an overall scale-dependent 
Yukawa coupling, which is renormalised in the $\overline{\rm MS}$ scheme. The \minnlo{} method, its extension to $Q\bar Q$F processes and its adaptations for $b\bar b H$ production are described in detail in \sct{sec:method}. The computation is exact, except for the double-virtual corrections that are approximated through the massification procedure of the bottom quarks outlined in \sct{sec:massification}. This allows us to exploit 
the two-loop amplitude for massless bottom quarks \cite{Badger:2021ega} instead of the full massive two-loop calculation, which is out of reach
with current technology, rendering the calculation of NNLO QCD corrections feasible.

Our \minnlo{} $b\bar{b}H$ generator has been implemented within the \POWHEGBOXRES{} framework \cite{Jezo:2015aia}. First, we have implemented a NLO+PS generator for
\bbH{} plus one jet ($b\bar bH$J) production using the \POWHEG{} method \cite{Nason:2004rx,Alioli:2010xd,Frixione:2007vw}.
For the evaluation of the tree-level and one-loop $b\bar bH$J amplitudes and tree-level \bbH{} plus two jets ($b\bar bH$JJ) amplitudes we 
employ \OpenLoops{}~\cite{Cascioli:2011va,Buccioni:2017yxi,Buccioni:2019sur}, using its interface within the \POWHEGBOXRES{} 
framework developed in \citere{Jezo:2018yaf}. 
In a second step, we have adapted the $b\bar bH$J NLO+PS implementation to reach NNLO QCD accuracy for \bbH{} production 
through (an extension of) the \minnlo{} approach described in the next section.


\section{NNLO+PS methodology}
\label{sec:method}
\subsection[Original \minnlo{} method]{Original \boldmath{\minnlo{}} method}
\label{sec:minnlops}
In the following, we summarise the \minnlo{} method for colour-singlet production, which was initially introduced in \citeres{Monni:2019whf,Monni:2020nks} and has been applied to several processes by now \cite{Lombardi:2020wju,Lombardi:2021rvg,Buonocore:2021fnj,Lombardi:2021wug,Zanoli:2021iyp,Gavardi:2022ixt,Haisch:2022nwz,Lindert:2022qdd,Biello:2024vdh}. In the following, we briefly review the method, and we refer to \citeres{Monni:2019whf,Monni:2020nks,Ebert:2024zdj} for further details.

Starting from a \POWHEG{} \cite{Nason:2004rx,Frixione:2007vw,Alioli:2010xd} NLO+PS calculation for colour-singlet (F)
production with an additional jet (FJ), the \minnlo{} master formula can be written as
\begin{align}
\label{eq:master}
      {\rm d}\sigma_{\rm\scriptscriptstyle F}^{\rm MiNNLO_{PS}}={\rm d}\Phi_{\scriptscriptstyle\rm FJ}\,\bar{B}^{\,\rm MiNNLO_{\rm PS}}\,\times\,\left\{\Delta_{\rm pwg}(\Lambda_{\rm pwg})+
      {\rm d}\Phi_{\rm rad}\Delta_{\rm pwg}(\ptrad)\,\frac{R_{\scriptscriptstyle\rm FJ}}{B_{\scriptscriptstyle\rm FJ}}\right\}\,,
\end{align}
where $B_{\scriptscriptstyle {\rm FJ}}$ and $R_{\scriptscriptstyle {\rm FJ}}$ are derived from the squared tree-level matrix elements for FJ and FJJ production, respectively. Here, $\Phi_{\scriptscriptstyle {\rm FJ}}$ represents the FJ phase space, $\Delta_{\rm pwg}$ is the \POWHEG{} Sudakov form factor, and $\Phi_{\tmop{rad}}$ and $\ptrad$  denote the phase space and transverse momentum of the second radiation. The \POWHEG{} $\bar B$ function is modified to achieve NNLO QCD accuracy for the Higgs production when QCD radiation is unresolved,
\begin{align}
 \label{eq:bbar}
  \bar{B}^{\text{\minnlo{}}}&=e^{-\tilde S(p_{T})}  \bigg\{ \frac{\alpha_s(\pt)}{2\pi} {\frac{\dd \sigma^{(1)}_{\text{\scalefont{0.77}FJ}}(\pt)}{\dd \Phi_{\text{\scalefont{0.77}FJ}}}}\bigg(1+\frac{\alpha_s(\pt{})}{2\pi}\tilde S^{(1)}\bigg) 
  +\bigg(\frac{\alpha_s(\pt)}{2\pi}\bigg)^2{\frac{\dd \sigma^{(2)}_{\text{\scalefont{0.77}FJ}}(\pt)}{\dd \Phi_{\text{\scalefont{0.77}FJ}}}}
  \nonumber \\&
  +\bigg[D(\pt) 
  -\frac{\alpha_s(\pt{})}{2\pi}
  \atpt{D^{(1)}}-\bigg(\frac{\alpha_s(\pt{})}{2\pi}\bigg)^2 \atpt{D^{(2)}}\bigg] 
  \times F^{\text{corr}} \bigg\}\,.
\end{align}
In the above equation, ${\rm d}\sigma^{(1,2)}_{\text{\scalefont{0.77}FJ}}$ denote the first- and second-order differential FJ cross sections. The remaining terms stem from the transverse-momentum ($\pt$) resummation formula,
\begin{align}
  \frac{\dd \sigma}{\dd \PhiB \dd \pt{}}&=\frac{\dd}{\dd \pt{}}\left\{e^{-\tilde S(\pt{})}\mathcal{L}(\pt{})\right\} 
 \label{eq:resum} \\&=e^{-\tilde S(\pt{})}\underbrace{\left\{-\mathcal{L}(\pt{})\frac{\dd}{\dd \pt{}}\tilde S(\pt{}) + \frac{\dd}{\dd \pt{}} \mathcal{L}(\pt{})\right\}}_{\eqqcolon\,D(\pt{})} \,,
\notag
\end{align}
where $e^{-\tilde S(\pt{})}$ is the Sudakov form factor, $\tilde S^{(1)}$ in \eqn{eq:bbar} is the corresponding $\mathcal{O}(\as)$ term 
in the expansion of the Sudakov exponent, and the function $D(\pt{})$ is defined in \eqn{eq:resum}.  The luminosity factor $\mathcal{L}(\pt{})$ 
in the above equation includes the squared virtual matrix elements for the colour-singlet process under consideration and the convolution of the parton densities with the collinear coefficient functions. In the \minnlo{} approach, renormalisation and factorisation scales are set to $\pt$, except for potential overall couplings at the Born level, whose scale can be chosen freely.

The last term in $\bar{B}$ in \eqn{eq:bbar}, which starts at order $\as^3(p_{\text{\scalefont{0.77}T}})$, adds the necessary (singular) terms to achieve NNLO accuracy \cite{Monni:2019whf}. Rather than truncating singular contributions from $D(\pt{})$ at $\as^3$, i.e., $\left(D-D^{(1)}-D^{(2)}\right)=D^{(3)}+\mathcal{O}(\as^4)$, as in the original \minnlo{} formulation of \citere{Monni:2019whf}, we follow the extension introduced in \citere{Monni:2020nks} by retaining the total derivative in \eqn{eq:resum} to keep subleading logarithmic contributions, which improves agreement with fixed-order NNLO predictions. The factor $F^{\rm corr}$ in \eqn{eq:bbar} spreads the Born-like $\left(D-D^{(1)}-D^{(2)}\right)$ contribution 
over the full FJ phase space to obtain a fully exclusive event generator at NNLO+PS accuracy \cite{Monni:2019whf}.

\subsection[Extension to ${Q\bar Q }$F processes]{Extension to \boldmath{${Q\bar Q }$F} processes}
The \minnlo{} approach is currently the only NNLO+PS method that also extends to processes involving colour charges in initial and final state, including heavy-quark pair ($Q\bar Q$) production \cite{Mazzitelli:2020jio,Mazzitelli:2021mmm,Mazzitelli:2023znt} and, very recently, the production of a heavy quark pair in association with colour-singlet particles ($Q\bar Q$F) \cite{Mazzitelli:2024ura}.\footnote{First steps towards the extension
to processes with light jets in the final state has been made in \citere{Ebert:2024zdj}.} 
In the following, we briefly recall these extensions of the \minnlo{} method, which exploits the knowledge of the singular structure of the cross section at small transverse momentum of the final-state system. The structure of large logarithmic contributions for the $Q\bar Q$F final state \cite{Catani:2021cbl} is very similar to the one valid for $Q\bar Q$ production at small $\pt{}$ \cite{Zhu:2012ts,Li:2013mia,Catani:2014qha,Catani:2018mei}. However, the
$Q\bar Q$F case involves more general kinematic configurations compared
to the $Q\bar Q$ case, where the heavy quarks are constrained to be back-to-back
to each other in the Born configuration. In either case, the starting point is the following factorisation theorem, which is expressed in the Fourier-conjugate
space to $\pt$ (so-called impact-parameter or $b$-space) \cite{Zhu:2012ts,Li:2013mia,Catani:2014qha,Catani:2018mei}:
\begin{align}\label{eq:facformula}
	\frac{\mathd\sigma}{ \mathd^2\vec{\pt}\, \mathd \Phi_{Q\bar{Q}{\rm F}}}&=\sum_{c=q,\bar{q},g}
  \frac{|M^{(0)}_{c\bar{c}}|^2}{2 m_{Q\bar{Q}{\rm F}}^2 }\int\frac{ \mathd^2\vec{b}}{(2\pi)^2} e^{i \vec{b}\cdot
  \vec{\pt} } e^{-S_{c\bar{c}} \left(\frac{b_0}{b}\right)}\sum_{i,j}\Tr({\mathbf H}_{c\bar{c}}{\mathbf \Delta})\,
 \,({C}_{ci}\otimes f_i) \,({C}_{\bar{c} j}\otimes f_j) \,.
\end{align}
Here, $m_{Q\bar{Q}{\rm F}}$, $\pt$ and $\Phi_{Q\bar{Q}{\rm F}}$ denote the invariant mass, transverse momentum and phase space of the $Q\bar Q$F system, respectively. The sum runs over all possible flavour configurations of the incoming partons, where the first particle has flavour \( c \) and the second one has flavour \( \bar{c} \).\footnote{To simplify the notation, we consider only the case in which the incoming partons have opposite flavours \( c \) and \( \bar{c} \) at LO, which is indeed the case for $b\bar{b}H$ production.} The Sudakov form factor $e^{-S_\ccbar}$ in \refeq{eq:facformula} resums logarithmic contributions from soft and collinear initial-state radiation and, hence, it has the exact same form as the one for colour-singlet production. Its exponent is defined as
\begin{align}
	\label{eq:sudakov}
  S_\ccbar\left(\frac{b_0}{b}\right) \equiv \int_{\frac{b_0^2}{b^2}}^{m_{Q\bar{Q}{\rm F}}^2}\frac{\mathd{}q^2}{q^2} \left[A_\ccbar(\alpha_s(q))\ln\frac{m_{Q\bar{Q}{\rm F}}^2}{q^2}+B_\ccbar(\alpha_s(q))\right]\,,
\end{align}
with $b_0=2 e^{-\gamma_\text{E}}$. Also the collinear coefficient functions $C_{ij}\equiv C_{ij}(z,p_1,p_2,\vec{b},\alpha_s)$ correspond to those of the colour-singlet case as they encode initial-state collinear radiation, which are universal ingredients convoluted with the parton distribution functions $f_i(z)$.
The colour-space operator \(\mathbf{H}_{c\bar c}\) can be expressed as $\mathbf{H}_{c\bar c} = |M_{c\bar c}\rangle \langle M_{c\bar c} | / |M^{(0)}_{c\bar c}|^2$, 
where $|M^{(0)}_{c\bar c}|^2$ is the corresponding Born squared matrix element and $M_{c\bar c}$ is the finite amplitude for $Q\bar{Q}$F production obtained in the following way.  We start by defining the finite remainder $\mathcal{R}_{c\bar c}$ as the minimal subtraction of infrared divergences in the dimensional regulator $\epsilon$, which is achieved through 
\begin{align}
  |\mathcal{R}^{}_{c\bar c}\rangle = \mathbf{Z}^{-1}_\ccbar(\Phi_{Q\bar{Q}{\rm F}},\mu,\epsilon) |\Mdiv_{c\bar c}\rangle \label{eq:ampvec}
\end{align}
using the operator $\mathbf{Z}$ introduced in \citeres{Becher:2009cu,Becher:2009qa}.
Here, $|\mathcal{M}^\textrm{IR-div}_{c\bar c}\rangle$ is the ultraviolet renormalised amplitude, and \(\mu\) denotes the scale at which the infrared poles are subtracted. The finite remainder admits a perturbative expansion,
\begin{align}
	|\mathcal{R}_{c\bar c} \rangle = \sum_i \left( \frac{\alpha_s(m_{Q\bar{Q}{\rm F}})}{2\pi}\right)^i |\mathcal{R}^{(i)}_{c\bar c}\rangle.
\end{align}
The connection to $M_{c \bar c}$ can be symbolically expressed as
\begin{align}
| M_{c \bar c} \rangle = \bar{\mathbf{h}} \, | \mathcal{R}_{c\bar c} \rangle\,,
\end{align}
where the explicit expression of the operator $\bar{\mathbf{h}}$ has been derived in \citere{Catani:2023tby} and extended to general $Q\bar Q$F kinematics in \citere{inprep:shark}. We note that the main difference between $M_{c \bar c}$ and $\mathcal{R}_{c\bar c}$ is that, while $\mathcal{R}_{c\bar c}$ is minimally subtracted, the subtraction of $M_{c \bar c}$ contains additional finite terms arising from soft-parton contributions.

Returning to \eqn{eq:facformula}, the crucial difference compared to the colour-singlet case is the presence of contributions originating from the operator ${\mathbf \Delta}$, which captures the resummation of single-logarithmic contributions that emerge from soft radiation connecting a heavy-quark line either with an initial-state parton or with the other final-state heavy quark. It can be written as $\mathbf{\Delta}=\mathbf{V}^\dagger \mathbf{D} \mathbf{V}$. The azimuthal operator $\mathbf{D}\equiv \mathbf{D}(\Phi_{Q\bar{Q}{\rm F}},\vec{b},\alpha_s)$ captures azimuthal correlations of the $Q\bar Q$F system in the small $\pt$ limit. Its average over the azimuthal angle $\phi$ 
is given by $\left[ \mathbf{D} \right]_\phi=\mathbb{1}$ . The operator $\mathbf{V}$, on the other hand, is obtained by the path-integral ordered exponentiation of the soft anomalous dimension $\mathbf{\Gamma}_t$ for the $Q\bar Q$F production,
\begin{align}
	\label{eq:V}
{\mathbf V} = {\cal
  P}\exp\left\{-\int_{b_0^2/b^2}^{m_{Q\bar{Q}{\rm F}}^2}\frac{\mathd{}q^2}{q^2}{\mathbf
  \Gamma}_t(\Phi_{Q\bar{Q}{\rm F}};\alpha_s(q))\right\}\,.
\end{align}
The matrix \(\mathbf{\Gamma}_t\) can be expanded in powers of \(\alpha_s(q)/(2\pi)\), with \(\mathbf{\Gamma}_t^{(1)}\) and \(\mathbf{\Gamma}_t^{(2)}\) representing the first- and second-order coefficients, respectively. For $Q\bar{Q}$F production up to NNLO, we can expand and isolate the \(\mathbf{\Gamma}_t^{(2)}\) term, moving it outside the path-ordering symbol. The isolated contribution can be absorbed into a redefinition of $B_{c\bar c}^{(2)}$. In general, \(\mathbf{\Gamma}_t^{(2)}\) includes non-trivial terms proportional to three-parton correlations. These terms have a vanishing expectation value with the LO matrix elements in the \(b \bar{b} H\) case, but must be retained for general $Q \bar{Q}$F processes \cite{Czakon:2013hxa}.
We remain with the NLL accurate operator,
\begin{align}
	\label{eq:V_NLL}
{\mathbf V}_{\rm NLL} = {\cal
 P}\Bigg[\exp\left\{-\int_{b_0^2/b^2}^{m_{Q\bar{Q}{\rm F}}^2}\frac{\mathd{}q^2}{q^2}\frac{\alpha_s(q)}{2\pi}{\mathbf
   \Gamma}^{(1)}_t\right\} \Bigg]\,.
\end{align}
Thus, the trace in colour space in \eqn{eq:facformula} is reduced to $\Tr({\mathbf H}_{c\bar{c}}{\mathbf \Delta}) =\langle M_{c\bar c}|\mathbf{V}_{\rm NLL} ^\dagger \mathbf{V}_{\rm NLL}  |M_{c\bar c}\rangle$. Following \citere{Mazzitelli:2020jio}, the all-order matrix elements in this expectation value can be simplified to the tree-level matrix elements by absorbing the difference at NNLO into a further redefinition of $B_{c\bar c}^{(2)}$.
The final replacement ${B}_\ccbar^{(2)} \to \hat B_\ccbar^{(2)}$ is then given by,
\begin{align}
	\label{eq:B2hat}
\hat{B}_\ccbar^{(2)} &= B_\ccbar^{(2)} + \frac{\langle
    M_{c\bar{c}}^{(0)} | {\mathbf \Gamma}^{(2)\,\dagger}_t +
  {\mathbf \Gamma}^{(2)}_t|M_{c\bar{c}}^{(0)}
  \rangle}{|M^{(0)}_{c\bar{c}}|^2} \notag\\
  &+ 2\,\text{Re}\left[\frac{\langle
    M_{c\bar{c}}^{(1)} | {\mathbf \Gamma}^{(1)\,\dagger}_t +
  {\mathbf \Gamma}^{(1)}_t|M_{c\bar{c}}^{(0)}
  \rangle}{|M^{(0)}_{c\bar{c}}|^2}\right] 
  - 2\,\text{Re}\left[\frac{\langle
    M_{c\bar{c}}^{(1)} |M_{c\bar{c}}^{(0)}
  \rangle}{|M^{(0)}_{c\bar{c}}|^2}\right]
\frac{\langle
    M_{c\bar{c}}^{(0)} | {\mathbf \Gamma}^{(1)\,\dagger}_t +
  {\mathbf \Gamma}^{(1)}_t|M_{c\bar{c}}^{(0)}
  \rangle}{|M^{(0)}_{c\bar{c}}|^2}\, ,
\end{align}
where the two terms in the second line account for the simplification in the matrix elements we just discussed, and the added term in the first line accounts for the \(\mathbf{\Gamma}_t^{(2)}\) contribution mentioned earlier.

After performing the Fourier and angular integrations, the factorisation formula in \eqn{eq:facformula} can be cast into a form similar to that of \eqn{eq:resum},
\begin{align}
	\frac{\mathd\sigma}{\mathd{} \pt\,\mathd{} \Phi_{Q\bar{Q}{\rm F}}}  = \frac{\mathd{}}{\mathd{}
  \pt}\Bigg\{\sum_{c}\,\Bigg[\sum_{i=1}^{n_c}{\cal C}^{[\gamma_i]}_\ccbar(\Phi_{Q\bar{Q}{\rm F}})e^{-\tilde{S}^{[\gamma_i]}_{\ccbar}(\pt)}\Bigg]{\cal L}_{c\bar{c}} (\Phi_{Q\bar{Q}{\rm F}},\pt)\Bigg\} + {\cal O}(\alpha_s^5)\,. \label{eq:minnloQQFmf}
\end{align}
Here, \(\gamma_i\) are the $n_c$ eigenvalues of ${\mathbf \Gamma}^{(1)}$. Eq.\,\eqref{eq:minnloQQFmf} has been derived by using the colour basis where ${\mathbf \Gamma}^{(1)}$ is diagonal, which thus leads to the following simplification of the expectation value:
\begin{align}
	e^{-\tilde{S}_\ccbar (\pt) } \langle M_{c\bar{c}}^{(0)} | \left({\mathbf V}_{\rm
    NLL}\right)^\dagger{\mathbf V}_{\rm NLL} |M_{c\bar{c}}^{(0)}
\rangle =|M^{(0)}_{c\bar{c}}|^2\sum_{i=1}^{n_c} {\cal C}^{[\gamma_i]}_\ccbar(\Phi_{Q\bar{Q}{\rm F}})e^{-\tilde{S}^{[\gamma_i]}_{\ccbar}(\pt)}\,,
\end{align}
where the eigenvalues  \(\gamma_i\) of ${\mathbf \Gamma}^{(1)}$ have been absorbed into the $B^{(1)}$ coefficient of the Sudakov
\begin{align}
	&B_{c\bar c}^{(1)} \rightarrow  B_{c\bar c}^{(1)}  + \gamma_i(\Phi_{Q\bar{Q}{\rm F}}),\,
\end{align}
while the complex coefficients ${\cal C}^{[\gamma_i]}_\ccbar$ are constructed numerically via the colour-decomposed scattering amplitudes from \OpenLoops{}.
We note that \(\gamma_i\) and \({\cal C}^{[\gamma_i]}_{\ccbar}\) have the same structure as in the \(Q\bar{Q}\) case after adapting the process-dependent tree-level matrix element. 
We refer to the appendix of \citere{Mazzitelli:2021mmm} for their explicit expressions. 

The luminosity ${\cal L}_{c\bar{c}} $ in \eqn{eq:minnloQQFmf} reads as
\begin{align}
	{\cal L}_{c\bar{c}} (\Phi_{Q\bar{Q}{\rm F}},\pt)\equiv \frac{|M^{(0)}_{c\bar{c}}|^2}{2 m_{Q\bar{Q}{\rm F}}^2}\sum_{i,j}\left[\Tr(\tilde{\mathbf H}_{c\bar{c}}{\mathbf D})\, \,(\tilde{C}_{ci}\otimes f_i)
  \,(\tilde{C}_{\bar{c} j}\otimes f_j)\right]_\phi \,.	\label{eq:luminosityQQF}
\end{align}
The product of $\Tr(\tilde{\mathbf H}_{c\bar{c}}{\mathbf D})$ and $(\tilde{C}_{ci}\otimes f_i)(\tilde{C}_{\bar{c} j}\otimes f_j)$, when averaged over the azimuthal angle, involves an implicit tensor contraction. This leads to a richer structure of azimuthal correlations, encoded in the $G$ functions, as discussed in \citere{Catani:2010pd}. For $Q\bar{Q}$ production, the contributions proportional to $\langle M_{gg}^{(0)} | \mathbf{D}^{(1)} | M_{gg}^{(0)}\rangle \times  G^{(1)}$ are analytically known. For the more general $Q\bar{Q}$F case, we extract them through a numerical integration over the azimuthal angle in $b$-space.

We recall that the definition of the coefficients in the Sudakov radiator $\tilde{S}$, the collinear coefficient functions $\tilde{C}_{ci}$, and 
the hard-virtual function 
\begin{align}
\tilde H_{c\bar c}\equiv \Tr(\mathbf{\tilde H}_{c\bar c})  = 1 + \frac{\alpha_s}{2\pi} H_{c\bar c}^{(1)} + \frac{\alpha_s^2}{(2\pi)^2} \tilde H_{c\bar c}^{(2)} + \mathcal{O}(\alpha_s^3),	\label{eq:Hsdefinition}
\end{align}
 as used in the previous equations, receive additional shifts within the \minnlo{} method,
indicated by the tilde above the symbols, which has been originally derived in \citere{Monni:2019whf}. For completeness, we provide these
shifts here as well:
\begin{align}
	B_{c\bar c}^{(2)} &\rightarrow  \tilde B_{c\bar c}^{(2)} = \hat B_{c\bar c}^{(2)} + 2\zeta_3 (A^{(1)}_{c\bar c})^2+2\pi \beta_0 H_{c\bar c}^{(1)}\,,\\
	C_{ci}^{(2)}(z)  &\to \tilde{C}_{ci}^{(2)}(z) = C_{ci}^{(2)}(z) -2\zeta_3 A_\ccbar^{(1)} \hat{P}_{ci}^{(0)}(z)\,,\\
		H_{c\bar{c}}^{(2)}  &\to \tilde{H}_{c\bar{c}}^{(2)}  = H _{c\bar{c}}^{(2)} - 2\zeta_3 A_\ccbar^{(1)} B_\ccbar^{(1)}\,.
\end{align}

The derivation of the modified \POWHEG{} \(\bar{B}\) function is now straightforward, thanks to the structure of the cross section in \eqn{eq:minnloQQFmf} that corresponds to a sum of terms each of which resembling the structure of the colour-singlet case, albeit with modified resummation coefficients. Therefore, the remaining steps in the derivation simply follow the same approach that was discussed in \sct{sec:minnlops}.

As a final remark, we note that obtaining the correct result for the IR-regulated amplitudes $H^{(n)}_{c\bar c}$ is a non-trivial task even with the knowledge of the corresponding IR-divergent counterparts.
This is due to the fact that the subtraction operator, and in particular its finite piece, needs to be adequately defined in order to obtain the correct N$^n$LO normalisation.
In the case of (associated) heavy-quark production, this operator receives contributions from soft emissions connecting the four hard partons. These soft-parton contributions have been computed for the case of heavy-quark production in \citere{Catani:2023tby}, and more recently have been extended to the general kinematics needed for $Q\bar{Q}$F processes \cite{inprep:shark}.

\subsection{Adaptation for Yukawa-induced processes}
\label{sec:yuk}
We now examine the scale dependence of the \minnlo{} formulae and coefficients when incorporating an overall $\overline{\rm MS}$-renormalised Yukawa coupling. In appendix D of \citere{Monni:2019whf}, the scale dependence in the original \minnlo{} framework was derived for cases where the Born-level process already involves the strong coupling constant to some power. However, for Higgs production in association with a bottom-quark pair in the 4FS, the cross section at Born level includes two overall powers of $\alpha_s$ and, in addition, the bottom-Yukawa coupling. To address this more general case, we have provided in \citere{Biello:2024vdh} all necessary formulae for a process with the following leading-order (LO) coupling structure:
\begin{align} \sigma_{\rm LO} \sim \alpha_s^{n_B} y_b^{m_B}\,,
\end{align} 
where both the strong coupling $\alpha_s$ and the bottom-Yukawa coupling $y_b$ appear with general powers, denoted as $n_B$ and $m_B$, respectively. Thus, in the case of Higgs production in association with bottom quarks it is $n_B = m_B = 2$.

The bottom-quark Yukawa coupling is defined as
\begin{align} 
\label{eq:Yuk} 
y_b = \frac{m_b}{v}\,,
\end{align} 
where $m_b$ is the bottom quark mass, and $v$ is the vacuum expectation value of the Higgs field. Given that the natural scale of the Yukawa coupling is much larger than the bottom-quark mass (typically around the Higgs mass) it is important to use the $\overline{\text{MS}}$ scheme. This scheme introduces a renormalisation scale $\muRy$ for the mass of the Yukawa coupling, which can be set appropriately. In the following, we review the relevant \minnlo{} formulae to implement the dependence on the strong coupling and the Yukawa coupling independently. For a detailed derivation, we refer to \citere{Biello:2024vdh}.

In the \minnlo{} framework, the scale-compensating terms arising from the variation of the overall Born couplings are implemented at the level of the hard-virtual coefficient function. By explicitly introducing the scales $\muRb$ and $\muRy$, the squared hard-virtual matrix element introduced in  \eqn{eq:Hsdefinition} can be written as
\begin{align}
\begin{split}
\tilde  H_{c\bar c} \equiv |\langle \mathcal{R}_{c\bar c} | \mathcal{R}_{c\bar c}  \rangle|^2 &=|{M}_\ccbar^{(0)}(\mQQF,\mQQF)|^2\,\bigg(1+\frac{\alpha_s(\pt{})}{2\pi}H_\ccbar^{(1)}+\frac{\alpha_s^2(\pt)}{(2\pi)^2}\tilde H_\ccbar^{(2)}\bigg) \\
&=|{M}_\ccbar^{(0)}(\muRb,\muRy)|^2\, 
\bigg(1+\frac{\alpha_s(\muR)}{2\pi} 
H_\ccbar^{(1)}(\KR,\tfrac{\muRb}{\mQQF},\tfrac{\muRy}{\mQQF})  
\\&
+\frac{\alpha_s^2(\muR)}{(2\pi)^2} 
 \tilde H_\ccbar^{(2)}(\KR,\tfrac{\muRb}{\mQQF},\tfrac{\muRy}{\mQQF})\bigg)+{\cal O}(\alpha_s^3)\,. \label{Hwithscales}
    \end{split}
\end{align}
Here, ${M}_\ccbar^{(0)}(\muRb, \muRy)$ represents the tree-level amplitude with the strong and the Yukawa coupling evaluated at $\muRb$ and $\muRy$, respectively, and note that ${M}_\ccbar^{(0)}\equiv \mathcal{R}_{c\bar c}$.
Additionally, we introduce a generic symbol $\muR$ in the second and third line of \eqn{Hwithscales} for the renormalisation scale of the extra powers of the strong coupling in the expansion of the hard function, which is set to $\muR=K_R\,\pt{}$ according to the \minnlo{} prescription.

Using the identity
 \begin{align}
  |{M}_\ccbar^{(0)}(\mQQF,\mQQF)|^2&=|{M}_\ccbar^{(0)}(\muRb,\muRy)|^2 
  \frac{\alpha_s^{n_B}(\mQQF)y^{{m_B}}_b(\mQQF)}{\alpha_s^{n_B}(\muRb) y^{m_B}(\muRy)},
\end{align}   
when incorporating the renormalisation group flow of the strong and the Yukawa coupling, the logarithmic scale-compensating terms can be absorbed into the hard-virtual coefficient function. It yields
\allowdisplaybreaks
\begin{align}
& H_\ccbar^{(1)}(\KR,\tfrac{\muRb}{m_{Q\bar{Q}{\rm F}}},\tfrac{\muRy}{\mQQF})=\,H_\ccbar^{(1)}+ n_B 2\pi\beta_0 \log\frac{(\muRb)^2}{\mQQF^2}+m_B \gamma_1 \log \frac{(\muRy)^2}{\mQQF^2}, \label{H1yuk}\\
& \tilde H_\ccbar^{(2)}(\KR,\tfrac{\muRb}{\mQQF},\tfrac{\muRy}{\mQQF})=\,\tilde H_\ccbar^{(2)}+ \left(2\pi\beta_0 \ln K_R^2+n_B 2\pi\beta_0 \log\frac{(\muRb)^2}{\mQQF^2}\right.\nonumber\\
 &\hspace{1cm}\left.+m_B \gamma_1 \log \frac{(\muRy)^2}{\mQQF^2}\right)H_\ccbar^{(1)} +n_B 4 \pi^2 \beta_1  \log \frac{(\muRb)^2}{\mQQF^2} 
 \nonumber\\	&\hspace{1cm}
  +\frac{1}{2} n_B(n_B-1) 4 \pi^2 \beta_0^2  \log^2 \frac{(\muRb)^2}{\mQQF^2} +n_B 4\pi^2 \beta_0^2 \log\frac{(\muRb)^2}{\mQQF^2} \log \KR^2 
  \nonumber\\& \hspace{1cm}
  + m_B \gamma_2 \log\frac{(\muRy)^2}{\mQQF^2} 
  +n_B 2\pi\beta_0 m_B \gamma_1\log \frac{(\muRy)^2}{\mQQF^2}\log\frac{(\muRb)^2}{\mQQF^2}    \nonumber  \\ &\hspace{1cm}
   {- m_B \pi \beta_0  \gamma_1 \log^2 \frac{(\muRy)^2}{\mQQF^2}} 
  + \frac{1}{2} m_B^2 \gamma_1^2 \log^2 \frac{(\muRy)^2}{\mQQF^2} 
  {+ m_B 2\pi\beta_0 \gamma_1 \log \frac{(\muRy)^2}{\mQQF^2} \log \KR^2}. \label{H2yuk}
\end{align}
Here, we have used $\beta_{0,1}$ and  $\gamma_{1,2}$ as first- and second-order coefficients of the QCD $\beta$ function and the anomalous dimension that governs the mass evolution, respectively. They admit the following perturbative expansion in $\alpha_s(\mu)$,
\begin{align}
&\beta(\alpha_s(\mu)) = - \sum_{r=0}^\infty \beta_r \left( \alpha_s(\mu) \right)^{r+2},   &&\text{ with } \quad \beta_0=\frac{33-2 n_f}{12\pi}  \quad \text{and} \quad \beta_1=\frac{153-19 n_f}{24\pi^2}\,,\\
&\gamma(\alpha_s(\mu)) =\sum_{r=1}^\infty \gamma_r \left(\frac{\alpha_s(\mu)}{2\pi}\right)^r, &&\text{ with } \quad \gamma_1=2 \quad \text{and} \quad \gamma_2=\frac{101}{6}-\frac{5}{9} n_f\,.
\end{align}
In our calculation, we set the number of light quark flavours \( n_f = 4 \).

As a result of the modification of $H^{(1)}$, the $B^{(2)}$ coefficient in the Sudakov factor also receives a $\muRb$ and $\muRy$ dependence. For completeness, we also provide the standard $\muR$ dependence of the coefficients in the Sudakov factor
\begin{align}
&A_\ccbar^{(2)}(\KR) =A_\ccbar^{(2)}+(2\pi \beta_0) A_\ccbar^{(1)} \log \KR^2 \,, \\ 
&\tilde B_\ccbar^{(2)}(\KR,\tfrac{\muRb}{\mQQF},\tfrac{\muRy}{\mQQF}) = \tilde B_\ccbar^{(2)}+(2\pi \beta_0)B_\ccbar^{(1)} \log \KR^2
+n_B (2\pi \beta_0)^2 \log \frac{(\muRb)^2}{\mQQF^2} \nonumber \\ 
&\hspace{1cm}+m_B 2 \pi \beta_0 \gamma_1 \log \frac{(\muRy)^2}{\mQQF^2}\,.
\end{align}
In \citere{Mazzitelli:2021mmm}, a resummation scale $Q=\KQ \mQQF$ was introduced in the modified logarithm, which controls the transition from the small to the large transverse-momentum region by gradually turning off resummation effects at large transverse momenta. Since there is an interplay with the Yukawa coupling scale $\muRy$, we present the full scale dependence of the hard-virtual coefficient function with respect to $\KQ$, $\muR$, $\muRb$, and $\muRy$ below.
The resummation-scale dependence is derived by splitting the integral in the Sudakov into two parts (one from $\pt$ to $Q$ and one from $Q$ to $\mQQF$), expanding the second part in $\alpha_s(\KR/\KQ\,\pt)$, absorbing logarithmic terms into $\tilde B^{(2)}$ and non-logarithmic terms into $H$ \cite{Mazzitelli:2021mmm}. In this case, the scale of the strong coupling in the expansion of the hard-virtual function is adjusted as
\begin{align}
 \muR=\KR \pt \rightarrow \muR=\frac{\KR}{\KQ}\pt\,.
\end{align}
The full-scale dependence of the expansion coefficients of $H$ is given by
 \begin{align}
&H^{(1)}_\ccbar(\KR,\tfrac{\muRb}{\mQQF},\tfrac{\muRy}{\mQQF},\KQ)= H_\ccbar^{(1)}(\KR,\tfrac{\muRb}{\mQQF},\tfrac{\muRy}{\mQQF})
\nonumber \\
&
+\bigg(-\frac{A_\ccbar^{(1)}}{2}\log\KQ^2+B_\ccbar^{(1)}\bigg) \log \KQ^2,\\
&\tilde H_\ccbar^{(2)}(\KR,\tfrac{\muRb}{\mQQF},\tfrac{\muRy}{\mQQF},\KQ)= \tilde H_\ccbar^{(2)}(\KR,\tfrac{\muRb}{\mQQF},\tfrac{\muRy}{\mQQF})+\frac{(A_\ccbar^{(1)})^2}{8}\log^4 \KQ^2 
\nonumber \\&
-\bigg(\frac{A_\ccbar^{(1)}B_\ccbar^{(1)}}{2}+\pi\beta_0\frac{A_\ccbar^{(1)}}{3}\bigg)\log^3 \KQ^2 
+\bigg(-\frac{A_\ccbar^{(2)}(\KR)}{2}+\frac{(B_\ccbar^{(1)})^2}{2}+\pi\beta_0 B_\ccbar^{(1)} 
\nonumber \\&
-n_B \pi \beta_0 A_\ccbar^{(1)} \log \frac{(\muRb)^2}{\mQQF^2} -\frac{1}{2}m_B \gamma_1 A_\ccbar^{(1)} \log \frac{(\muRy)^2}{\mQQF^2} \bigg)\log^2 \KQ^2 
\nonumber \\&
+\bigg(\tilde B_\ccbar^{(2)}(\KR,\tfrac{\muRb}{\mQQF},\tfrac{\muRy}{\mQQF})+2 n_B \pi \beta_0 B_\ccbar^{(1)} \log \frac{(\muRb)^2}{\mQQF^2}+m_B \gamma_1 B_\ccbar^{(1)} \log \frac{(\muRy)^2}{\mQQF^2}\bigg) \log \KQ^2  \nonumber \\&
+ \bigg( B_\ccbar^{(1)} \log \KQ^2 -\frac{A_\ccbar^{(1)}}{2}\log^2 \KQ^2 -2\pi\beta_0 \log \KQ^2\bigg) H_\ccbar^{(1)}(\KR,\tfrac{\muRb}{\mQQF},\tfrac{\muRy}{\mQQF}).
\end{align}   
We refrain from discussing the factorisation-scale ($\muF = \KF\pt$) dependence, which is absorbed into the collinear coefficient functions and has no direct connection with $\muRy$. For the detailed formulae see \citere{Mazzitelli:2021mmm}.

The invariant mass $\mQQF$ refers to the invariant mass of the \bbH{} system, denoted as $m_{\bbH}$ for this process. To evaluate the theoretical uncertainty of our \minnlo{} predictions, we can vary $\muR$, $\muRb$, and $\muRy$ around their central values, either simultaneously by a common factor or independently. Our default choice and its impact on the \bbH{} process in the 4FS will be discussed in detail in \sct{sec:pheno}.

\section{Approximation of the two-loop amplitude}
\subsection{Massification procedure} \label{sec:massification}
The process-dependent component contributing to the hard-virtual coefficient function is the finite remainder up to two loops. The one-loop amplitude, which enters $H^{(1)}_\ccbar$, and the squared one-loop amplitude, which enters $\tilde{H}^{(2)}_\ccbar$, are obtained from \OpenLoops{}. While tree-level and one-loop contributions are computed exactly, only the two-loop finite remainder, which enters $\tilde{H}^{(2)}_\ccbar$, is calculated using an approximation, since the calculation
of the exact two-loop amplitude with massive bottom quarks is well beyond the current technology for five-point two-loop amplitudes. 
Instead, we employ the small bottom-mass limit, which captures
all logarithmically enhanced and constant terms, while neglecting power corrections in $m_b$, as follows:
\begin{align}
  \frac{2 \text{Re} \langle \mathcal{R}^{(0)}_{c\bar c}| \mathcal{R}^{(2)}_{c\bar c} \rangle }{\langle \mathcal{R}_{c\bar c}^{(0)}| \mathcal{R}_{c\bar c}^{(0)}\rangle } = \sum_{i=0}^4 \kappa_{c\bar c, i} \log^i \left(\frac{m_b}{\muR}\right) +
  \frac{2 \text{Re} \langle \mathcal{R}^{(0)}_{0,c\bar c}| \mathcal{R}^{(2)}_{0,c\bar c} \rangle }{\langle \mathcal{R}_{0,c\bar c}^{(0)}| \mathcal{R}_{0,c\bar c}^{(0)}\rangle }+\mathcal{O}\left(\frac{m_b}{\muQ}\right).	\label{eq:kappamass}
\end{align}
Given that the bottom mass is generally much smaller than the typical scale of the $b\bar bH$ process, this should serve as an excellent approximation.
Here $\mathcal{R}_{0,c\bar c}^{(i)}$ is the $i$-th order coefficient in an expansion in $\alpha_s/(2\pi)$ of the finite remainder of the $c\bar c\to b\bar bH$ 
amplitude with massless bottom quarks, $\muR$ is the renormalisation scale and $\muQ$ is a characteristic hard scale of the process. 
The process-dependent coefficients $\kappa_{c\bar{c},i}$ are derived via a massification procedure. This approach connects the IR collinear poles in the massless amplitudes to logarithmic $m_b$-dependent terms in the massive amplitudes.

The first massification of a massless amplitude was performed in the context of QED corrections for the Bhabha scattering in \citere{Penin:2005kf}. The procedure was extended for non-abelian gauge theories in \citere{Mitov:2006xs}. The derivation of the massification technique relies on the factorisation properties of QCD amplitudes. The un-renormalised amplitude with massless bottom quarks can be decomposed in colour space as,
\begin{align}
  |\Mdiv_{0,\ccbar}\rangle = \mathcal{J}_{0,\ccbar}\left(\frac{\muQ^2}{\mu^2}, \alpha_s(\mu^2),\epsilon \right) \mathcal{S}_{0,\ccbar}\left( \left\{\tilde p_i \right\}, \frac{\muQ^2}{\mu^2},\alpha_s(\mu^2),\epsilon \right) |\mathcal{H}_{0,\ccbar}\rangle. \label{eq:masslessfactorisation}
\end{align}
$\Mdiv_0$ denotes the amplitude before removing the IR divergences through the operator $\mathbf{Z}_0$ in the minimal way with massless heavy-flavour quarks~\cite{Becher:2009cu,Becher:2009qa},
\begin{align}
  |\mathcal{R}^{}_{0,\ccbar}\rangle = \mathbf{Z}^{-1}_{0,\ccbar}(\{\tilde p_i\},\mu,\epsilon) |\Mdiv_{0,\ccbar}\rangle\,,
\end{align}
where we introduce a set of massless momenta $\{\tilde p_i\}$. The phase-space point $\{p\}$ generated in the code with massive bottom quarks must be mapped into a set of momenta with bottom quarks in the massless shell. This mapping is arbitrary, and its choice is beyond the accuracy in the small-mass limit. However, care must be taken with the mapping to avoid infrared divergent regions of phase space with massless bottom quarks that could compromise the accuracy of the approximation. The mappings adopted in this work are discussed in appendix~\ref{sec:massif_mapping}. In \eqn{eq:masslessfactorisation} $\mathcal{J}_0$ is the massless jet functions that capture the collinear divergences, $\mathcal{S}_0$ is the soft function that encodes the soft singularities and depends on the momenta of the external partons $\left\{\tilde p_i\right\}$, $\mathcal{H}_{0}$ encodes the short-distance process-dependent dynamics. Here, \(\muQ^2\) represents the hard scale of the process, which is of the order of the invariant mass of the partonic event, and \(\epsilon\) denotes the dimensional regulator.

The key idea of the massification procedure is to consider the massive amplitude in the small-mass limit, $m_b^2\ll \muQ^2$, and connect the logarithmic terms in $m_b$ to the collinear poles of the massless amplitude. This matching can be understood as a change in the renormalisation scheme. In the small-mass limit, the massive amplitude obeys a similar decomposition,
\begin{align}
  |\Mdiv_{\ccbar}\rangle = \mathcal{J}_\ccbar \left(\frac{\muQ^2}{\mu^2}, \frac{m_b^2}{\mu^2},\alpha_s(\mu^2),\epsilon \right) \mathcal{S}_\ccbar \left( \left\{p_i \right\}, \frac{\muQ^2}{\mu^2},\alpha_s(\mu^2),\epsilon \right) |\mathcal{H}_{\ccbar}\rangle+\mathcal{O}\left(\frac{m_b}{\muQ}\right), \label{eq:massivefactorisation}
\end{align}
where $\mathcal{J}$, $\mathcal{S}$, and $\mathcal{H}$ are, respectively, the jet, soft and hard functions for massive bottom quarks. We stress that $\Mdiv_{\ccbar}$ is the amplitude before the subtraction of the IR divergences in the small-mass limit, performed in the minimal way via the operator $\mathbf{Z}_{m_b\ll\muQ,\ccbar}$ in order to obtain the finite remainder $|\mathcal{R}_\ccbar\rangle=\mathbf{Z}^{-1}_{m_b\ll\muQ,\ccbar}|\mathcal{M}_\ccbar\rangle$. There is a freedom in organising subleading soft terms into the jet and soft functions. The following requirement can completely fix this ambiguity,
\begin{align}
  \mathcal{J}_{\ccbar}\left(\frac{\muQ^2}{\mu^2}, \frac{m_b^2}{\mu^2},\alpha_s(\mu^2),\epsilon \right)&= \prod_{i=c,\bar c,b,\bar b} \mathcal{J}_{i}\left(\frac{\muQ^2}{\mu^2}, \frac{m_b^2}{\mu^2},\alpha_s(\mu^2),\epsilon \right)\\
  &=\prod_{i=c,\bar c,b,\bar b} \left[\mathcal{F}_{m_b,i}\left(\frac{\muQ^2}{\mu^2}, \frac{m_b^2}{\mu^2},\alpha_s(\mu^2),\epsilon \right)\right]^{\frac{1}{2}}. \label{eq:fixJ}
\end{align}
Here, $i$ runs over the entire set of coloured ingoing and outgoing asymptotic states: we have introduced the jet function $\mathcal{J}_i$ related to a specific leg. \(\mathcal{F}_{i}\) denotes the space-like form factor for a state \(i\), which spans over all possible states — such as a gluon, a light quark, or a heavy-flavor state — depending on the specific partonic interaction. A similar decomposition as in \eqn{eq:fixJ} can be done for the massless jet function in terms of massless form factors for the bottom-quark legs. The soft singularities are the same as in \neqn{eq:masslessfactorisation} and \eqref{eq:massivefactorisation}, while the jet function encodes all the mass dependence from quasi-collinear singularities. For this reason, the hard function $\mathcal{H}_\ccbar$ differs from the massless counterpart $\mathcal{H}_{0,\ccbar}$ only for power corrections in $m_b$ that are neglected in this approach. The previous observations naturally lead to a simple connection between the two amplitudes,
\begin{align}
  |\Mdiv_\ccbar\rangle=\left[\mathcal{Z}_{b}\left(\frac{m_b^2}{\mu^2},\alpha_s(\mu^2),\epsilon \right)\right]^{n_b/2}|\Mdiv_{0,\ccbar}\rangle+\mathcal{O}\left(\frac{m_b}{\muQ}\right), \label{eq:MitovMoch}
\end{align}
where $n_b=2$ is the number of quark legs that must be promoted from massless to massive lines. In the above equation, we have introduced the massification factor $\mathcal{Z}_{b}$ which is the ratio of the bottom-quark form factors in massive and massless cases,
\begin{align}
  \mathcal{Z}_{b}\left(\frac{m_b^2}{\mu^2},\alpha_s(\mu^2),\epsilon \right)=\frac{\mathcal{F}_{m_b,b}\left(\frac{\muQ^2}{\mu^2}, \frac{m_b^2}{\mu^2},\alpha_s(\mu^2),\epsilon \right)}{\mathcal{F}_{0,b}\left(\frac{\muQ^2}{\mu^2},\alpha_s(\mu^2),\epsilon \right)}. \label{eq:defZ}
\end{align}
At its current level of development, the approximation technique enables the massification of internal bottom-quark loops within the two-loop amplitudes. The master formula was firstly derived for QED corrections in Bhabha scattering~\cite{Becher:2007cu}, and recently, the needed ingredients for QCD amplitudes are computed~\cite{Wang:2023qbf}. The massive form factors in \eqn{eq:defZ} must be computed by including internal massive quark loops. However, a non-trivial soft function appears once vacuum polarisation diagrams with massive particles are considered at NNLO. An extension of the factorised formula in \eqn{eq:MitovMoch} is required, as pointed out in \cite{Becher:2007cu},
\begin{align}
  |\Mdiv_\ccbar\rangle=&\prod_{i=c,\bar c, b, \bar b} \left[\mathcal{Z}_{i}\left(\frac{m_b^2}{\mu^2},\alpha_s^{(n_f)}(\mu^2),\epsilon \right)\right]^{1/2} \nonumber \\
  &\,\mathbf{S}\left(\frac{m_b^2}{\mu^2},\left\{\tilde p_i\right\}, \alpha_s^{(n_f)}(\mu^2),\epsilon\right)|\Mdiv_{0,\ccbar}\rangle+\mathcal{O}\left(\frac{m_b}{\muQ}\right)\,. \label{eq:BecherMelnikov}
\end{align}
We stress the presence of a massification factor for each external parton due to the internal massive bottom-quark loops that affect all the form factors. The strong coupling is renormalised according to the total number of flavours, including the bottom one. As the most recent applications of this procedure, we want to apply the approximation by connecting the finite remainder in the small-mass limit with the massless IR-finite counterpart. It yields
\begin{align}
  |\mathcal{R}_\ccbar \rangle =&\, \mathbf{Z}_{m_b\ll\mu_q,\ccbar}(\{p_i\},\mu,\epsilon) \prod_{i} \left[\mathcal{Z}_{i}\left(\frac{m_b^2}{\mu^2},\alpha_s^{(n_f)}(\mu^2),\epsilon \right)\right]^{1/2}\mathbf{S}\left(\frac{m_b^2}{\mu^2},\left\{\tilde p_i\right\}, \alpha_s^{(n_f)}(\mu^2),\epsilon\right)  \nonumber \\
  &\times \mathbf{Z}^{-1}_{0,\ccbar}(\{\tilde p_i\},\mu,\epsilon) |\mathcal{R}_{0,\ccbar} \rangle +\mathcal{O}\left(\frac{m_b}{\muQ}\right)\\
  =&\, \mathcal{\bar F}_\ccbar \left(\frac{m_b^2}{\mu^2},\alpha_s^{(n_f)}(\mu^2) \right) \mathbf{\bar S}\left(\frac{m_b^2}{\mu^2},\left\{\tilde p_i\right\}, \alpha_s^{(n_f)}(\mu^2)\right)|\mathcal{R}_{0,\ccbar}\rangle +\mathcal{O}\left(\frac{m_b}{\muQ}\right)\,. \label{eq:masterformulagenmass}
\end{align}
Here we have introduced the function $\mathcal{\bar F}_\ccbar$ and the matrix $\mathbf{\bar S}$ which are free from $\epsilon$ poles, and they admit a perturbative expansion in the strong coupling constant $\alpha_s^{(n_f)}$,
\begin{align}
	\mathcal{\bar F}_\ccbar \left(\frac{m_b^2}{\mu^2},\alpha_s^{(n_f)}(\mu^2) \right)=1+ \left(\frac{\alpha_s^{(n_f)}}{2\pi}\right) \mathcal{\bar F}^{(1)}_\ccbar+\left(\frac{\alpha_s^{(n_f)}}{2\pi}\right)^2	\mathcal{\bar F}^{(2)}_\ccbar+ \mathcal{O}(\alpha_s^3). \label{eq:Fbar}
\end{align}
We stress that the soft function,
\begin{align}
	\mathbf{S}\left(\frac{m^2}{\mu^2},\left\{\tilde p_i\right\}, \alpha_s^{(n_f)}(\mu^2),\epsilon\right)=1+\left(\frac{\alpha_s^{(n_f)}}{2\pi}\right)^2 \mathbf{C}_d \mathcal{S}^{(2)} + \mathcal{O}(\alpha_s^3)\,, \label{eq:Sdef}
\end{align}
acts as an operator in the colour space and depends on the standard dipole,
\begin{align}
  \mathbf{C}_d=-\sum_{(i,j)}\frac{\mathbf{T}_i\cdot \mathbf{T}_j}{2}\log\left(\frac{-(\tilde p_i+\tilde p_j)^2}{\mu^2}\right).
\end{align}
Here, the sum runs over all the pairs of external partons. The coefficient $\mathcal{S}^{(2)}$ is reported in appendix~\ref{sec:massif_ingredients} together with massification coefficient factors $\mathcal{\bar F}^{(i)}_\ccbar$.

The starting point of this procedure constitutes amplitudes with massless bottom quarks in the loops. Therefore, we must match the massless results into a massive 4FS calculation. For this reason, it is required to apply a finite renormalisation shift for the strong coupling to match the decoupling scheme,
\begin{align}
  \alpha_s^{(n_f)}=\alpha_s^{(n_l)}\left\{ 1+\left(\frac{\alpha_s^{(n_l)}}{2\pi}\right)\frac{2}{3}\ell_b +\left(\frac{\alpha_s^{(n_l)}}{2\pi}\right)^2 \left[\frac{4}{9}\ell_b^2 +\frac{19}{3}\ell_b +\frac{7}{6}\right] +\mathcal{O}(\alpha_s^3)\right\}
\end{align}
with $\ell_b=\log(m_b/\muR)$ and $n_l=n_f-1$ is the number of light fermions in 4FS. In addition, we need to apply a similar decoupling relation to the Yukawa coupling,
\begin{align}
	y_b^{(n_f)}=y_b^{(n_l)}\left\{ 1- \left(\frac{\alpha_s^{(n_l)}}{2\pi}\right)^2 \left[ \frac{4}{3}\ell_b^2 + \frac{10}{9}\ell_b+\frac{89}{108}  \right] +\mathcal{O}(\alpha_s^3)  \right\}.
\end{align}
The first non-trivial shift of the Yukawa vertex starts at $\mathcal{O}(\alpha_s^2)$ as it is the first order in $\alpha_s$ where the internal quark loops start to affect the renormalisation of this coupling. By applying the decoupling shift and expanding \eqn{eq:masterformulagenmass}, we obtain the coefficients $\kappa_{\ccbar, i}$ of \refeq{eq:kappamass} in terms of the massless finite remainders at tree and one-loop level. The non-logarithmic contribution includes the two-loop finite remainder, as explicitly indicated in \eqn{eq:kappamass}. In the next section, we discuss the computational aspects of this process-dependent two-loop contribution.

The massification procedure outlined here has already been employed in other processes involving heavy quarks at the Born level. The approximation based on \eqn{eq:MitovMoch} was used in \citere{Buonocore:2022pqq}. The first application of the approximation, directly applied to the finite remainder and based on \eqn{eq:masterformulagenmass}, is described in \citere{Mazzitelli:2024ura}. More recently, this refined approach has been used to estimate the double-virtual contribution in associated $t\bar t H$ production in the small top-mass limit~\cite{Devoto:2024nhl}.

\subsection{Numerical implementation of the massless two-loop amplitude}
We have implemented a \texttt{C++} library for a fast numerical evaluation of two-loop virtual corrections in the leading-colour approximation with massless bottom quarks.~\footnote{In the very final stages of this work, the full-colour amplitude with massless bottom quarks was presented in~\citere{Badger:2024awe}. The implementation of subleading contributions in our generator is disccused in Appendix~\ref{app:FC}.} We have used the analytic results of \citere{Badger:2021ega}, where the authors provided the finite remainder $\mathcal{F}^{(2)}$,
\begin{align}
 |\mathcal{F}^{(2)}_{c\bar c} \rangle = (\mathbb{1}-\mathbf{I}_{0,\text{LC}}) | \mathcal{M}^{\textrm{IR-div},(2)}_{0,c\bar c, \text{LC}} \rangle, 
\end{align}
after the subtraction of the poles in terms of the Catani operator~\cite{Catani:1998bh} in the leading-colour approximation, $\mathbf{I}_{0,\text{LC}}$. The colour structure is trivial in this approximation, therefore the operator is proportional to the identity in the colour space. The library computes the finite remainder in the minimal subtraction scheme~\cite{Becher:2009cu,Becher:2009qa} as follows,
\begin{align}
  2 \text{Re} \langle \mathcal{R}^{(0)}_{c\bar c}| \mathcal{R}^{(2)}_{0,c\bar c} \rangle &= 2 \text{Re} \langle \mathcal{R}^{(0)}_{c\bar c} | \mathbf{Z}^{-1}_0 | \mathcal{M}^{\textrm{IR-div},(2)}_{0,c\bar c} \rangle \nonumber \\
  &=2 \text{Re} \langle \mathcal{R}^{(0)}_{c\bar c} | \mathbf{Z}^{-1}_{0,\text{LC}} | \mathcal{M}^{\textrm{IR-div},(2)}_{0,c\bar c, \text{LC}} \rangle + \mathcal{O}\left(\frac{1}{N_C}\right) \nonumber \\
  &=2 \text{Re} \langle \mathcal{R}^{(0)}_{c\bar c} | \mathbf{Z}^{-1}_{0,\text{LC}} (\mathbb{1}-\mathbf{I}_{0,\text{LC}})^{-1} | \mathcal{F}^{(2)}_{c\bar c} \rangle +\mathcal{O}\left(\frac{1}{N_C}\right).	\label{eq:masslessLCamp}
\end{align}
$| \mathcal{M}^{\textrm{IR-div},(2)}_{0,c\bar c, \text{LC}} \rangle $ is the leading-colour IR-unrenormalised amplitude, while the operators $\mathbf{Z}_{0,\text{LC}}$ extract the poles in the minimal way. We stress the importance of expanding the Catani operator up to $\mathcal{O}(\epsilon^2)$ in order to compute its inverse at $\mathcal{O}(\epsilon^0)$.

We perform a change of basis for the Master Integrals (MIs) to express the amplitudes in terms of Pentagon Functions~\cite{Chicherin:2021dyp}. The phase-space point in \POWHEG{} is passed to the two-loop library via a \texttt{Fortran-C++} interface. We evaluate the Mandelstam and momentum-twistor variables in the library. A crucial feature of our two-loop library for a stable numerical performance is the ability to evaluate amplitudes in a normalised phase-space point, where invariants are of order one. Instead of computing two-loop finite remainders for LHC-like phase-space points produced from \POWHEG\,, we compute them with rescaled momenta and then normalise the result with the Born amplitudes,
\begin{align}
  |\mathcal{F}^{(2)}( \{\tilde p_i\},\muR )\rangle  = \frac{\langle {M}^{(0)}( \{\tilde p_i\}) | {M}^{(0)}( \{\tilde p_i\} ) \rangle }{\langle {M}^{(0)}( \{\bar p_i\} )| {M}^{(0)}( \{\bar p_i\} ) \rangle } |\mathcal{F}^{(2)}( \{\bar p_i\},1 )\rangle ,\,\hspace{0.5cm}\,\bar p_i=\frac{\tilde p_i}{\muR}\,. \label{eq:rescmomenta}
\end{align}
Since the renormalisation scale $\muR$ is the invariant mass of the colour-singlet system, $m_{b\bar bH}$, we call the library for the calculation of amplitudes with momenta of $\mathcal{O}(1)$ instead of the typical LHC energy. Using rescaled momenta, all the kinematic variables, specifically momentum twistors, are of the same order of magnitude for generic phase space points in the library. We verified that both approaches give the same result for stable PS points, while we saw improvements for unstable points in the gluon channel. This shows a clear improvement in the numerical stability for the evaluation of amplitudes when using rescaled momenta. Other precautions are taken into account. For instance, we evaluate the coefficients in quadruple precision as default, using the \texttt{qd} library. On the other hand, the MIs are computed in double precision as the default setting. However, the library switches to quadruple precision for MIs when the gram determinant defined in eq.\,(2.6) of \citere{Chicherin:2021dyp} is positive for double Mandelstam variables or not in agreement with a direct evaluation in terms of momenta via the pseudo-scalar invariant defined in eq.\,(2.6) of \citere{Badger:2021ega}.

We have conducted cross-checks with an independent implementation described in \citere{Devoto:2024nhl}, using the Yukawa coupling renormalised in the on-shell scheme. After pointwise validation of the massless amplitudes in \eqn{eq:masslessLCamp} at selected random phase-space points for LHC energies in both the gluon and quark channels, we have successfully compared the two-loop finite remainder in the small-mass limit with the generalised approach outlined in \eqn{eq:masterformulagenmass}. Additional cross-checks have been performed using the public library described in~\citere{Badger:2024awe}, where we compared the numerical results of the full-colour amplitude in the large $N_C$ limit with those obtained from our leading-colour implementation.

\section{Results in the four-flavour scheme} \label{sec:pheno}
\subsection{Setup} \label{sec:setup}

We provide numerical predictions for Higgs boson production in association with a massive bottom-quark pair at 13 TeV centre-of-mass energy at the LHC. The Higgs boson is kept stable, except for \sct{sec:HH} where we include Higgs decays to photons in the narrow-width approximation. We set the mass of the Higgs boson to $m_H=125$\,GeV and use a Higgs width of $\Gamma_H=0$\,GeV when considering its decay. Since the calculation is carried out in the 4FS, we set the number of light quark flavours $n_f=4$ and renormalise the bottom quark mass in the on-shell scheme with $m_b^{\text{OS}}=4.92$\,GeV. By contrast, the bottom-Yukawa coupling $y_b$ is computed in the $\overline{\text{MS}}$ scheme,
which is derived from an input value $\hat m_b\equiv m_b(m_b)=4.18$\,GeV and evolved to its respective central hard scale, $\muRy$, via four-loop running, based on the solution of the Renormalisation Group Equation (RGE) \cite{Harlander:2003ai, Baikov_2014}.  We note that the bottom-Yukawa coupling strongly depends on the value of $\alpha_s(\hat m_b)$ used in the RGE evolution. We evaluate it using methods aligned with those adopted in modern PDF fits with $n_f=4$. We start from the 5FS value $\alpha_s(m_Z)=0.118$ and evolve it down to the bottom mass via $n_f=5$ running. At this scale, we apply the decoupling relation \cite{Vogt:2004ns} in order to obtain $\alpha_s(\hat m_b)$ in the 4FS. This value also corresponds to the boundary condition for evaluating the strong coupling in the 4FS at any other scale via a running with $n_f=4$. The scale variation of the bottom-Yukawa coupling is performed via a three-loop running. In our phenomenological analysis, the hard scale $\muRy$ is either fixed to the Higgs mass or to the following dynamically:
\begin{align}
  \muRy=\frac{H_T}{4}=\frac{1}{4} \sum_{i=H,b,\bar b} \sqrt{m^2(i)+\pt^2(i)},
\end{align}
where $m(i)$ and $\pt(i)$ denote the invariant mass and transverse momentum
of particle $i$, respectively. The strong-coupling factors at the Born level are always evaluated at $\muRb=H_T/4$,
which is an appropriate dynamical scale throughout the phase space, both at small and large transverse momenta.
We recall that contributions proportional to the top-Yukawa coupling are switched off in our calculation, since 
we focus on the gauge-invariant set of  contributions proportional to $y_b^2$
in this paper.

For the parton densities, we use the NNLO set of NNPDF4.0 \cite{NNPDF:2021njg} with four active flavours via the LHAPDF interface \cite{Buckley:2014ana} (LHAID=334300). The central factorisation and renormalisation scales are set according to the \minnlo{} method. Scale uncertainties are estimated using the envelope of the conventional 7-point variations, which involves varying the factors $\KR$ and $\KF$ independently by a factor of 2 with the constraint \mbox{$1/2 \leq \KR/\KF \leq 2$}. The scale variation of the Born couplings is synchronised with the variation of $\KR$. Moreover, we choose $\KQ=0.5$ for the resummation scale factor, while we have checked that $\KQ=0.25$ leads to similar results.

We use \PYTHIA{8} \cite{Bierlich:2022pfr} with the A14 tune for all parton-shower predictions. Unless stated otherwise, the Higgs boson is treated as a stable particle and the effects of hadronisation, multi-parton interactions (MPI) and QED radiation are kept off.

We define jets by clustering all partons (i.e.\ including bottom quarks) using the  anti-$k_t$ algorithm~\cite{Cacciari:2008gp} as implemented in {\tt FastJet} \cite{Cacciari:2011ma} using $R=0.4$. Jets are classified as bottom-flavoured jets ($b$-jets) if they contain at least one bottom-flavored quark/hadron and meet the following criteria for the transverse momentum and the pseudo-rapidity of the $b$-jet:
\begin{align}
p_{\text{T},b\text{-jet}}>30\,\text{GeV},\hspace{1cm}|\eta_{b\text{-jet}}|<2.4\,.	\label{eq:fidreg}
\end{align}
In the $1\,bj_{\text{EXP}}$ and $2\,bj_{\text{EXP}}$ categories of our analyses, a Higgs boson and at least one or two $b$-jets are required in the fiducial phase space, respectively. The definition of the $b$-jets used here is close to the experimental criteria and can be implemented theoretically as the bottom quarks are treated as massive particles. For massless bottom quarks, care must be taken to ensure the IR safety of the theoretical predictions, either by reshuffling
the bottom momenta into massive ones or by applying an appropriate algorithm
to define the jet flavour. A discussion on different flavour algorithms will follow in \sct{sec:bjetstudy}.

\subsection{Inclusive cross section}
\begin{table}[ht!]
  \vspace*{0.3ex}
  \begin{center}
	   \renewcommand{\arraystretch}{1.3}
    \begin{tabular}{|c|c|c|}
      \hline
      Generator ($\muRy$) & $\sigma_{\rm total}$ [pb] & ratio to NLO+PS \\
      \hline \hline
      4FS NLO+PS ($m_H$)\;& \; $0.354(6)_{-16\%}^{+20\%}$\,& 1.000 \\
      4FS \minlo{} ($m_H$)\;& \; $0.271(1)_{-27\%}^{+45\%}$\,& 0.765\\
      4FS \minnlo{} ($m_H$) \;& \; $0.466(0)_{-14\%}^{+16\%}$\, & 1.314 \\
      \hline
      4FS NLO+PS ($H_T/4$)  \;& \; $0.385(3)_{-14\%}^{+16\%}$\,& 1.000\\
      4FS \minlo{} ($H_T/4$)\;& \; $0.299(3)_{-27\%}^{+42\%}$\,& 0.777\\
      4FS \minnlo{} ($H_T/4$)  \;& \;  $0.494(6)_{-14\%}^{+16\%}$\,& 1.284\\
      \hline
      5FS NLO+PS ($m_H$)  \;& \; $0.645(5)_{-10\%}^{+11\%}$\,& 1.000\\
      5FS \minlo{} ($m_H$)\;& \; $0.571(1)_{-23\%}^{+17\%}$\,& 0.885 \\
      5FS \minnlo{} ($m_H$)  \;& \;  $0.509(1)_{-5.3\%}^{+2.9\%}$\,& 0.790\\
      \hline
    \end{tabular}
  \end{center}
  \vspace{-1em}
  \caption{
	  Total $b \bar b H$ cross section in the 4FS and 5FS at NLO+PS, and for \minlo{} and \minnlo{}. The scale in brackets indicates the choice for the central value of the renormalisation scale for the Yukawa coupling. The strong couplings at
	  Born level in 4FS are always evaluated at the dynamical scale $H_T/4$. \label{tab:XS}
    The quoted errors represent scale uncertainties, while the numbers in brackets are numerical uncertainties on the last digit. }
\end{table}

We begin the phenomenological analysis of the \bbH{} process by studying the total inclusive cross section in \tab{tab:XS} at NLO+PS (\POWHEG{}) and NNLO+PS (\minnlo{}) in the 4FS and the 5FS, where we consider two scale choices for the bottom-Yukawa coupling, $\muRy=m_H$ and $\muRy=H_T/4$ in the 4FS. To facilitate this analysis, we have developed a private NLO+PS generator in \POWHEGBOXRES{} for the $y_b^2$ contribution in the 4FS.\footnote{To ensure consistency, the NLO+PS implementation has been cross-checked against the public version in \POWHEGBOXVTWO{}~\cite{Jager:2015hka}.} The NLO+PS predictions are obtained using the same setup of the \minnlo{} as discussed earlier, except for the scale variation of the Yukawa coupling, which is performed using the two-loop running. The factorisation and renormalisation central scales are set to $H_T/4$ in the NLO+PS generator. Using the setup of \minnlo{}, we also provide \minlo{} predictions, which are formally NLO accurate for this observable, by turning off the term $(D-D_1-D_2)$ in \eqn{eq:bbar}.

Looking at  \tab{tab:XS}, we first notice that the \minlo{} result is significantly smaller than the NLO+PS cross section for both Yukawa scales and fails to provide an accurate prediction, as already noticed for $b\bar{b} \ell^+\ell^-$ production in \citere{Mazzitelli:2024ura}. This behaviour is due to a mis-cancellation of large logarithmic corrections in $m_b$. Indeed, \minlo{} results contain some NNLO corrections from real (double-real and real-virtual) radiation, but not the ones encoded in the ($D-D_1-D_2$) term, including the double-virtual contributions. 
The quasi-collinear logarithmic terms arise from the presence of the bottom quark mass, which acts as a regulator for both the real phase-space integration and the loop integration. These contributions are expected to largely cancel between the real and virtual amplitudes. This cancellation can be understood by examining the 5FS, where these logarithms manifest as $1/\epsilon$ poles, which are eliminated by the KLN theorem~\cite{Kinoshita:1962ur,Lee:1964is}. In the \minlo{} approach, the relative $\mathcal{O}(\alpha_s^2)$ contribution is incomplete because it includes only the real amplitudes. The associated logarithmic terms introduce a numerically significant negative effect, as seen from \tab{tab:XS}. We have checked that incorporating the logarithmic corrections in the double-virtual amplitudes — calculated using the massification procedure detailed in \ref{sec:massification} — restores the expected cancellation and yields a positive $\mathcal{O}(\alpha_s^2)$ correction. However, due to this unphysical effect, we have opted not to include the \minlo{} results in the remainder of this article.

Based on the \minnlo{} predictions in \tab{tab:XS}, the NNLO corrections increase the NLO cross section by $30\%$ for both Yukawa scales, making them essential for achieving precise predictions in the 4FS. Furthermore, the \minnlo{} prediction has a relatively small sensitivity to the considered scales of the bottom-Yukawa coupling, especially considering that its square is an overall factor to the cross section, highlighting a reduced dependence on scale choice at NNLO. 
For comparison, we also show 5FS predictions (i.e.\ the process
$b\bar b \rightarrow H+X$) in \tab{tab:XS}.\footnote{The NLO+PS 5FS results presented in \tab{tab:XS} are obtained using the same setup as detailed in \citere{Biello:2024vdh}, incorporating four-loop running to obtain the central bottom-Yukawa coupling and two-loop running for scale variations. Similarly, the \minnlo{} predictions in the 5FS are generated using the identical setup as described in \citere{Biello:2024vdh}}. The cross section in 5FS at NNLO+PS exhibits a smaller theoretical uncertainty from scale variation compared to the 4FS prediction, which provides a more conservative estimate. Different sources contribute to the distinct scale uncertainties of the two predictions, such as the presence of Born-level strong couplings and the sizable NNLO corrections in the massive calculation. The \minnlo{} results clearly demonstrate agreement between the two schemes within scale uncertainties. This highlights that the long-standing discrepancy between 4FS and 5FS predictions is resolved by the newly-computed NNLO corrections in the 4FS. A detailed comparison between the massive and massless schemes at differential level is presented in \sct{sec:4FSvs5FS}.

\subsection{Differential distributions}
We now focus on differential distributions and discuss different aspects of the results.
\subsubsection{Shower effects}
We start by comparing predictions obtained after including only the \POWHEG{} radiation, namely at Les-Houches-Event (\texttt{LHE}) level, with those obtained after showering with \PYTHIA{8} (\texttt{PY8}). Since we have found 
that observables inclusive over radiation show practically no effects from the shower,
we refrain from showing them here. Nevertheless, we would like to point out
that this is in contrast to the 5FS predictions presented in \citere{Biello:2024vdh},
where, in particular the Higgs transverse-momentum spectrum receives significant corrections from the parton shower, which are absent in the 4FS
predictions presented here, when employing the local dipole recoil 
in the parton shower \cite{Cabouat:2017rzi}. This stability in the 4FS
can be explained by the higher multiplicity present already 
at the Born level, which ensures that observables related to the 
$b\bar{b} H$ final state, like the Higgs transverse-momentum spectrum, are
genuinely NNLO accurate, which is not the case in the 5FS, where at large transverse momenta of the Higgs boson the predictions are 
effectively only NLO accurate.

\begin{figure*}[t]
  \begin{center}
    \includegraphics[width=.49\textwidth]{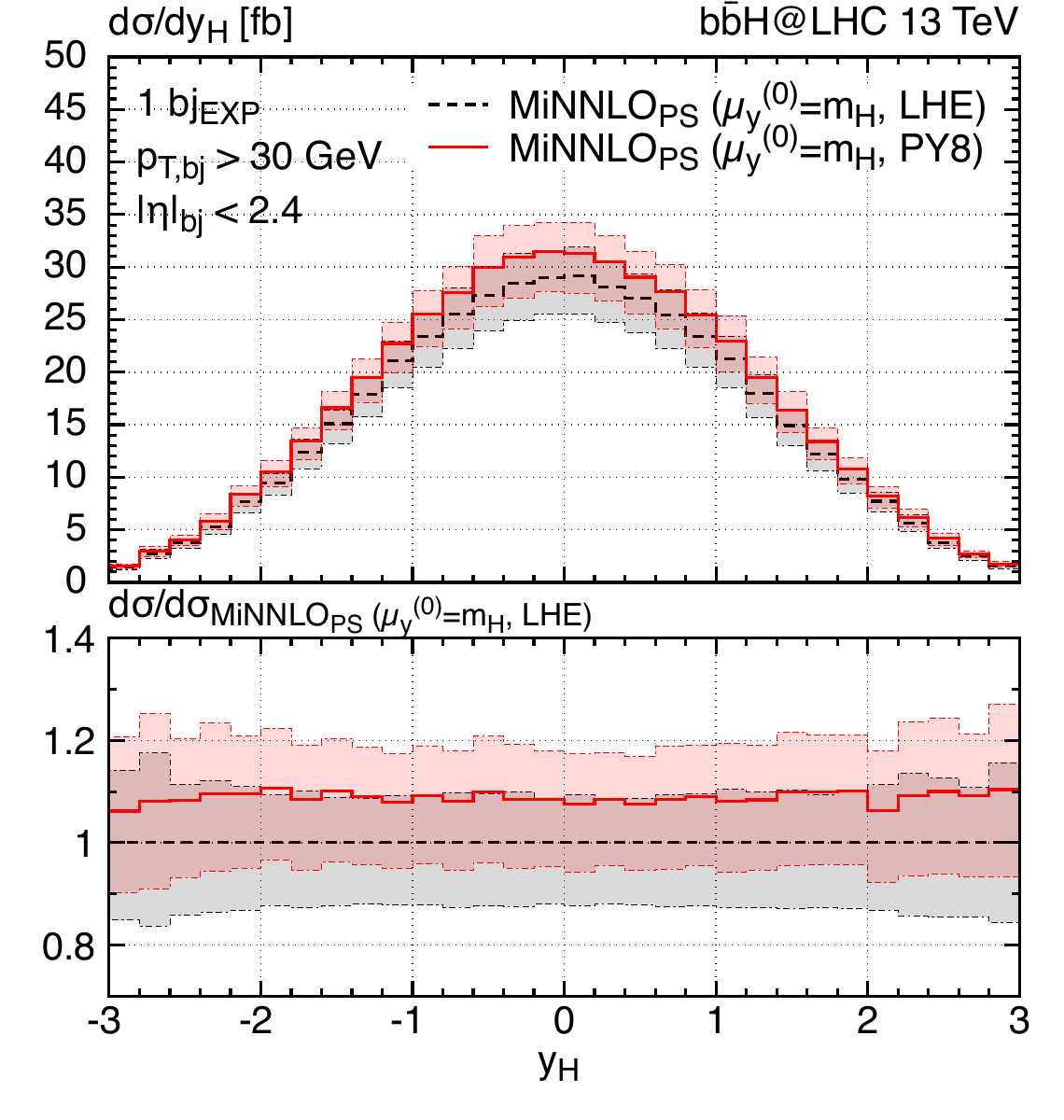}
     \includegraphics[width=.49\textwidth]{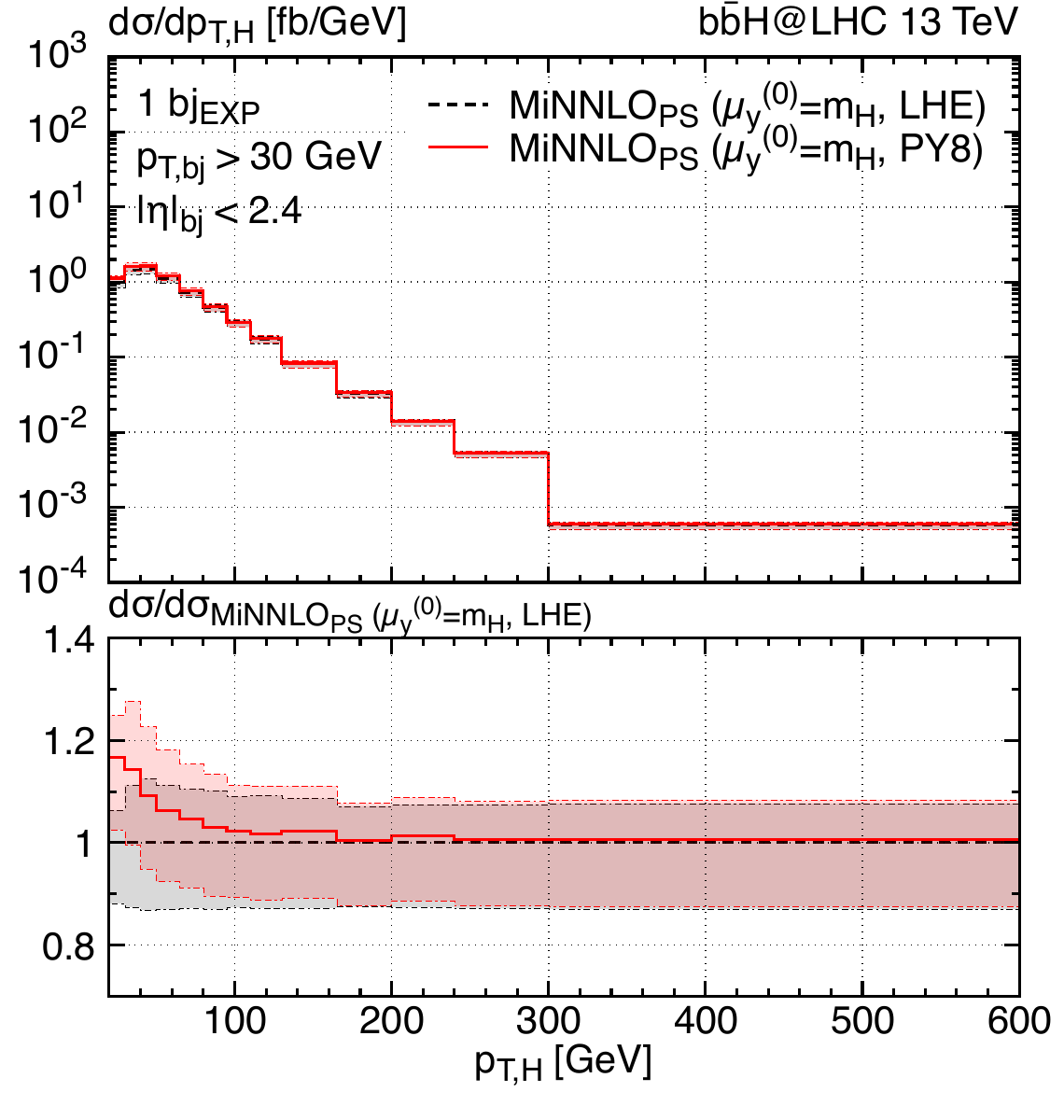}
	  \caption{Comparison of \minnlo{} predictions in the 4FS at LHE (black, dashed) and including \PYTHIA{8} parton shower (red, solid) 
	  for the Higgs rapidity (left plot) and Higgs transverse momentum (right plot) with at least one $b$-jet.}
    \label{fig:LHEvsPY8}
  \end{center}
\end{figure*}

The plots in \fig{fig:LHEvsPY8} show \minnlo{} predictions before and after parton shower, requiring at least one $b$-jet.
The parton shower increases the cross section for Higgs observables when at least one $b$-jet ($1\,bj_{\text{EXP}}$) is required. The corrections
are essentially flat in angular observables, like the Higgs rapidity ($\yh$) shown in the left plot of \fig{fig:LHEvsPY8}. In the Higgs transverse-momentum ($\pth$) spectrum in the right plot, on the other hand, the shower effects the spectrum only towards small $\pth$.
When the Higgs is produced with high transverse momentum, the recoiling bottom quarks are typically hard enough 
so that one hard $b$-jet is always present.
 In what fallows, all \minnlo{} predictions will be presented after matching with the parton shower.

\subsubsection{Impact of the scale choice for the bottom Yukawa}

\begin{figure*}[h!]
  \begin{center}
    \includegraphics[width=.49\textwidth]{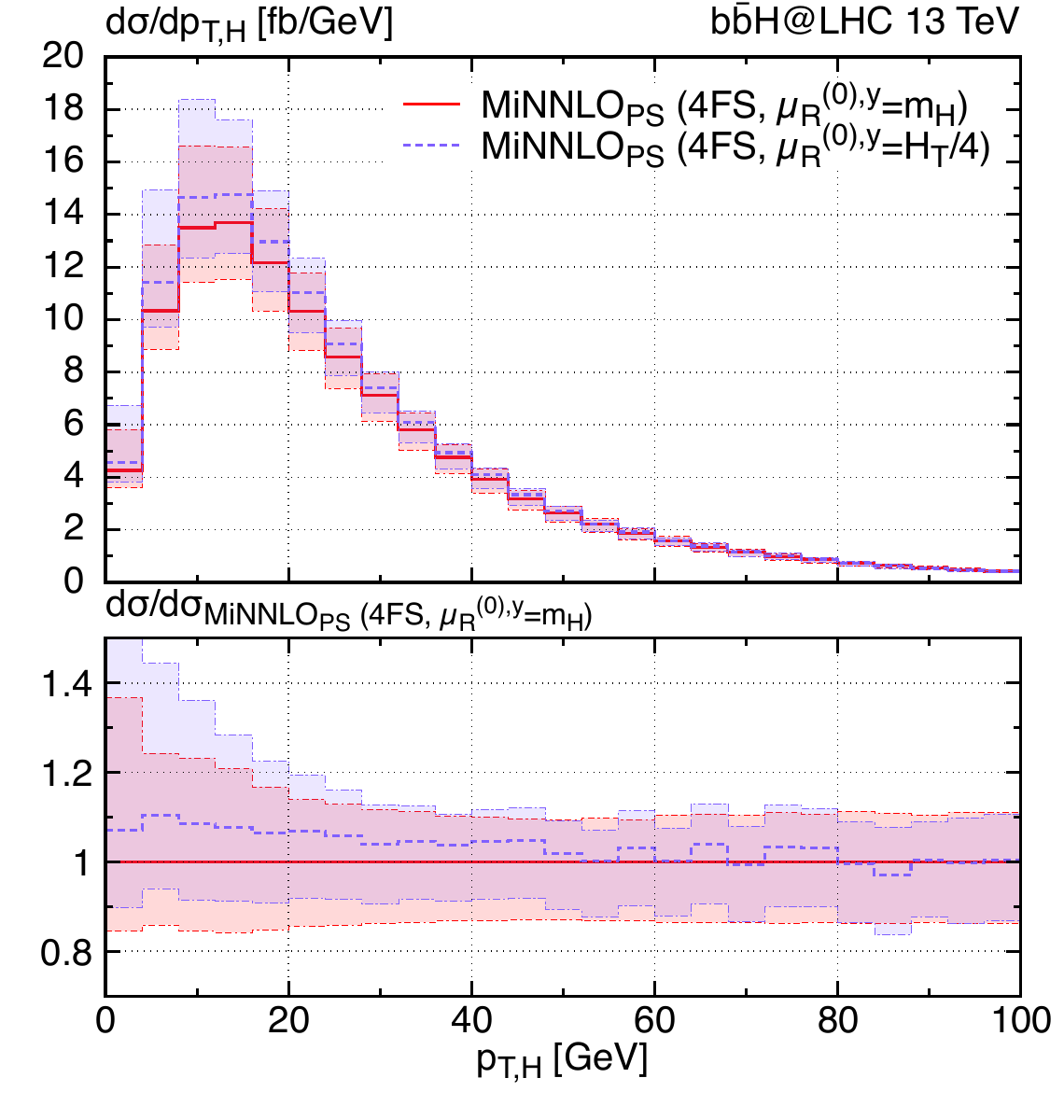}
    \includegraphics[width=.49\textwidth]{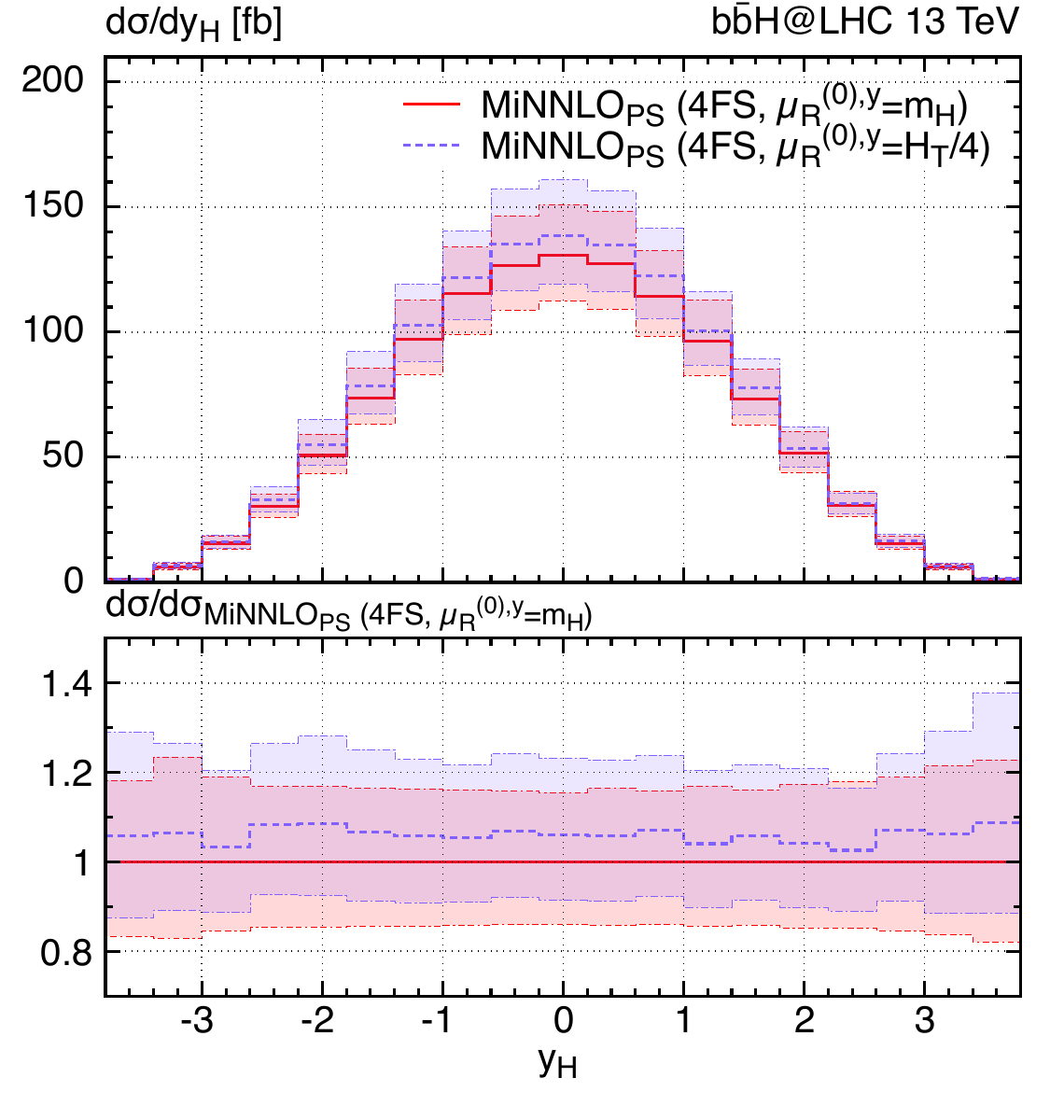}
    \includegraphics[width=.49\textwidth]{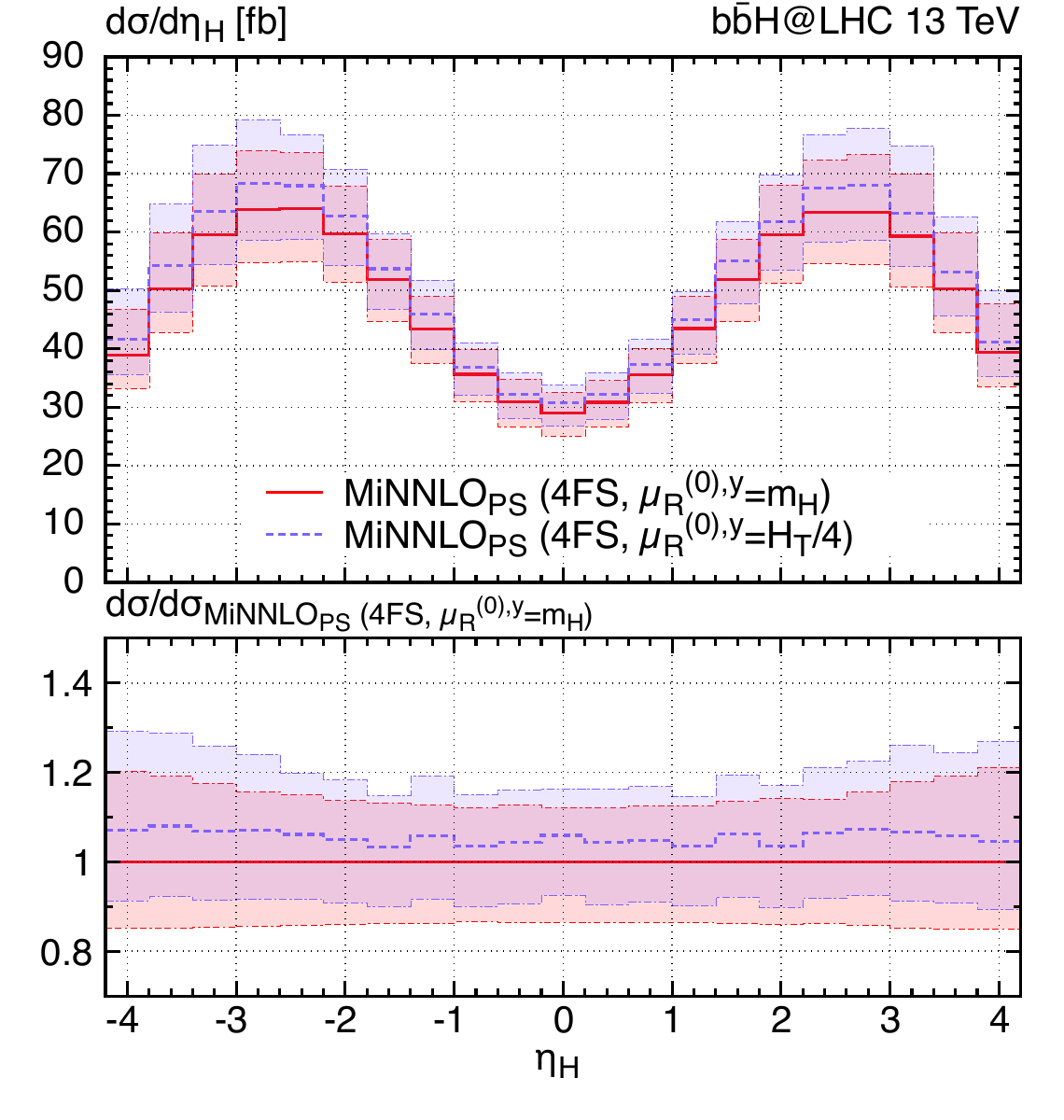}
    \includegraphics[width=.49\textwidth]{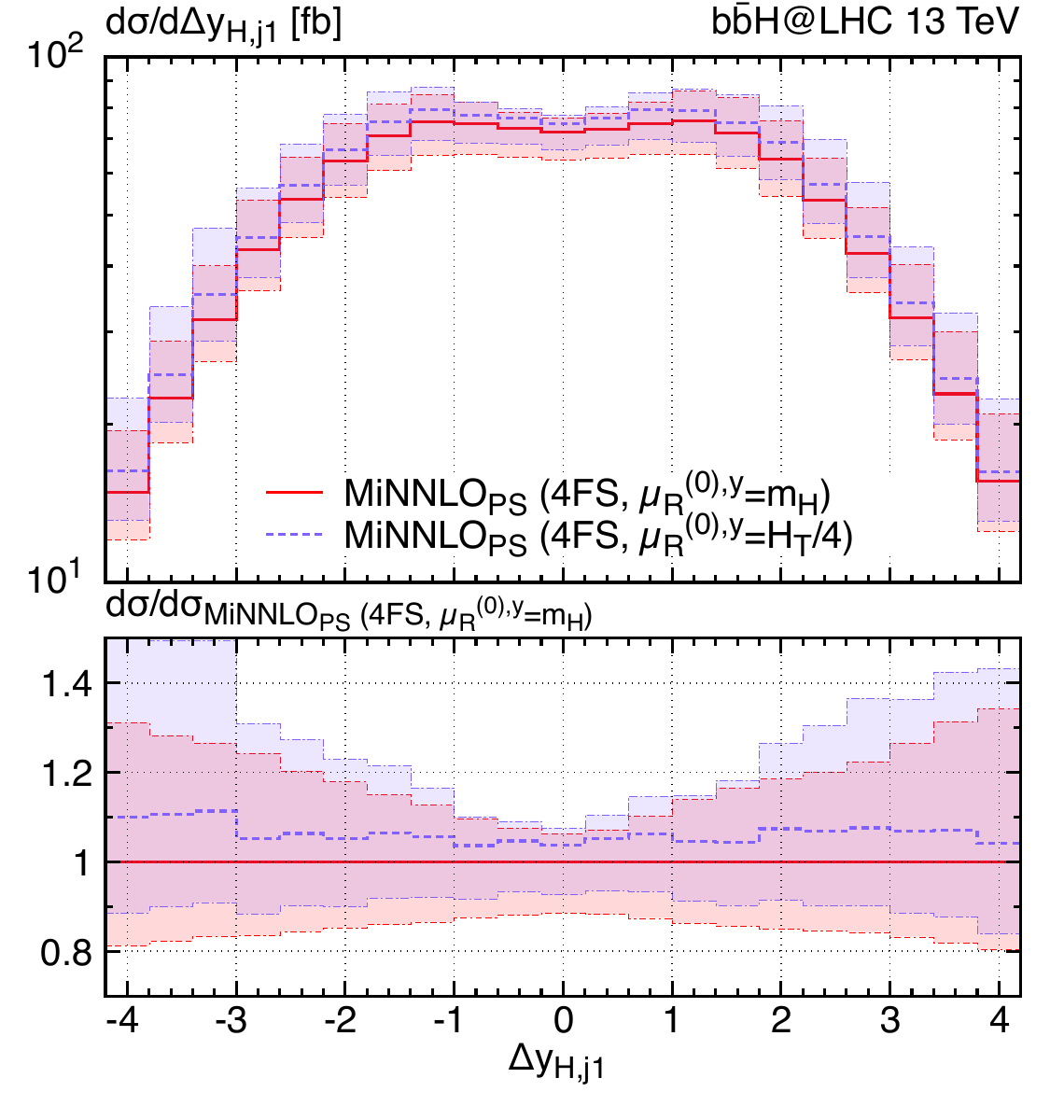}
    \caption{Comparison of two different scale choices $\muRy=m_H$ (red, solid)  and $\muRy=H_T/4$ (violet, dashed) of the bottom-Yukawa coupling in \minnlo{} predictions for Higgs observables.}
    \label{fig:Higgs_obs}
  \end{center}
\end{figure*}

\begin{figure*}[t]
  \begin{center}
    \includegraphics[width=.49\textwidth]{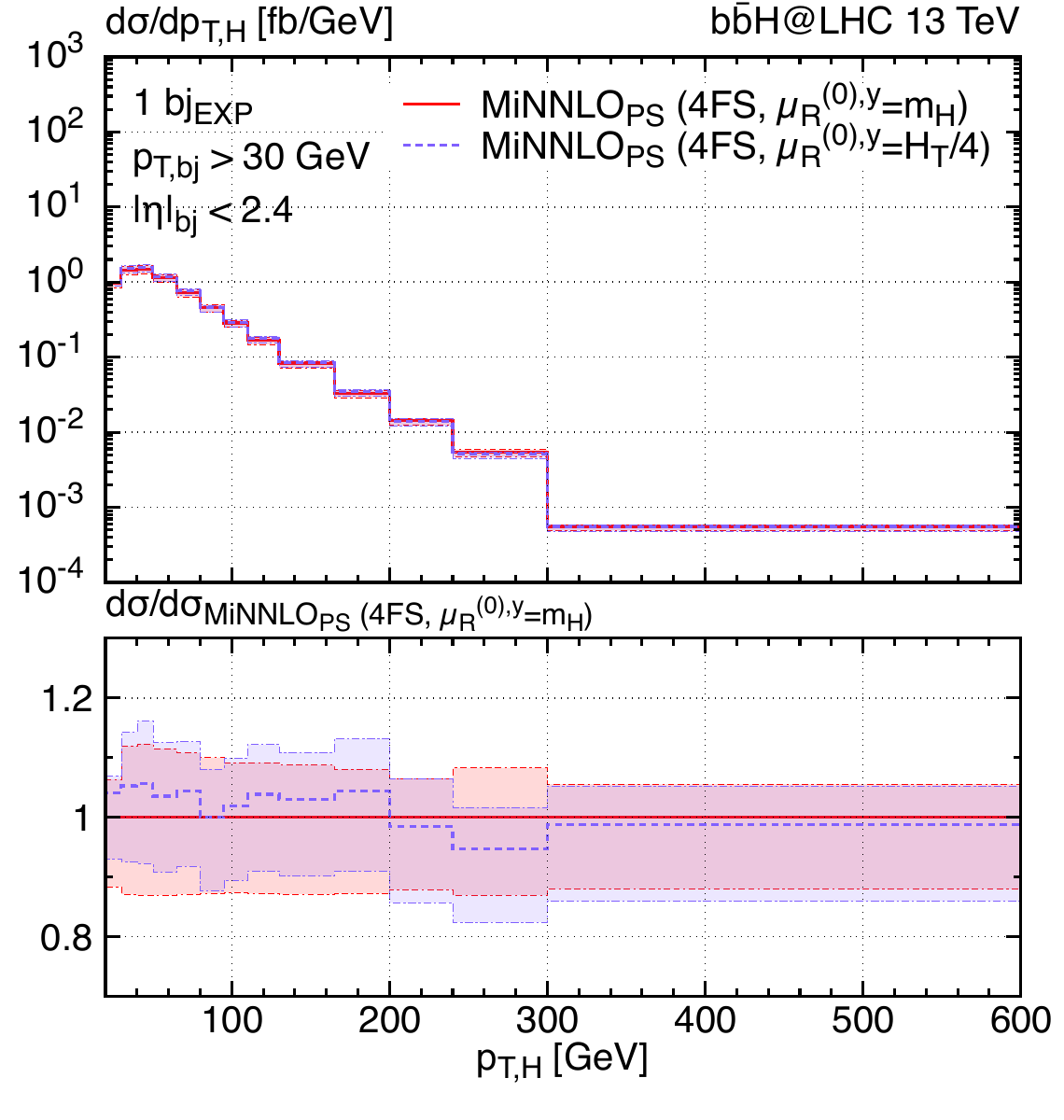}
    \includegraphics[width=.49\textwidth]{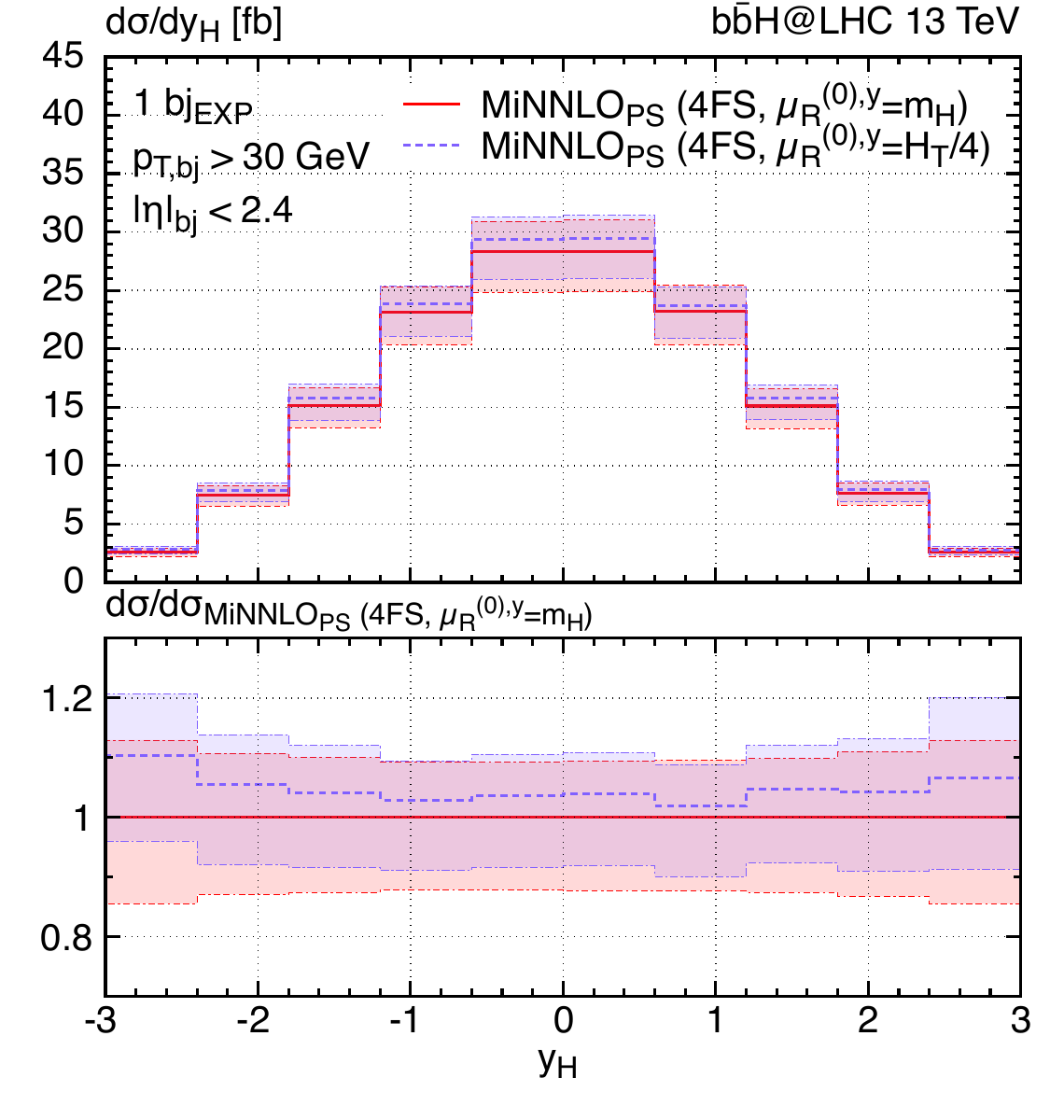}\\
    \includegraphics[width=.49\textwidth]{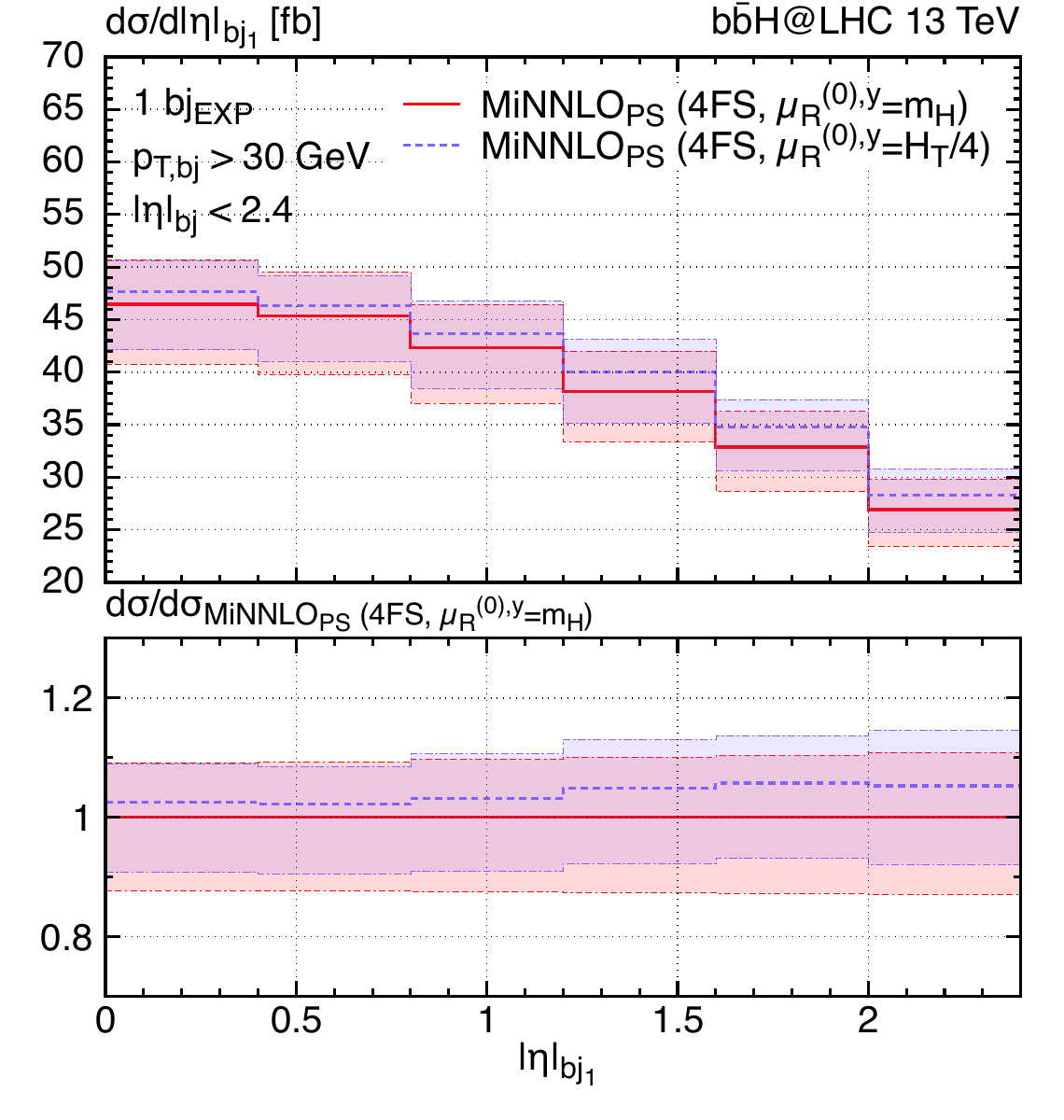}
    \includegraphics[width=.49\textwidth]{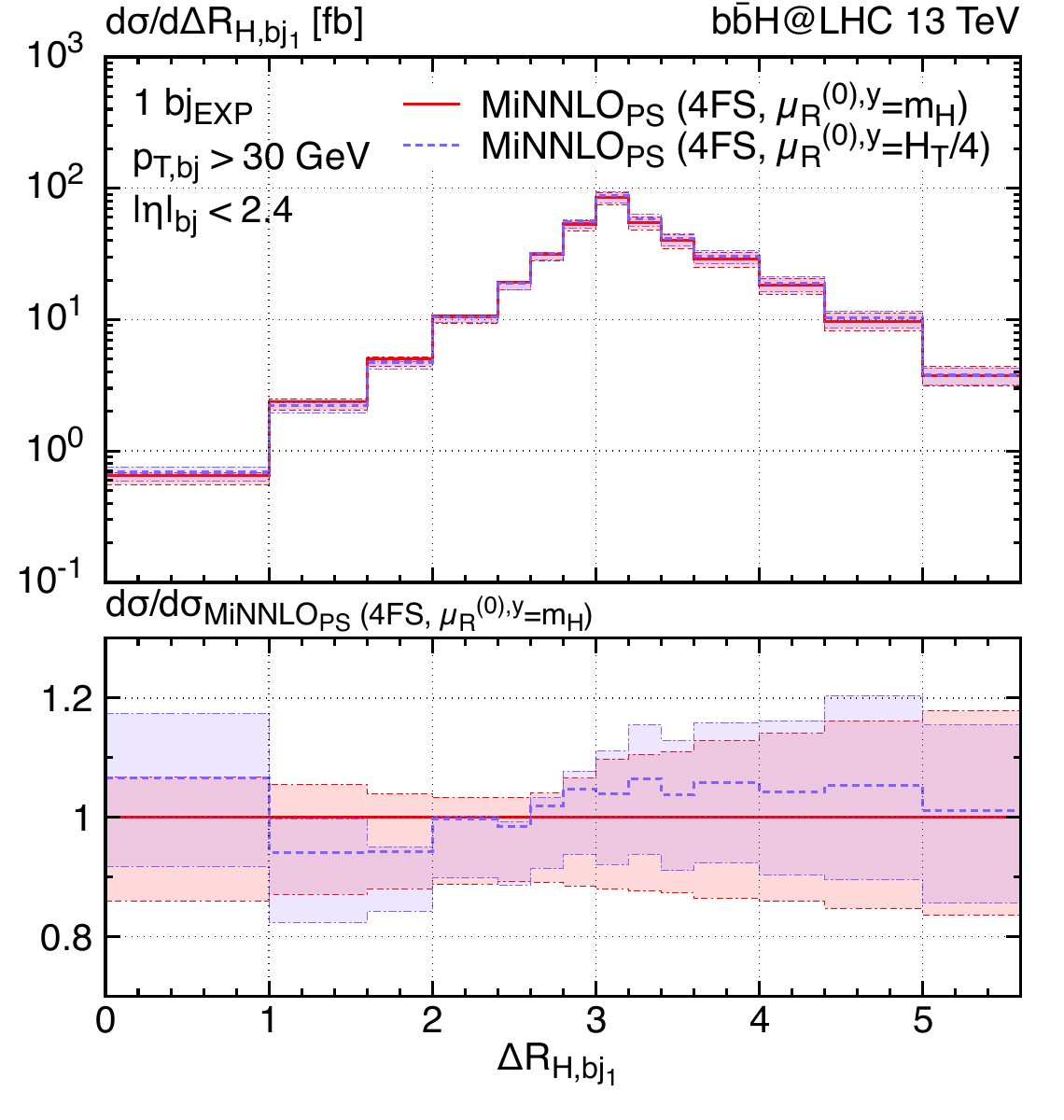}
	  \caption{Comparison of two different scale choices $\muRy=m_H$ (red, solid)  and $\muRy=H_T/4$ (violet, dashed) of the bottom-Yukawa coupling in \minnlo{} predictions $b$-jet observables.} 
\label{fig:bjetsmuY}
  \end{center}
\end{figure*}

We continue our phenomenological analysis by studying the impact of different scale choices for the bottom-Yukawa coupling 
at the differential level. To begin, we discuss the differential Higgs observables shown in \fig{fig:Higgs_obs}. 
The first plot, shows the Higgs transverse-momentum spectrum, where the two \minnlo{} predictions are relatively close,
especially in the high $\pth$ region. At low $\pth$ they differ up to 10\%.
The other plots in \fig{fig:Higgs_obs} show angular observables, including the Higgs rapidity ($y_H$) and 
pseudo-rapidity ($\eta_H$) distributions.\footnote{Note that the Higgs pseudo-rapidity is not defined 
at zero transverse momentum of the Higgs boson, but these
events have zero phase-space measure in a 4FS calculation.}
Notice that these two distributions exhibit completely different shapes due to the Higgs boson being a massive particle. 
For massless particles, these observables coincide. 
However, the introduction of a mass creates a difference between the two distributions arising from the Jacobian factor
\begin{align}
	\frac{\dd \sigma}{\dd \etah} = \frac{\ptarg{H} \cosh \etah}{\sqrt{\marg{H}^2+\ptarg{H}^2 \cosh^2\etah}} \frac{\dd \sigma}{\dd \yh}.
\end{align}
Therefore, while a Higgs boson with zero pseudo-rapidity also has zero rapidity, the Jacobian factor alters the distribution near 
the peak, resulting in a maximum at a non-zero pseudo-rapidities. 
In both cases, the choice of the bottom-Yukawa coupling scale has a relatively small impact, inducing an effect of about 5\%--10\%,
which is completely flat in these observables.
The rapidity difference between the Higgs and the leading jet in the last plot of \fig{fig:Higgs_obs} shows exactly the same relative behaviour
between the two scale choices.

Figure\,\ref{fig:bjetsmuY} shows differential cross sections with the $1\,bj_{\text{EXP}}$ requirement. The upper panel shows again 
the Higgs transverse momentum and rapidity distributions. The relative difference between the two scale choices is even smaller 
than in the fully inclusive case. However, the scale uncertainties are slightly reduced in the Higgs rapidity distribution
when using the dynamical scale choice.
In the first plot of the lower panel in \fig{fig:bjetsmuY}, we show the absolute pseudo-rapidity spectrum of the leading $b$-jet ($|\eta|_{bj_1}$). 
Also here the two scale choices lead to fully consistent \minnlo{} predictions. Nevertheless, there is a very small difference in shape
towards larger $|\eta|_{bj_1}$,  although the effect remains below about $5\%$.
In the last plot of \fig{fig:bjetsmuY}, we show the distance between the Higgs boson and the leading $b$-jet in the $\eta$-$\phi$-plane 
($\Delta R_{H, bj_1}$),  where $\phi$ is the azimuthal angle and $\eta$ is the pseudo-rapidity in the laboratory frame. 
As expected, the distributions peaks around $\pi$. 
For $\Delta R_{H, bj_1}$ more pronounced shape differences between the two scale choices 
can be observed, especially below the peak, with effects up to 10\%. Still, the predictions remain consistent within scale uncertainties.

While it has become customary to use the dynamical scale choice for the Yukawa coupling $\muRy = H_T/4$ in NLO(+PS) calculations
in the 4FS, mostly because this results in a larger cross sections, as already noted in \tab{tab:XS}, which are closer to the 5FS results,
we adopt a fixed scale $\muRy = m_H$ as our default setting in the remainder of the paper. Not only does a setting of the scale of
the Yukawa coupling of the order of the Higgs mass appear to be more appropriate, it also 
ensures consistency with the scale setting used for the bottom-Yukawa coupling in our 5FS predictions, 
enabling a more direct comparison at NNLO+PS level. Moreover, after having achieved NNLO QCD accuracy in both the 4FS and the 
5FS, it is less relevant to tune the scales of these calculations in order for them to be in better agreement.
Since the residual effects of changing the scale reduces significantly at higher orders, either scale choice is sufficient
to achieve agreement between 4FS and 5FS predictions when NNLO QCD corrections are included.

\subsubsection{Comparison with the NLO predictions}
\begin{figure*}[ht!]
  \begin{center}
    \includegraphics[width=.49\textwidth]{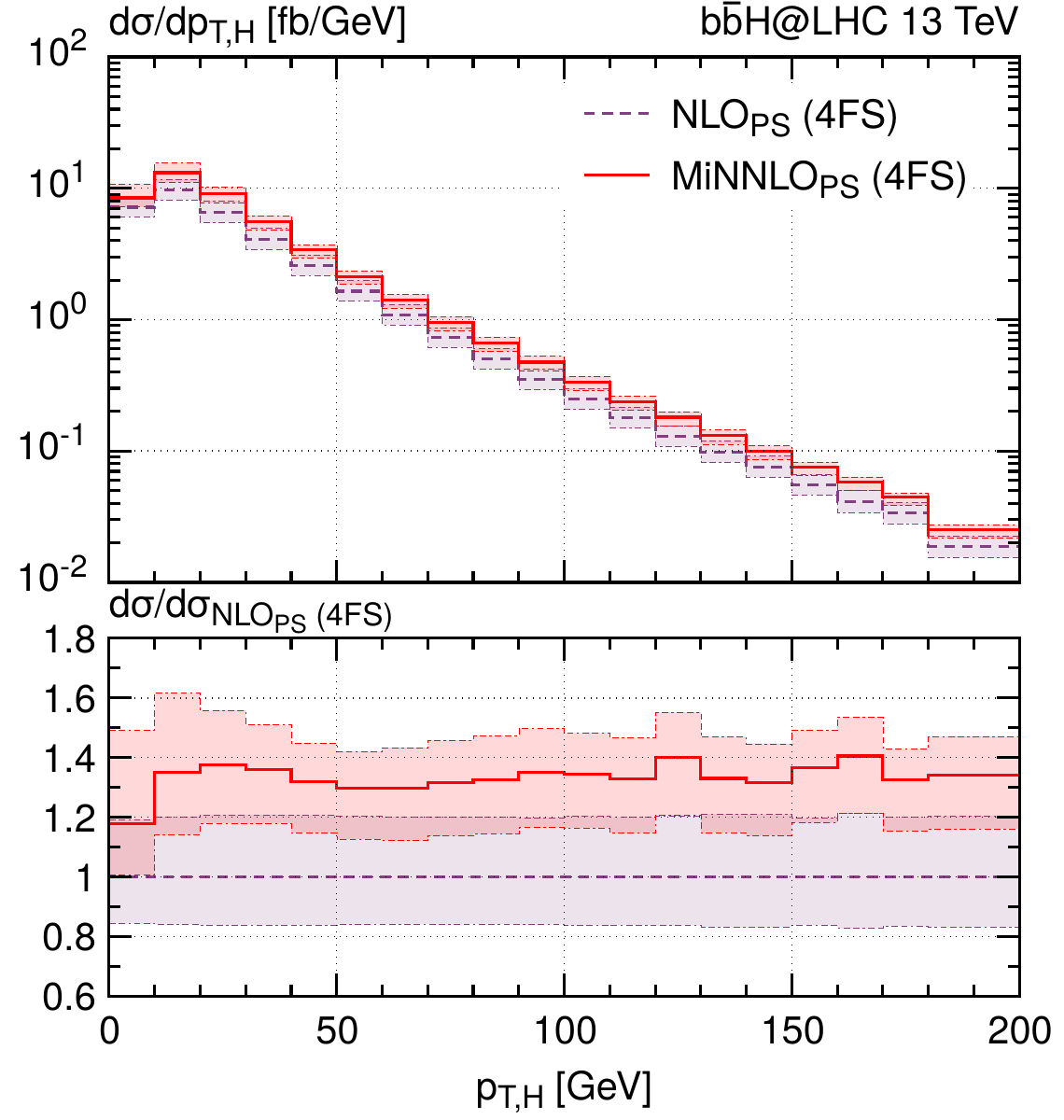}
     \includegraphics[width=.49\textwidth]{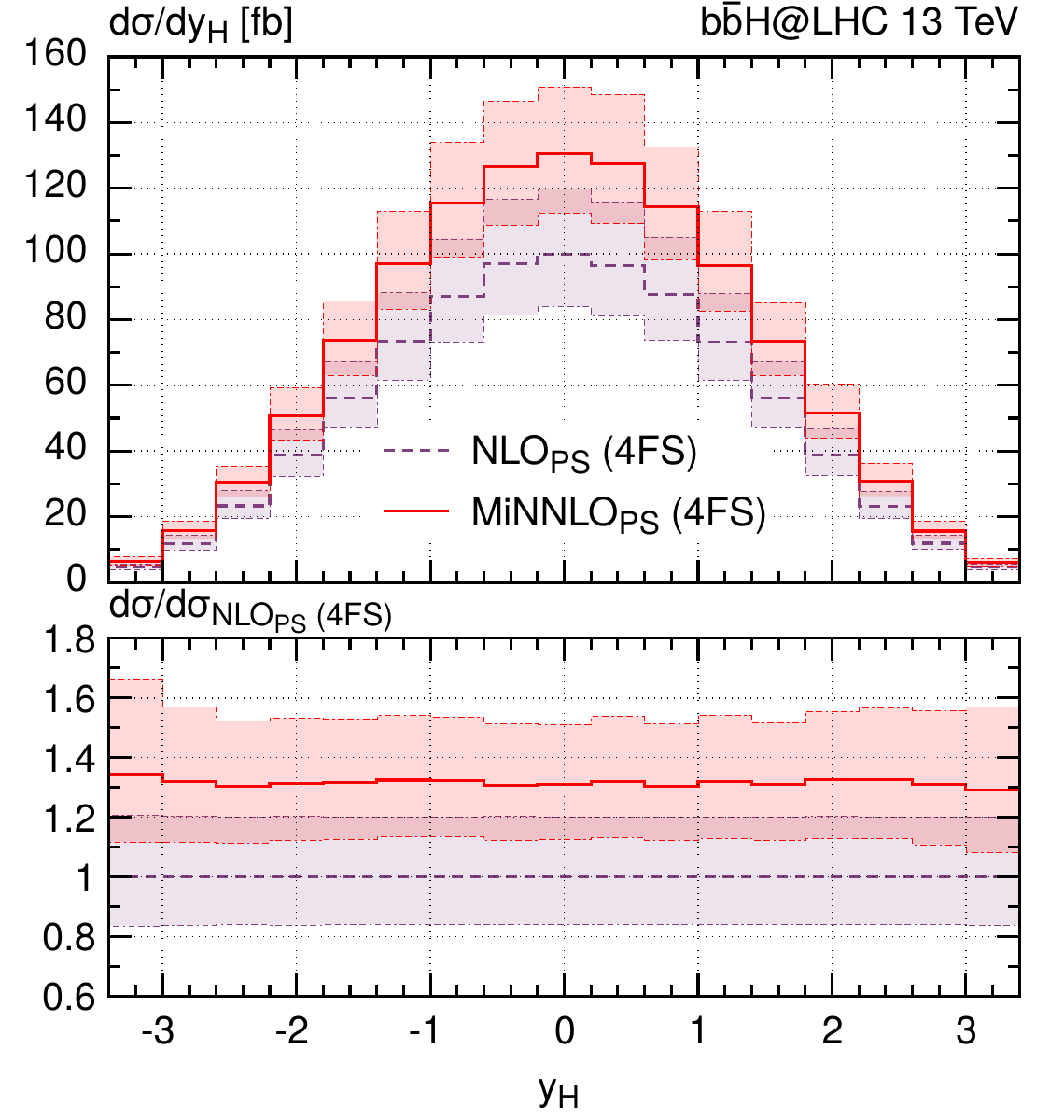}
    \caption{Comparison of NLO+PS (purple, dashed) and \minnlo{} (red, solid) predictions in the 4FS for Higgs observables.}
    \label{fig:NLOvsMiNNLOH0}
  \end{center}
\end{figure*}

\begin{figure*}[h!]
  \begin{center}
    \includegraphics[width=.49\textwidth]{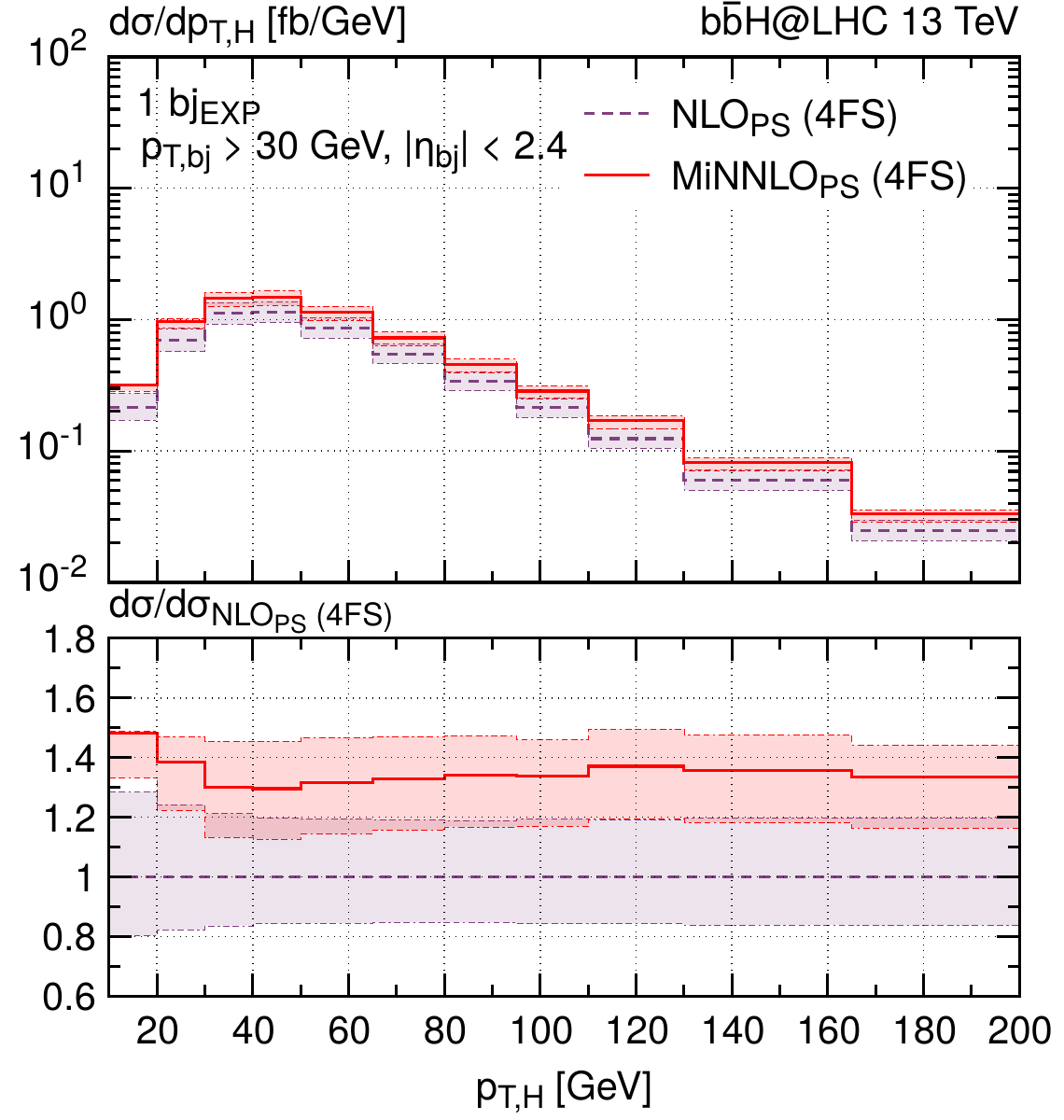}
    \includegraphics[width=.49\textwidth]{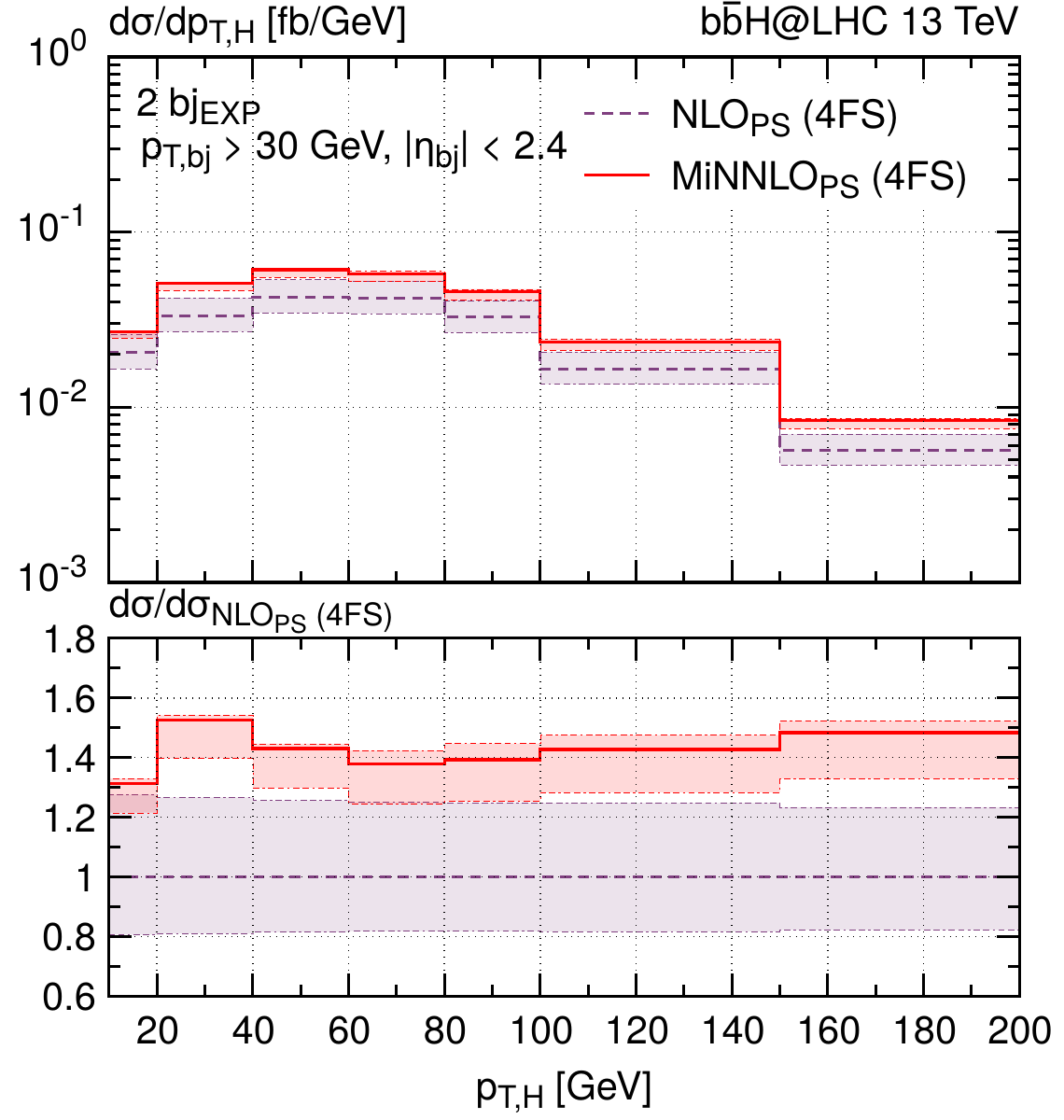}
	  \caption{Comparison of NLO+PS (purple, dashed) and \minnlo{} (red, solid) predictions in the 4FS for the Higgs transverse-momentum spectrum with the requirement of at least one (left) or at least two $b$-jets (right).}
    \label{fig:NLOvsMiNNLOH1}
  \end{center}
\end{figure*}
\begin{figure*}[h!]
  \begin{center}
    \includegraphics[width=.49\textwidth]{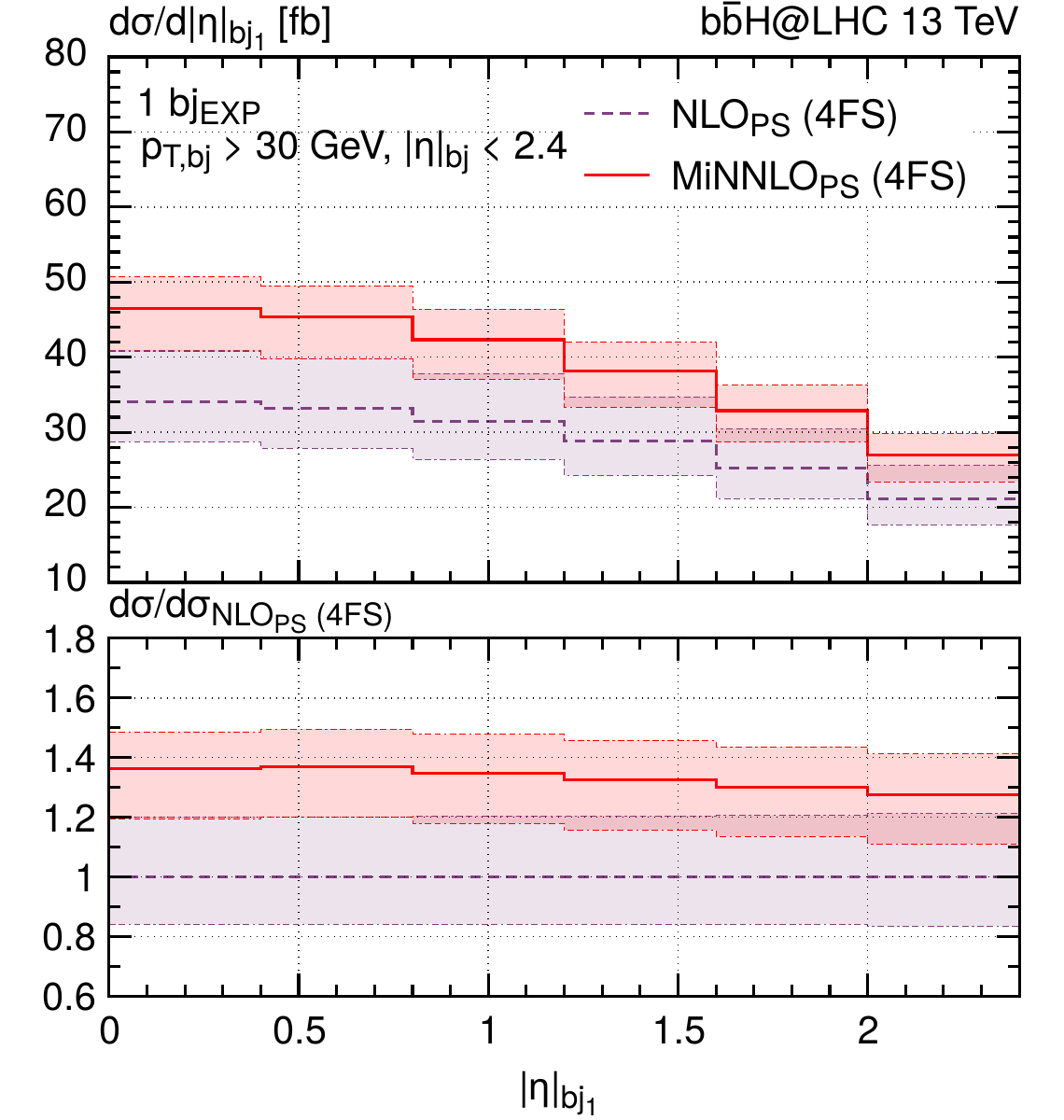}
    \includegraphics[width=.49\textwidth]{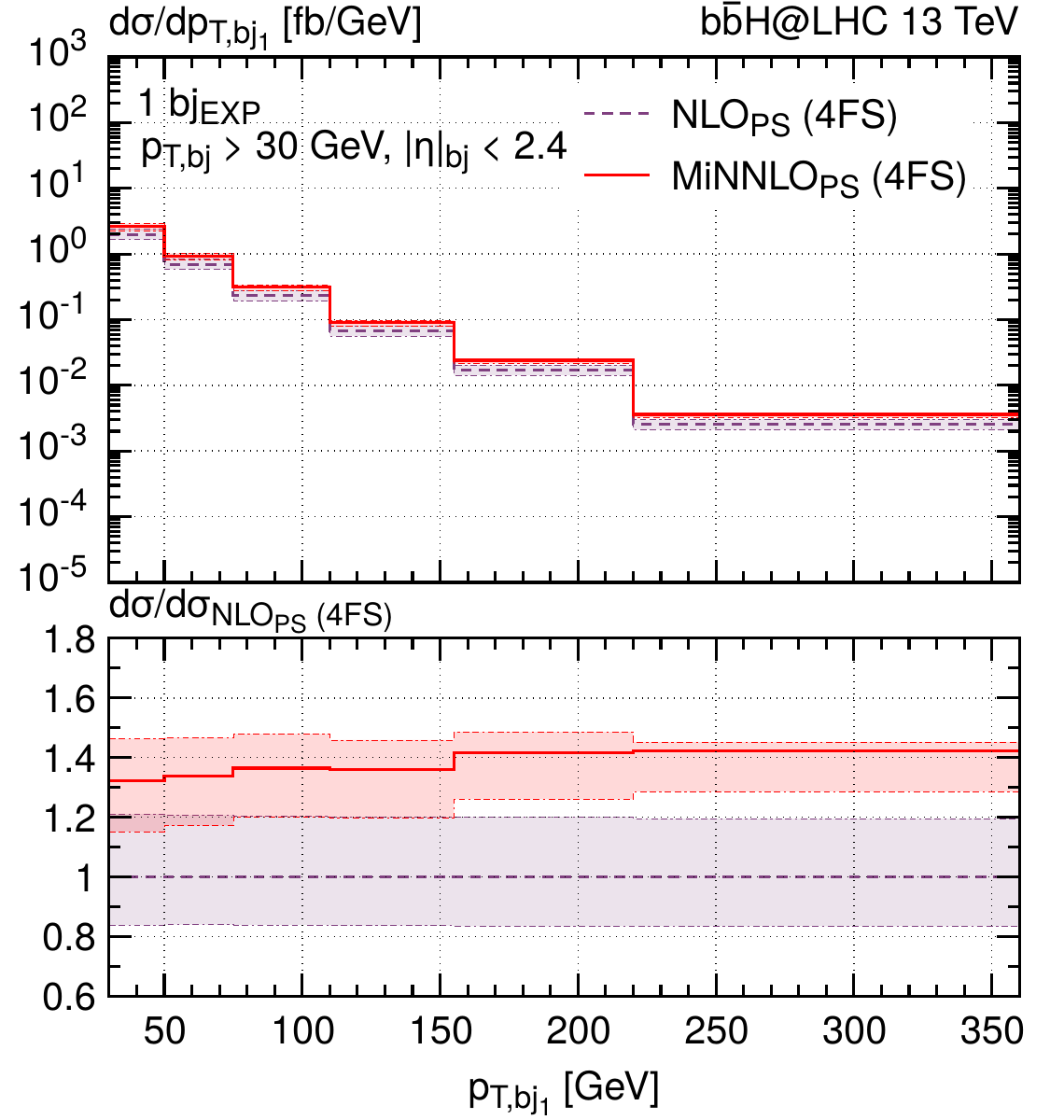}
    \includegraphics[width=.49\textwidth]{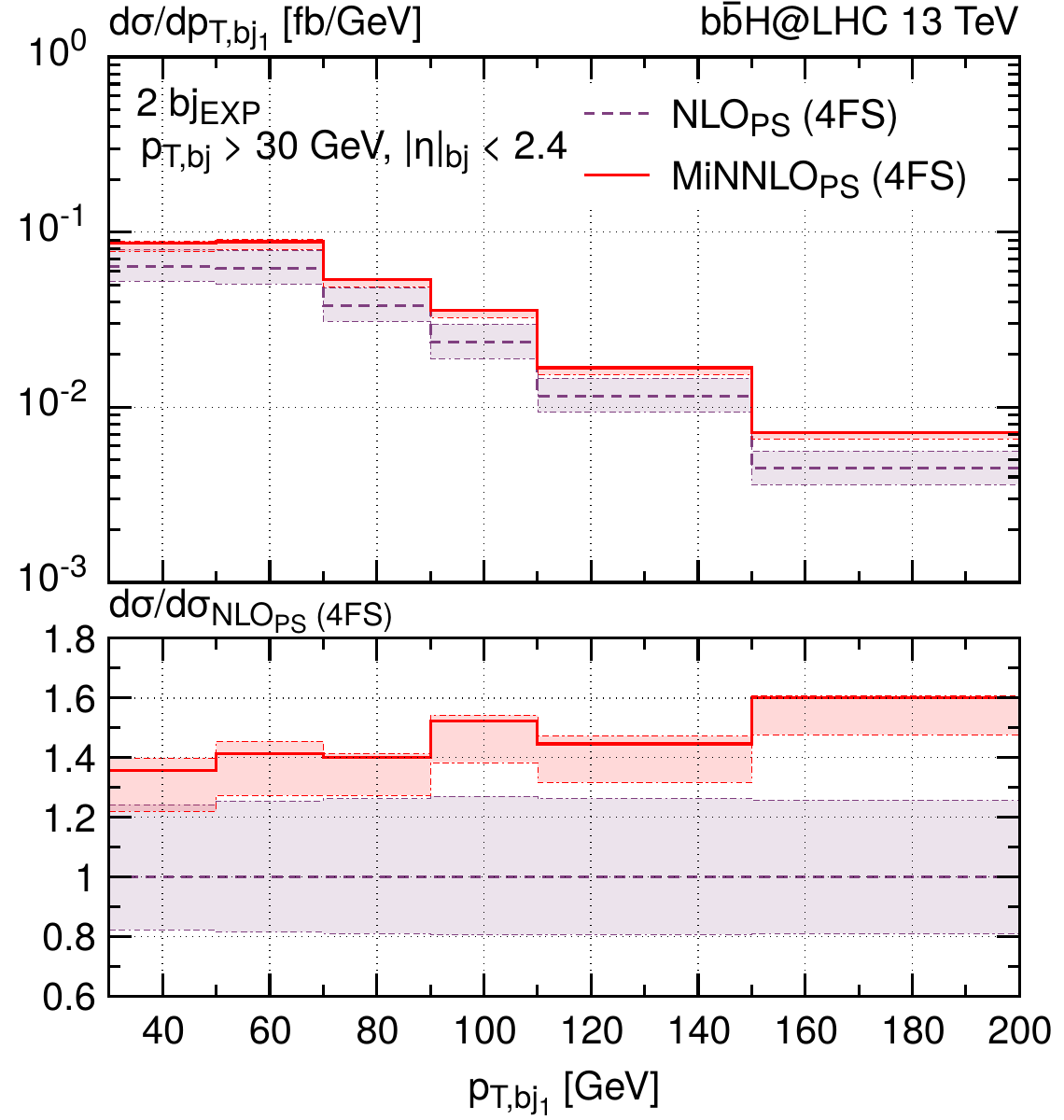}
    \includegraphics[width=.49\textwidth]{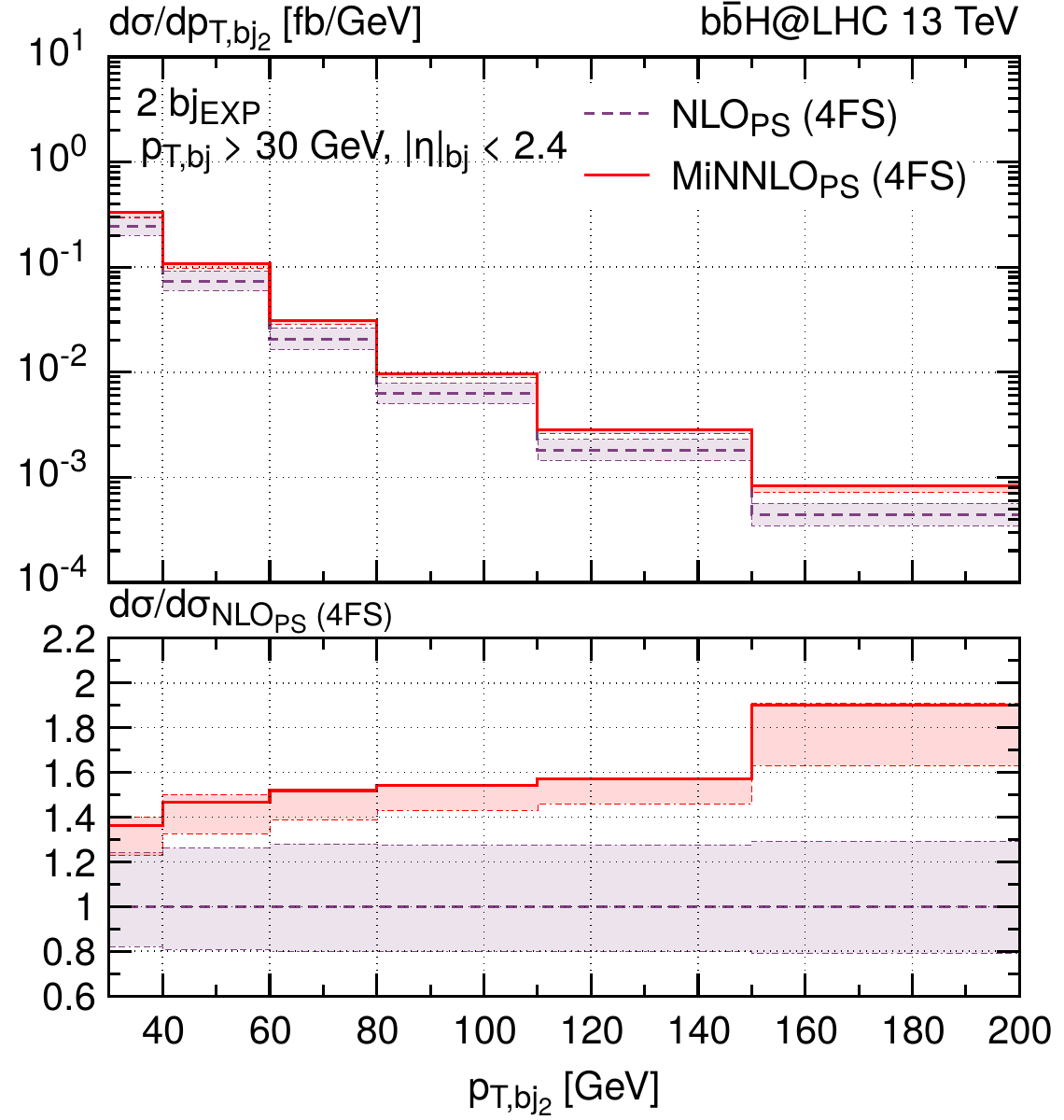}
    \caption{Comparison of NLO+PS (purple, dashed) and \minnlo{} (red, solid) predictions in the 4FS for the b-jet observables with the requirement of at least one (upper plots) or at least two $b$-jets (lower plots).}
    \label{fig:NLOvsMiNNLOb1}
  \end{center}
\end{figure*}

We now compare our \minnlo{} predictions with NLO+PS results to assess the relevance of NNLO QCD corrections in the 4FS. 
We can observe from \fig{fig:NLOvsMiNNLOH0} that NNLO corrections increase the NLO distributions in the Higgs transverse 
momentum and rapidity by about 30\%, which shows a slight dependence at small transverse momenta, but is completely flat
in the rapidity distribution. Indeed, we noticed the substantial effect of the NNLO corrections 
already at the level of the total inclusive cross section in \tab{tab:XS}. The scale variations at NLO+PS do not cover the 
central \minnlo{} result and their bands barely touch. Due to the large corrections the scale uncertainties only reduce mildly,
which can be considered as a good sign, as it makes the \minnlo{} scale uncertainties more robust.

Next, we stay with the transverse-momentum spectrum of the Higgs boson, but include a requirement on the minimal number of $b$-jets,
shown in \fig{fig:NLOvsMiNNLOH1}. We see that $\pth$ spectrum becomes broader with its peak shifting towards large values
as more $b$-jets are required. Moreover, the relative correction at NNLO slightly increases to about $+40$\% in both $b$-jet
categories and still with a very mild dependence on the exact $\pth$ value.
We also notice that the scale uncertainties are reduced in the \minnlo{} predictions compared to the NLO+PS ones, especially in the case 
where at least two identified $b$-jets are required. In that case, also the scale bands do not overlap any longer in several bins.
This shows that NLO QCD accuracy in the 4FS is insufficient to provide reliable predictions for $b\bar bH$ production.

Finally, we consider $b$-jet observables in \fig{fig:NLOvsMiNNLOb1}, where the effects become significantly more drastic 
for observables with at least two $b$-jets required. Looking at the upper plots in \fig{fig:NLOvsMiNNLOb1}, which show the pseudo-rapidity
and transverse-momentum distribution of the leading $b$-jet in the $\ge$1-$b$-jet category, we find similar results to before: NNLO 
corrections increase the NLO+PS cross section by 30\%--40\%, the dependence of these corrections on the observables is rather mild, 
and scale uncertainties decrease slightly, with largely overlapping bands, except at high transverse momenta of the leading $b$-jet.
By, contrast in the leading and subleading $b$-jet transverse-momentum spectra with the requirement of at least two $b$-jets, there
is a substantial increase in the NNLO corrections towards large transverse momenta, reaching up to a factor of two. As for the 
Higgs transverse-momentum spectrum in the $\ge$2-$b$-jet category, the \minnlo{} scale uncertainties are much smaller than the 
NLO+PS ones. Moreover, NLO+PS predictions completely fail in describing the cross section at large transverse momentum.

\section{Comparison against the five-flavour scheme} \label{sec:4FSvs5FS}
This section aims at providing a thorough comparison of the \minnlo{} generators in the 4FS and 5FS at the differential level. 
In \tab{tab:XS}, we have already compared the fully inclusive cross sections in both schemes. Besides distributions in the inclusive
phase-space, we will study observables requiring at least one or two identified \(b\)-jets in the final state. 
In the 4FS, the experimental definition of \(b\)-jets, as described in \sct{sec:setup}, can be directly applied, with infrared safety ensured by the 
finite bottom mass. However, in the 5FS, using an experimental definition of \(b\)-jets leads to IR-unsafe observables for massless bottom quarks.
In principle, this can be adjusted in a parton-shower matched simulation by reshuffling the massless momenta to massive ones. 
Alternatively, an IR-safe definition of the jet flavour can be employed. Therefore, before comparing 4FS and 5FS results involving $b$-jets, 
we first explore different \(b\)-jet definitions within the \minnlo{} 5FS predictions in the following subsection.

 \subsection{Definition of $b$-jets in the massless case} \label{sec:bjetstudy}
In recent years, several attempts have been made to extend the anti-$k_t$ jet clustering algorithm 
to provide an infrared-and-collinear (IRC)-safe definition of heavy-flavour jets, when the respective quark
is treated as massless.  
Various proposals have been recently formulated, including \textit{flavoured anti-}$k_t$~\cite{Czakon:2022wam}, 
\textit{Flavour Dressing}~\cite{Gauld:2023zlv} and \textit{Interleaved Flavour Neutralisation}~\cite{Caola:2023wpj}. See also \citeres{Buckley:2015gua,Caletti:2022hnc,Caletti:2022glq} for alternative approaches to defining the jet flavour.

These algorithms address issues in flavour tagging, specifically the mismatch between virtual and real contributions 
when a flavour algorithm is applied to a theory prediction at fixed order in a massless scheme. This alignment is essential 
for ensuring an infrared-safe definition of observables involving flavoured jets. The potentially dangerous configurations 
involve either the splitting of a gluon into a bottom-quark pair within the same jet or soft wide-angle emissions of bottom quarks that are clustered 
with another hard parton. Both of these (potentially divergent) mechanisms alter the jet flavour if the algorithm is not properly defined. 

These issues arise in the experimental approach for $b$-jet tagging, referred to as \texttt{EXP} in the following, as used in the previous 
section for our massive predictions.\footnote{It shall be noted that, in principle, even in a scheme where the quark is treated as being massive, 
logarithms in the quark mass appear in the \texttt{EXP} jet-flavour definition 
that can be potentially large and deteriorate the perturbative convergence. For bottom quarks, 
and in particular the $b\bar bH$ process, this can be neglected though, as it may happen only in rather extreme
(physically not relevant) regions of phase space.}
The challenge posed by a gluon splitting into a collinear bottom-quark pair that both end up in the same jet 
can be addressed with a straightforward solution: applying a modulo-2 condition on the number of bottom-flavoured quarks/hadrons within the same jet.
This \textit{naive} approach, labelled \texttt{NAI} in the following, classifies a jet as a $b$-jet if it contains only an odd number of bottom-flavoured quarks/hadrons. This solution does not solve the potential divergences from soft wide-angle emissions, but it captures the potentially
more problematic and more frequent configurations. In addition, we consider in our analysis one of the more sophisticated IRC-safe approaches, precisely the Interleaved Flavour Neutralisation (\texttt{IFN}) \cite{Caola:2023wpj}. The choices of the parameters in the definition of the neutralisation distance in IFN are the suggested ones: $\alpha=2$ and $\omega=1$. We developed a \texttt{Fortran-C++} interface to enable the use of the \texttt{Fastjet} plugin within our \POWHEG{} analyses. 
This general-purpose interface is applicable to all processes implemented in \POWHEG{}.

\begin{figure*}[t]
  \begin{center}
    \includegraphics[width=.49\textwidth]{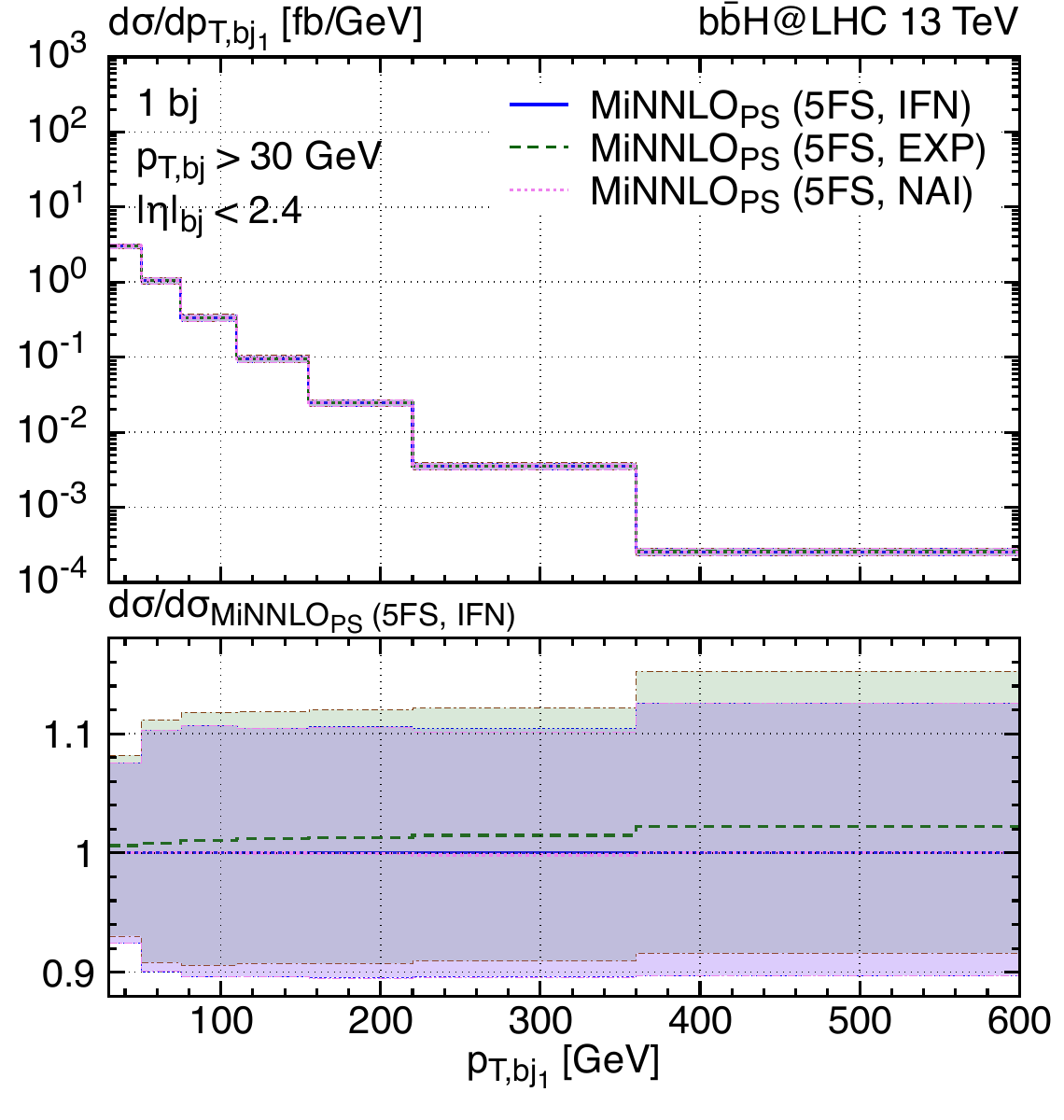}
    \includegraphics[width=.49\textwidth]{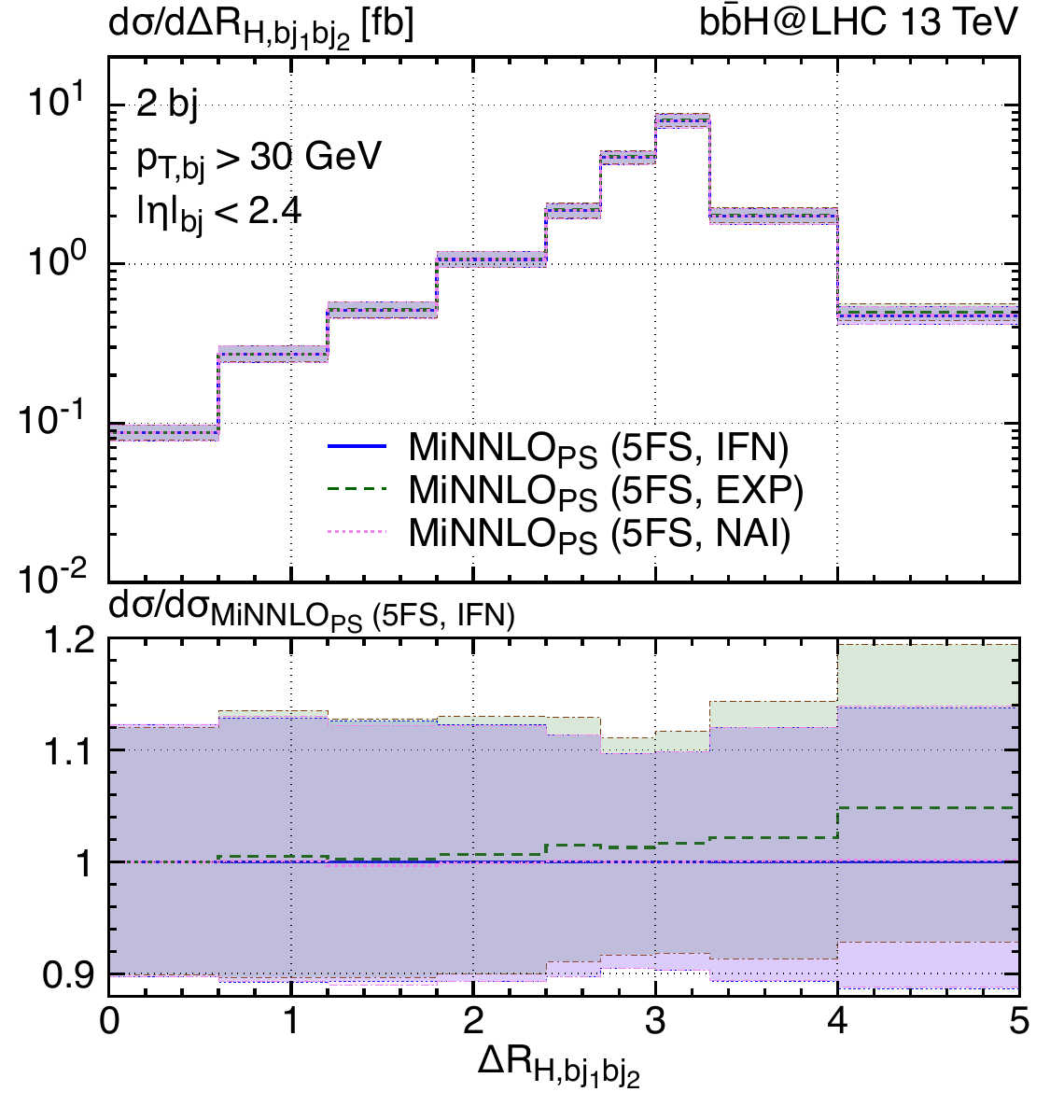}\\
    \includegraphics[width=.49\textwidth]{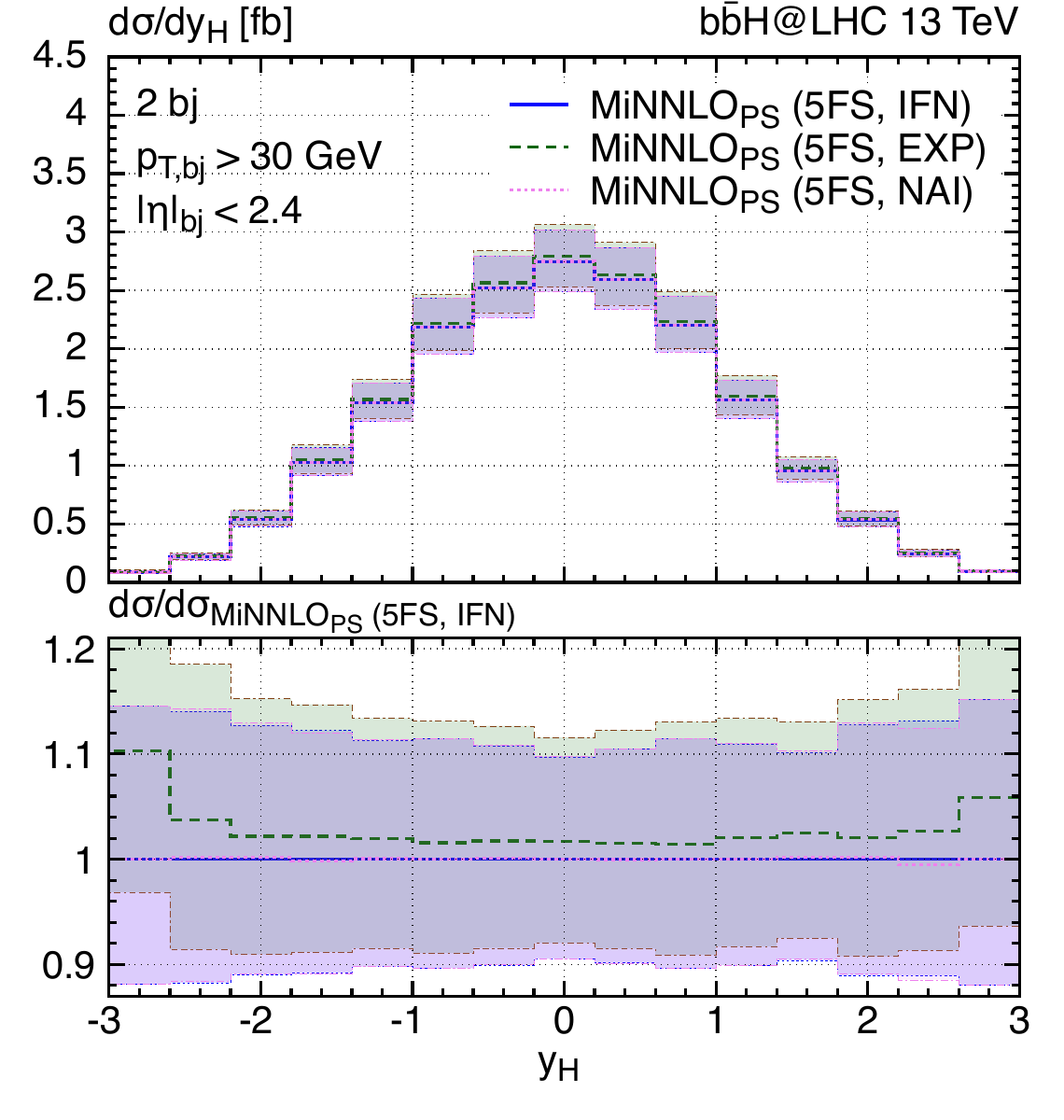}
    \includegraphics[width=.49\textwidth]{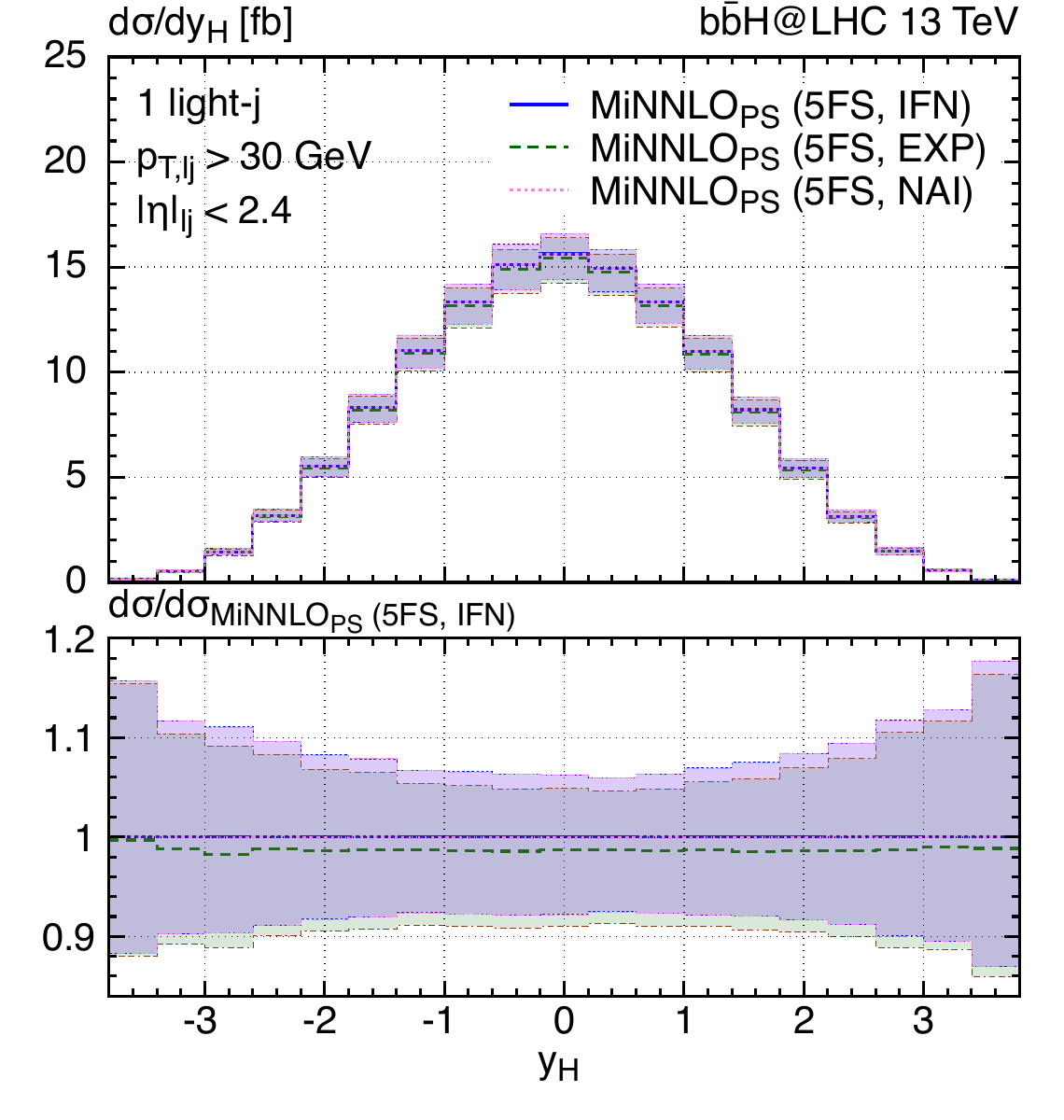}
    \caption{Comparison of different jet-flavour algorithms for $b$-jet predictions obtained with \minnlo{} in the 5FS for the {\tt EXP} (green, dashed), {\tt NAI} (pink, dotted) and {\tt IFN} (blue, solid) $b$-jet definitions.}
    \label{fig:jetalgocomparison}
  \end{center}
\end{figure*}
The \minnlo{} 5FS calculation is divided into several stages (corresponding to the ones in \POWHEG{}). Firstly, a fixed-order type prediction is obtained at the so-called \texttt{stage-2} of \POWHEG{}. At this stage, the bottom quarks are massless, and we have verified that the \texttt{EXP} prediction depends on the technical cut-off present in the generation of the final-state radiation. The effect is small and visible only when the channel induced by bottom and anti-bottom quarks is selected.
 Although the effect is minor, IRC-unsafety is numerically evident due to the cut-off dependence. We, therefore, proceed with including the \POWHEG{} radiation at \texttt{stage-4} according to the master formula in \eqn{eq:master} to produce LHE events, where the massless bottom quarks are mapped into massive states. Finally, we attach the shower radiation (that includes massive bottom quarks) to the LHE events for a physical description.
 
 The massless-to-massive mapping introduces only power corrections in the quark mass, as long as the observable is infrared-safe and no collinear effects are screened by the mass. In addition, the \POWHEG{} matching is formally derived only for IRC-safe observables. In fact, in the formulation of the $\bar B$ function in \POWHEG{} in every event, the virtual and real contributions are combined into the same event weight. This combination
 cannot be split a posteriori by an IRC-unsafe $b$-tagging algorithm.
As a result, the  \texttt{EXP} or \texttt{NAI} $b$-jet 
tagging yield finite results 
in the \minnlo{} 5FS calculation (the same being true for any
parton-shower matched prediction, e.g.\ any NLO+PS one).
However, this poses the question whether such predictions can be trusted
in providing a physical description of $b$-jet observables that
are formally not IRC-safe in the 5FS or whether the finite results are an 
artificial remnant of the matching method. Although from a 
theoretical viewpoint it appears to us that only IRC-safe definitions, 
such as \texttt{IFN}, 
yield sensible results even in a matched parton-shower calculation
for massless bottom quarks, 5FS predictions have been employed with 
the standard experimental definition for years in comparison to data,
without being obviously flawed. Moreover, we have not found a
way to unambiguously show numerically that matched parton-shower 
calculations fail to provide physical results in practice. On the contrary,
we have not found any sensitivity to technical cut-offs so far that are present 
in the generation of LHE events at {\tt stage-4}, unlike our findings
for {\tt stage-2} discussed above.
Therefore, we consider it beyond the scope of this paper to provide 
a final answer to this question, which we leave to future considerations.

Figure\,\ref{fig:jetalgocomparison} shows a comparison of the three jet-flavour definitions for $b$-jet observables obtained with the \minnlo{} generator in the 5FS and including shower radiation. We note that we have selected the
observables that show the largest differences.  
The first plot shows the transverse-momentum spectrum of the leading $b$-jet. Here, a difference between \texttt{EXP} and the other two definitions is evident in the large transverse-momentum region, while no differences between 
\texttt{IFN} and \texttt{NAI} are visible. This indicates that the differences of the \texttt{EXP} results are due to gluon splittings into collinear bottom quarks. 
A similar trend is observed for the rapidity separation between the Higgs and the dijet system formed by the two leading $b$-jets in the second plot
in \fig{fig:jetalgocomparison}.

In the second row of \fig{fig:jetalgocomparison} the Higgs rapidity spectrum is shown. In the left plot, we require the presence of at least two $b$-jets, while in the right plot, the event is accepted if it contains at least one jet that is not a $b$-jet, with the same requirements on the jet transverse momentum and rapidity. As expected, the \texttt{IFN} and \texttt{NAI} definitions lead to a smaller cross section compared to \texttt{EXP} in the two-$b$-jet region, as the flavour of some of the \texttt{EXP} $b$-jets is neutralised to ensure IRC safety. In contrast, due to number conservation, the \texttt{EXP} result, when at least one light jet is present, is smaller than the predictions by the other two definitions.

In general, the discrepancy between the experimental jet-flavour definition and IRC-safe algorithms like \texttt{IFN} is quite small in this process and well within scale uncertainties, as pathological configurations in the short-distance physics are suppressed by the bottom-quark PDFs. A small adjustment to the experimental approach, using the \texttt{NAI} definition, leads to predictions that are essentially identical to the theoretically robust \texttt{IFN} approach, with differences indistinguishable compared to the numerical uncertainties. Furthermore, we observe similar effects in the 4FS implementation using the different definitions of $b$-jets discussed here. We refer to \citere{Gauld:SOON} for more in-depth studies of different jet algorithms in the context of bottom-quark production. We continue with the flavour scheme comparison using the \texttt{IFN} tagging, noting that the \texttt{NAI} definition leads to indistinguishable results and even the \texttt{EXP} definition to very similar results.

\subsection{Integrated results}

\begin{table}[h!]
  \vspace*{0.3ex}
  \begin{center}
	             \renewcommand{\arraystretch}{1.4}
\begin{tabular}{|c|c|c|c|c|}
  \hline
    Fiducial region & Generator & $\sigma_{\rm integrated}$ [fb]   & \begin{tabular}{@{}c@{}} Ratio to \vspace{-2mm}\\  NLO+PS\end{tabular} & \begin{tabular}{@{}c@{}} Ratio to \vspace{-2mm}\\  5FS results\end{tabular} \\
  \hline \hline
      \multirow{4}{*}{$pp\rightarrow H+ 0\,b$ jets}
                                & 5FS NLO+PS  &$557.(2)_{-10\%}^{+11\%}$ & 1.000 & 1.000  \\
                                  & 5FS \minnlo{}  & $ 404.(2)_{-9.0\%}^{+5.9\%}$ & 0.692 & 1.000  \\
                \cline{2-5} & 4FS NLO+PS  &$285.(2)_{-25\%}^{+29\%}$  & 1.000 & 0.488 \\
                                & 4FS \minnlo{}  & $373.(9)_{-20\%}^{+22\%}$ & 1.311 & 0.925  \\
   \hline
  \multirow{4}{*}{$pp\rightarrow H+\ge 1\,b$ jets}
                                & 5FS NLO+PS  & $88.(1)_{-10\%}^{+11\%}$  & 1.000 & 1.000 \\
                                  & 5FS \minnlo{}  &$104.(9)_{-8.7\%}^{+8.8\%}$  & 1.182 & 1.000\\
                \cline{2-5} & 4FS NLO+PS  &$69.(4)_{-16\%}^{+21\%}$ & 1.000 & 0.784  \\
                                & 4FS \minnlo{}  &$92.(1)_{-12\%}^{+9.8\%}$ & 1.327  & 0.878 \\
   \hline
   \multirow{4}{*}{$pp\rightarrow H+\ge 2\,b$ jets}
                                & 5FS NLO+PS  &$4.7(8)_{-11\%}^{+11\%}$  & 1.000 & 1.000 \\
                                  & 5FS \minnlo{}  & $7.6(4)_{-10\%}^{+11\%}$  & 1.617 & 1.000\\
                \cline{2-5} & 4FS NLO+PS  &$ 4.4(9)_{-18\%}^{+25\%}$  & 1.000 & 0.936 \\
                                & 4FS \minnlo{}  & $6.4(4)_{-9.1\%}^{+1.5\%}$  & 1.434 & 0.843 \\
   \hline
    \multirow{4}{*}{$pp\rightarrow H+\ge 1\,\ell$ jets}
                                & 5FS NLO+PS  &$31.(1)_{-9.7\%}^{+12\%}$ & 1.000 & 1.000  \\
                                  & 5FS \minnlo{}  & $ 53.(0)_{-7.6\%}^{+6.9\%}$ & 1.710 & 1.000  \\
                \cline{2-5} & 4FS NLO+PS  &$31.(4)_{-18\%}^{+26\%}$  & 1.000 & 1.000 \\
                                & 4FS \minnlo{}  & $38.(3)_{-12\%}^{+2.0\%}$  & 1.220 & 0.722 \\
   \hline 
\end{tabular}
  \end{center}
  \vspace{-1em}
  \caption{
   Cross section rates for $b\bar bH$ production without any $b$-jets in the events ($pp \rightarrow H + 0\,b$ jets) and with tagging of at least one $b$-jet ($pp \rightarrow H + \ge 1\,b$ jets), 2 $b$-jets ($pp \rightarrow H + \ge 2\,b$ jets), and at least one light jet ($pp \rightarrow H + \ge 1\,\ell$ jets) using the {\tt IFN} jet-flavour algorithm. \label{tab:XS-bjets}}
\end{table}

We compare 4FS and 5FS predictions both at NLO+PS and NNLO+PS levels to assess the improvements in the consistency of the two 
schemes for the \bbH{} process when higher-order corrections are included. The bottom-Yukawa coupling is evaluated at the scale of the 
Higgs mass throughout. We note that LO results are completely off and their uncertainties vastly underestimate the actual size of higher-order
correction, which is why we refrain from including them in this comparison.

We start by studying fiducial rates with different requirements on $b$-jets and flavour-less jets, which are defined via the \texttt{IFN} jet-flavour algorithm,
 reported in \tab{tab:XS-bjets}. Requiring at least one $b$-jet reduces the cross section by roughly a factor of 4--5, while
 requiring a second $b$-jet reduces the cross section by another factor of roughly 10. Compared to the cross section with at least one
 light flavour-less jet ($\ell$-jet), the cross section with at least one $b$-jet is three times larger. This is easy to understand, as the LO process
 contains two bottom quarks, while the light quarks are generated only through radiation at higher orders.

Looking at the column with the ratio to the NLO+PS predictions in \tab{tab:XS-bjets}, one notices that
NNLO corrections in either scheme are significant in all fiducial categories, 
ranging from 13\% to 60\%. Comparing the predictions in the two flavour schemes, we find that for the 0-$b$-jet and $\ge$1-b-jet cases, the cross sections at NLO+PS 
are not compatible with each other, whereas the \minnlo{} predictions 
improve the comparison substantially. The 0-$b$-jet rates agree within 8\% and
the $\ge$1-$b$-jet rates agree within 13\%, fully compatible within the respective
scale uncertainties at NNLO+PS.

Interestingly, we notice that for the $\ge$2-$b$-jet and $\ge$1-$\ell$-jet
the NLO+PS results in 4FS and 5FS are very close, 
even closer than for the \minnlo{} predictions. These observables become
successively less accurate in the 5FS, which at Born level does not feature
bottom quarks in the hard matrix elements. Therefore, we conclude that
the agreement 
at NLO+PS for these observables that are less accurate in the 5FS 
is completely accidental. Indeed, we will see when considering differential 
distributions in the next subsection that the shapes in 4FS and 5FS are
vastly different at NLO+PS, which confirms the accidental agreement at 
the integrated cross section level.

\subsection{Differential results}

\begin{figure*}[t]
  \begin{center}
    \includegraphics[width=.49\textwidth]{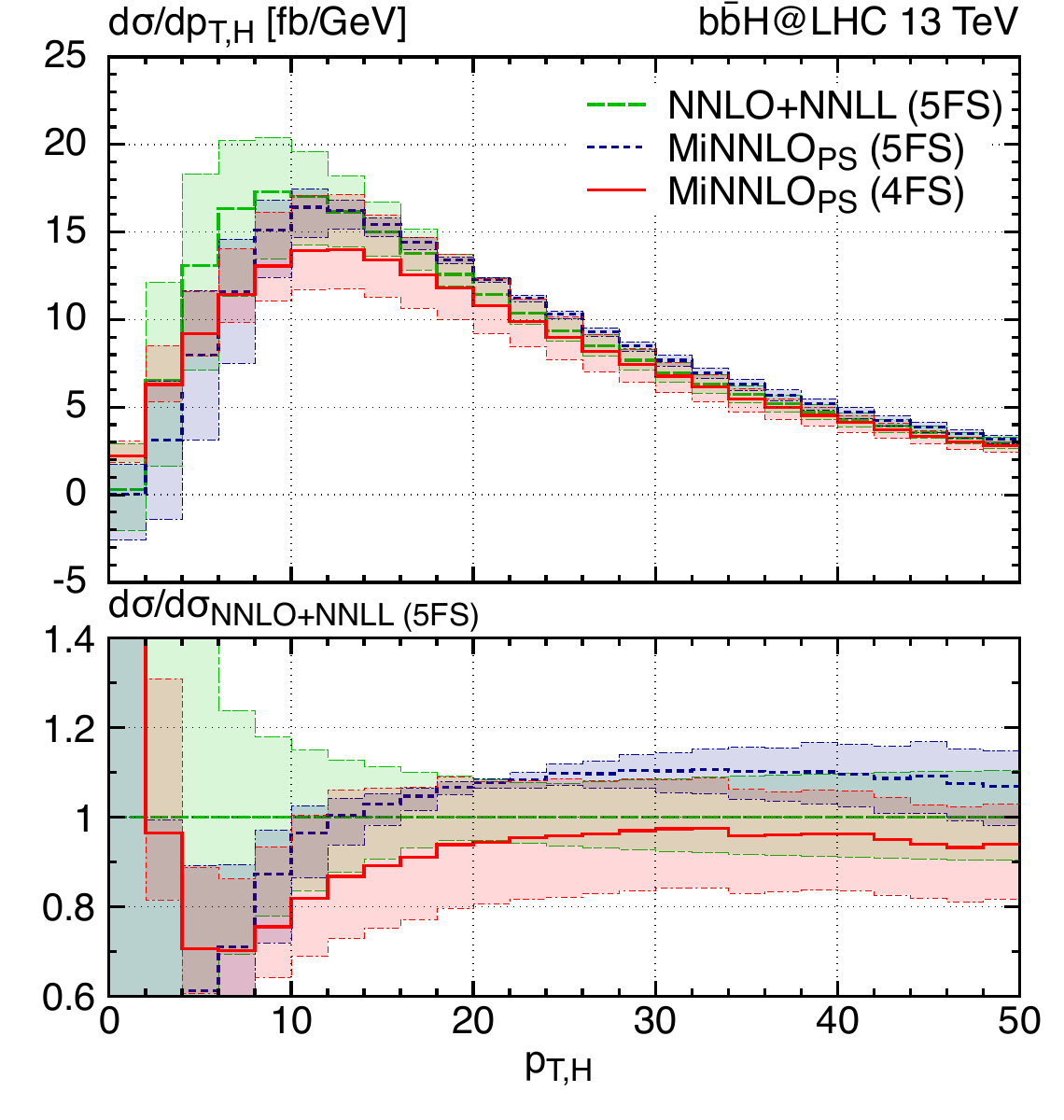}
    \includegraphics[width=.49\textwidth]{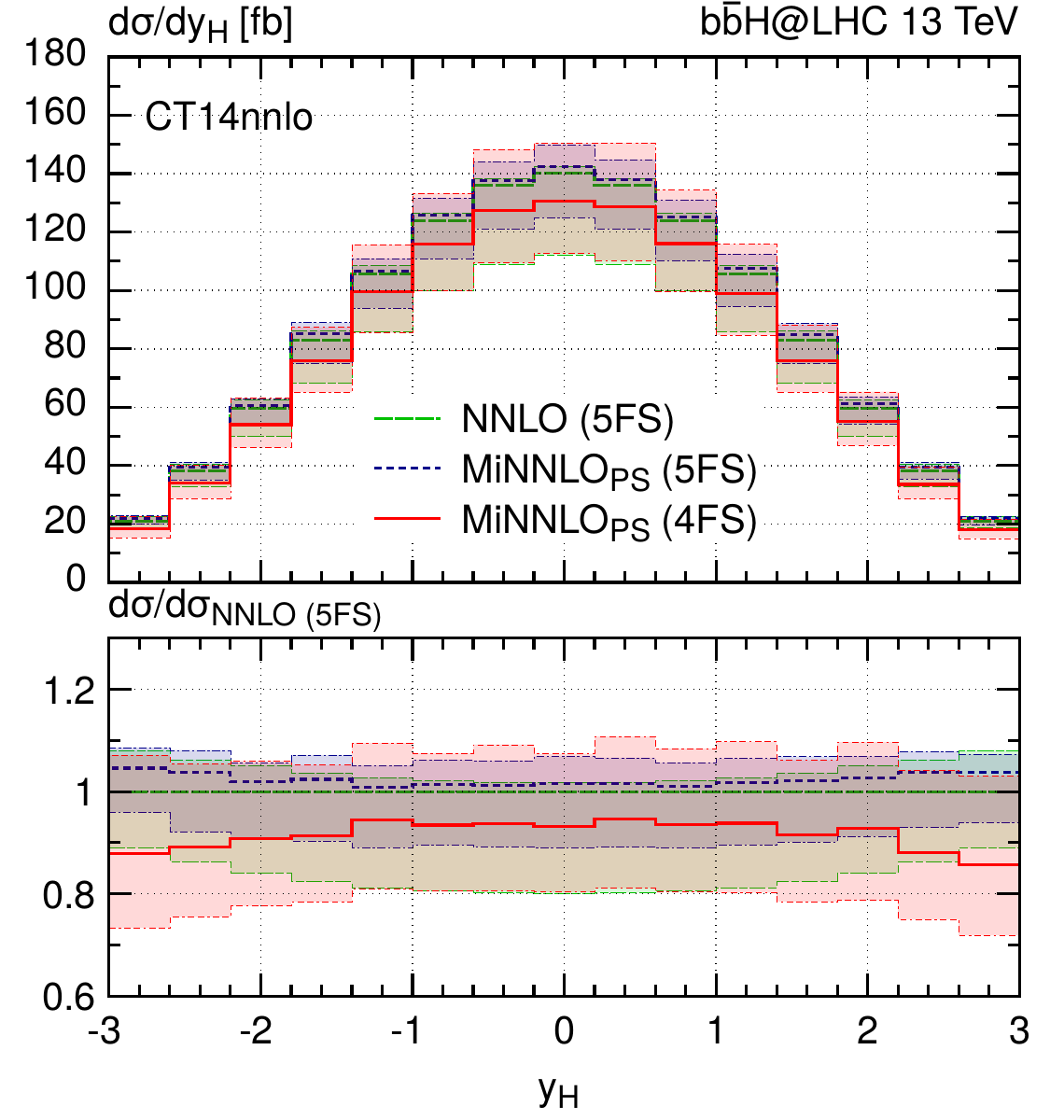}
	  \caption{Comparison of the two \minnlo{} generators in the 5FS and 4FS with analytic results in the 5FS. The Higgs $\pt$ spectrum (left) is compared to the NNLO+NNLL predictions of \citere{Harlander:2014hya} using NNLO NNPDF 4.0 sets, while the Higgs rapidity spectrum (right) is compared with the NNLO fixed-order result of \citere{Mondini:2021nck} based on NNLO CT14 sets. For the right plot, we used NNLO CT14 sets in our \minnlo{} generators, both in the 5FS and 4FS.} 
\label{fig:ptzoomNNLL}
  \end{center}
\end{figure*}

\begin{figure*}[h]
\vspace{-1cm}
 \begin{center}
    \includegraphics[width=.49\textwidth]{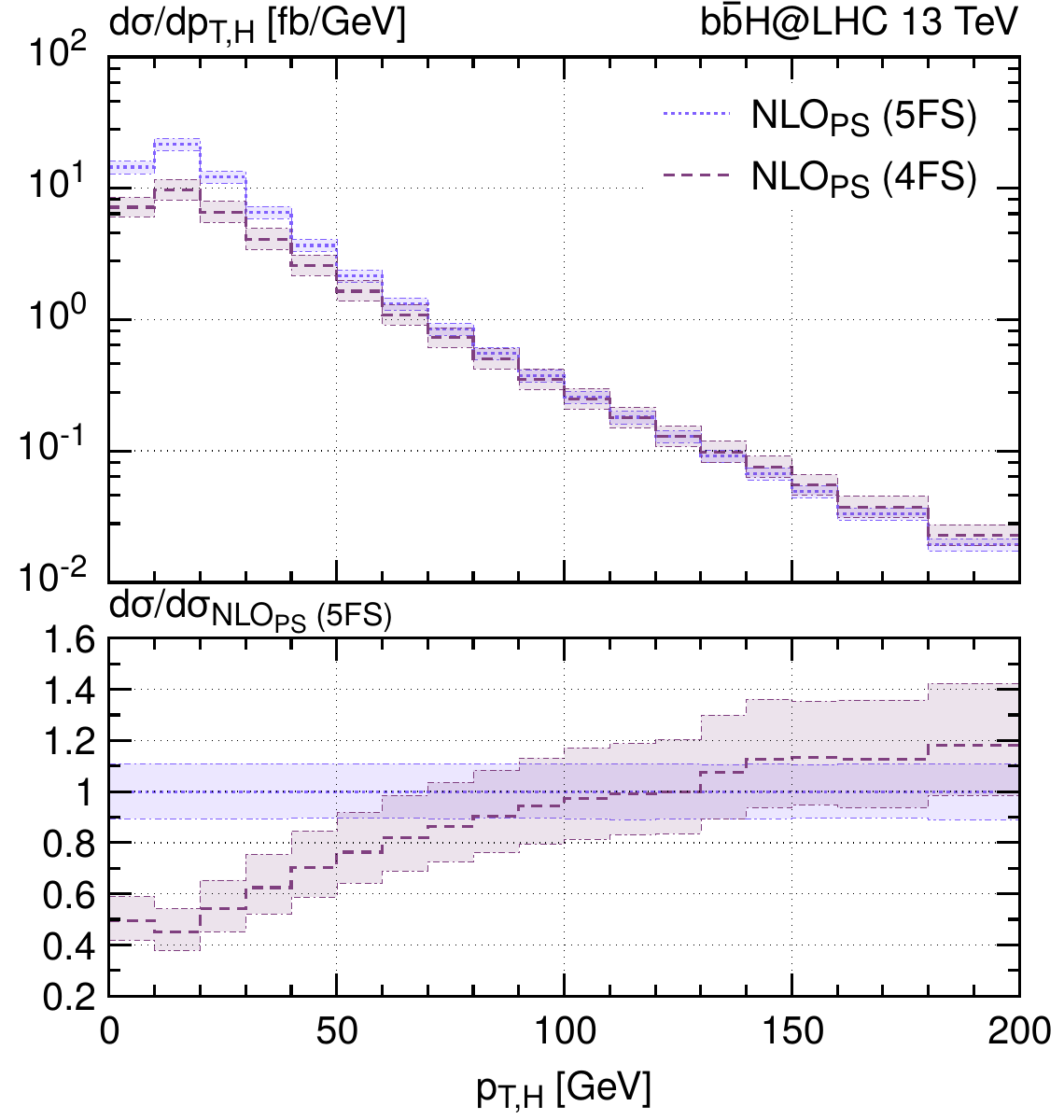}
    \includegraphics[width=.49\textwidth]{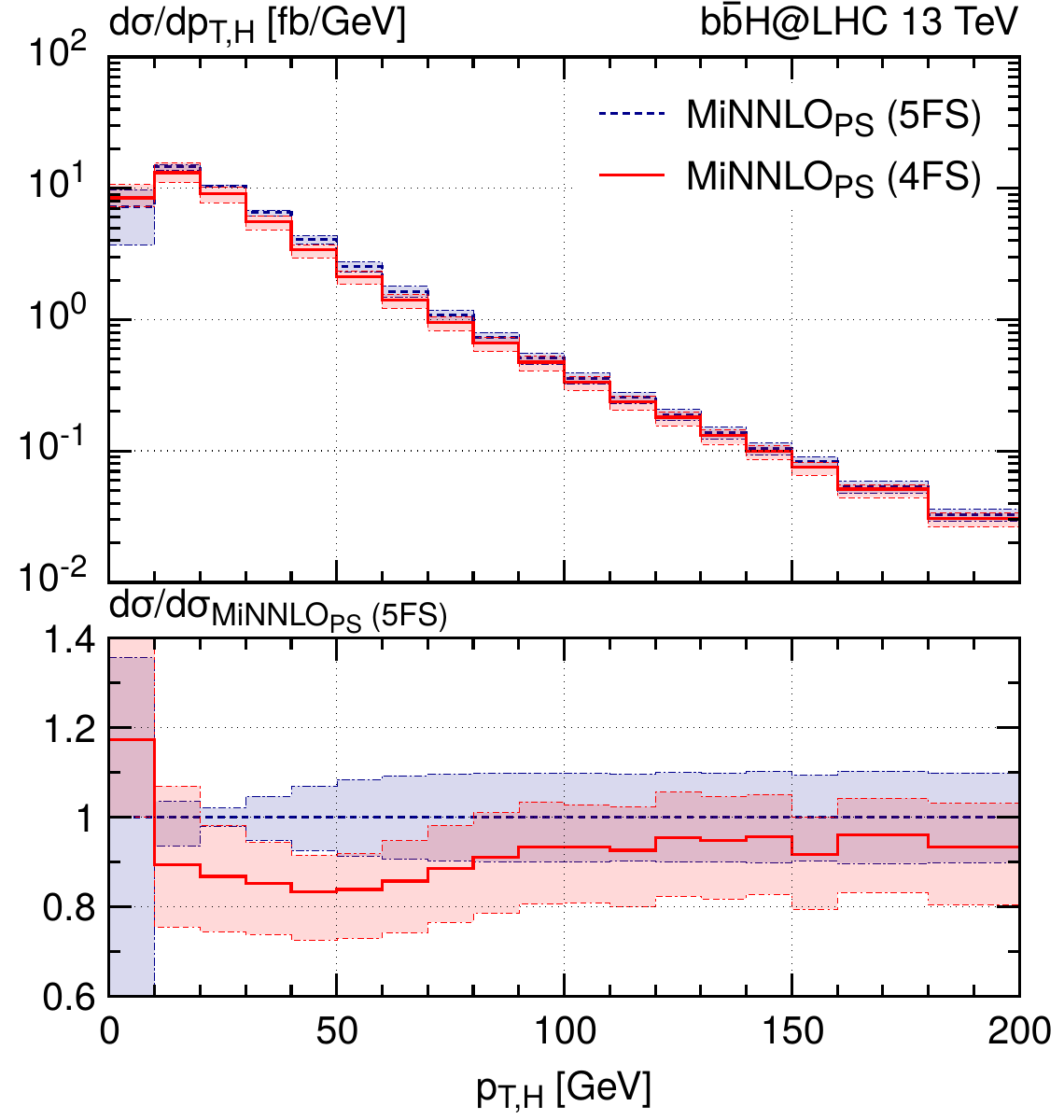}
    \includegraphics[width=.49\textwidth]{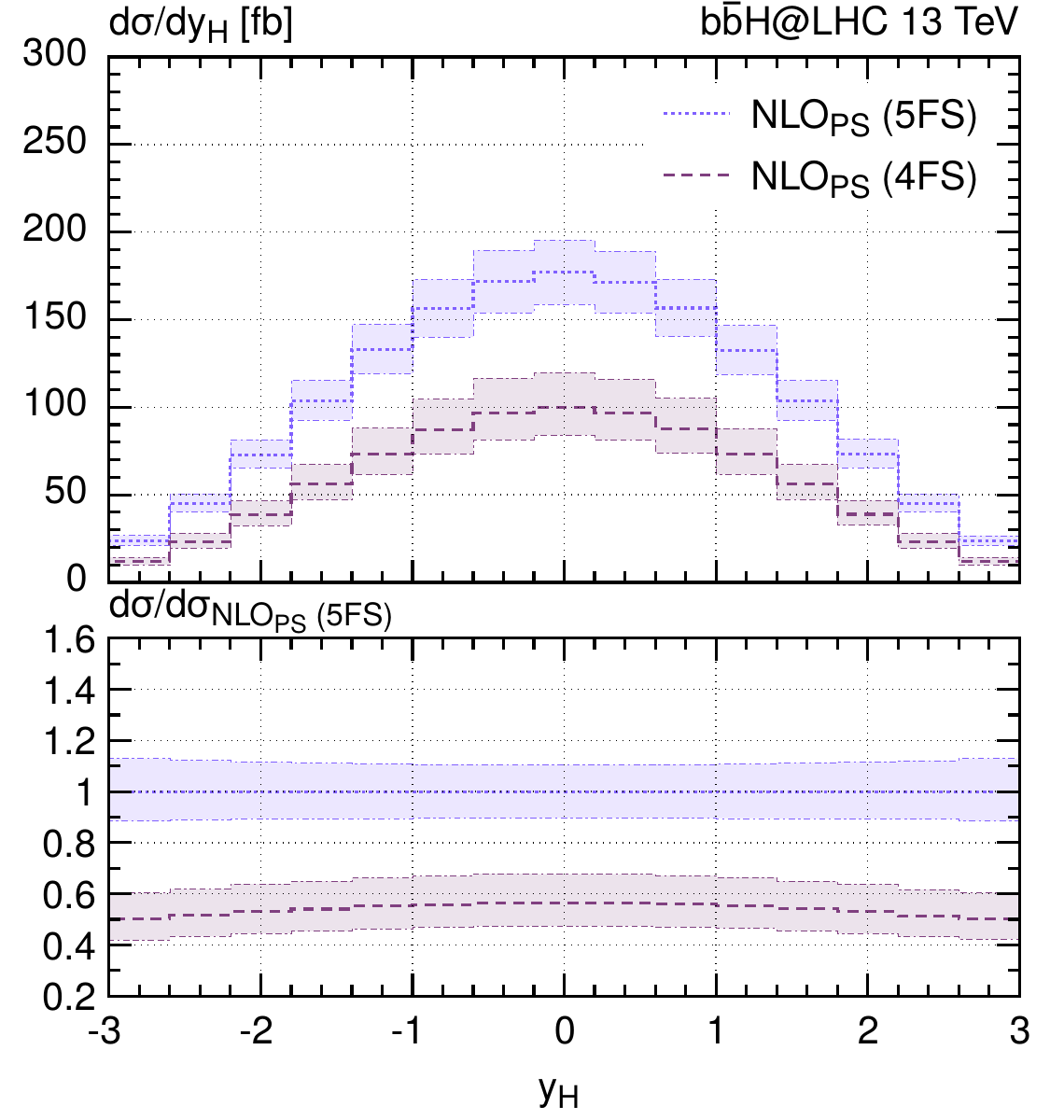}
    \includegraphics[width=.49\textwidth]{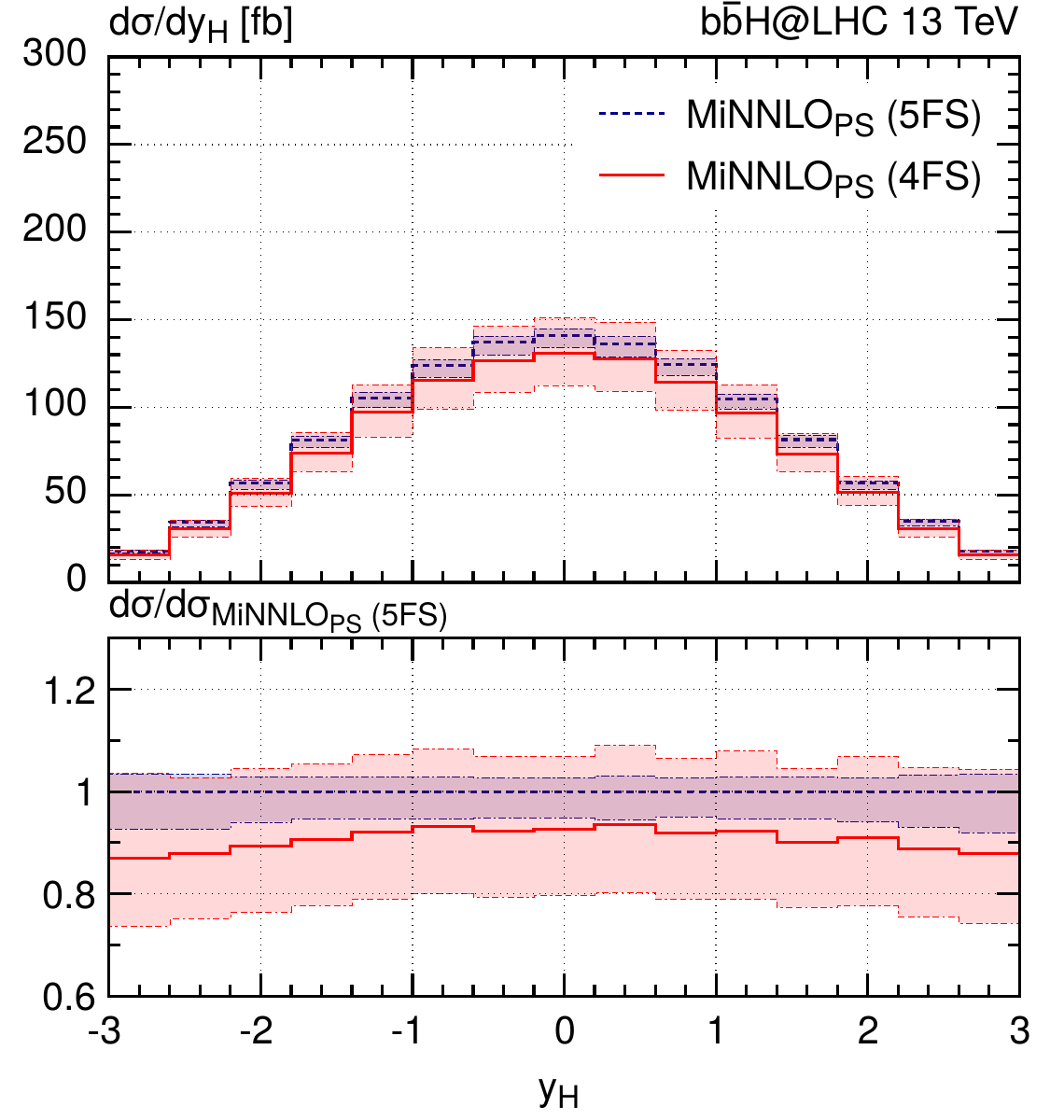}
	 \caption{Comparison of 5FS and 4FS predictions at NLO+PS (left) and \minnlo{} (right) for Higgs observables with $\mu_y^{(0)}=m_H$ as fixed-scale of the Yukawa coupling.}
 \label{fig:5FSvs4FSH0}
 \end{center}
 \end{figure*}
\afterpage{\clearpage}

\begin{figure*}[h]
\vspace{-1cm}
 \begin{center}
    \includegraphics[width=.49\textwidth]{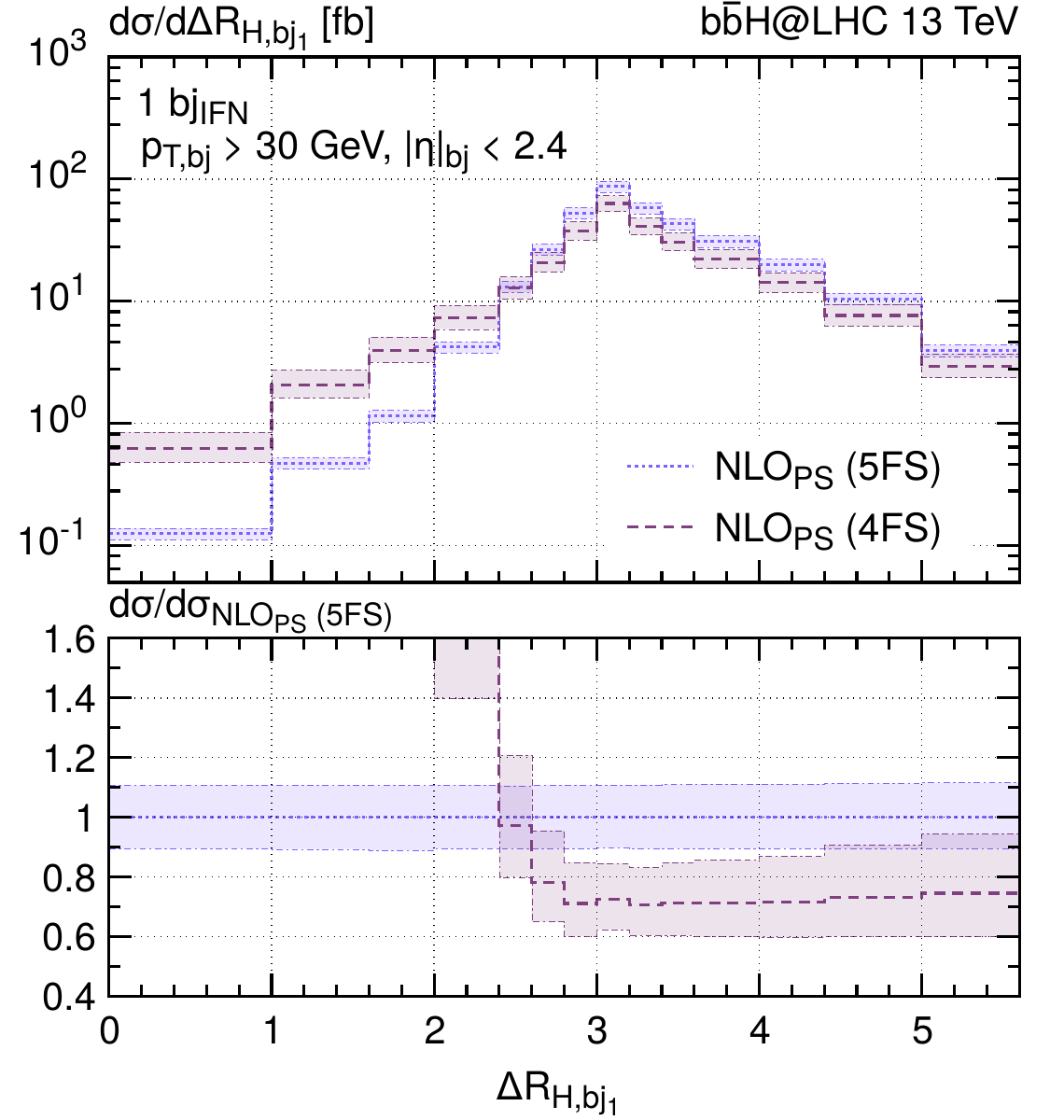}
     \includegraphics[width=.49\textwidth]{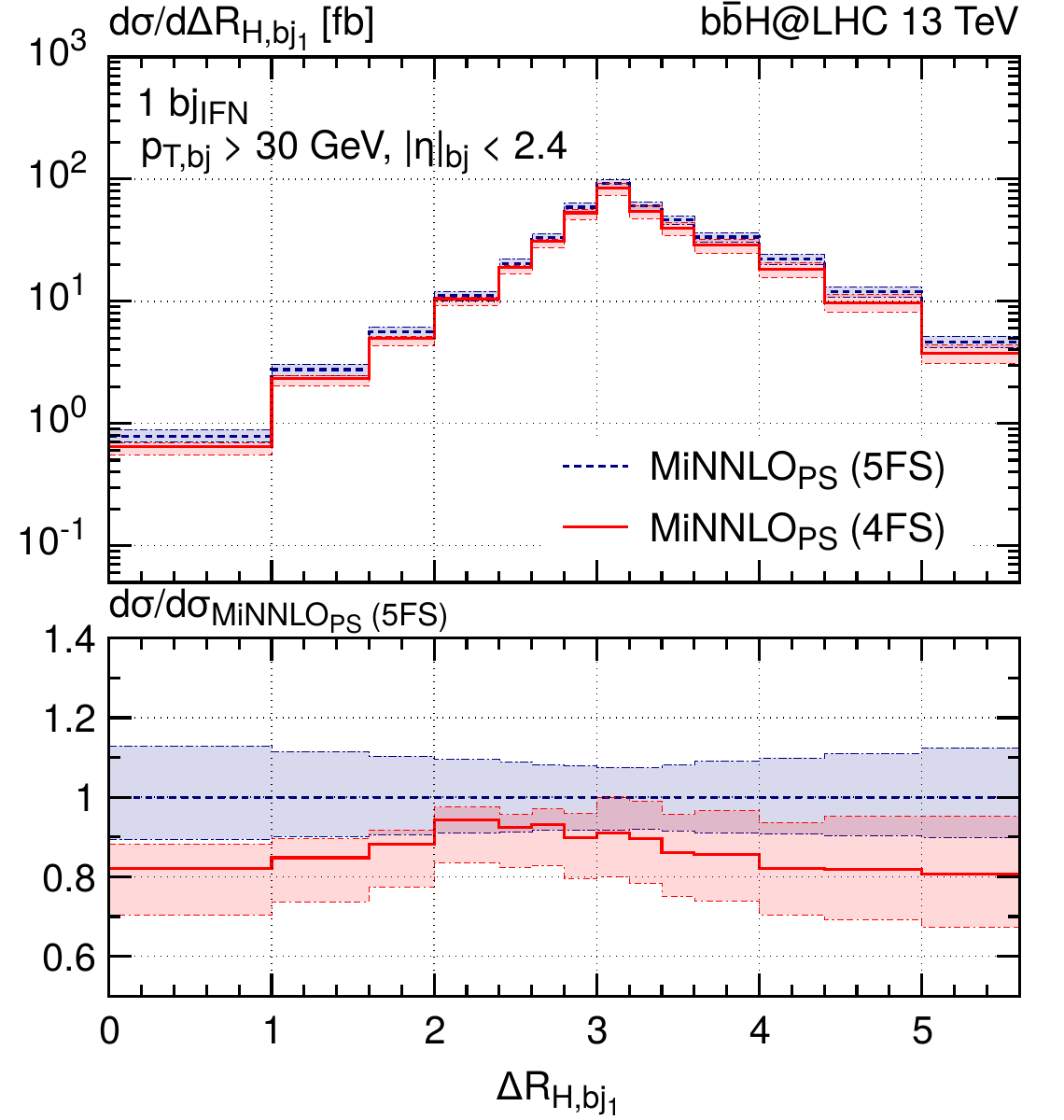}
     \includegraphics[width=.49\textwidth]{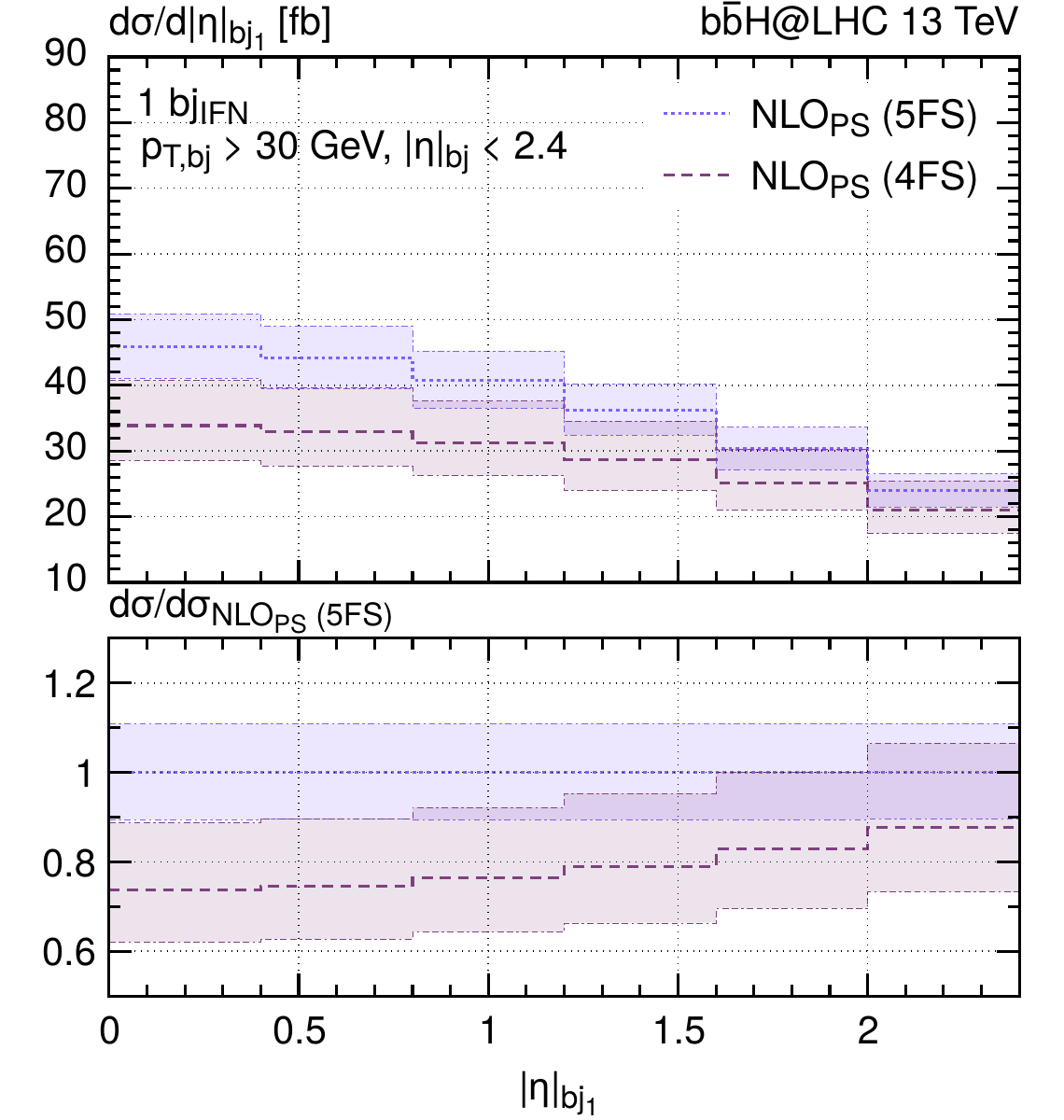}
     \includegraphics[width=.49\textwidth]{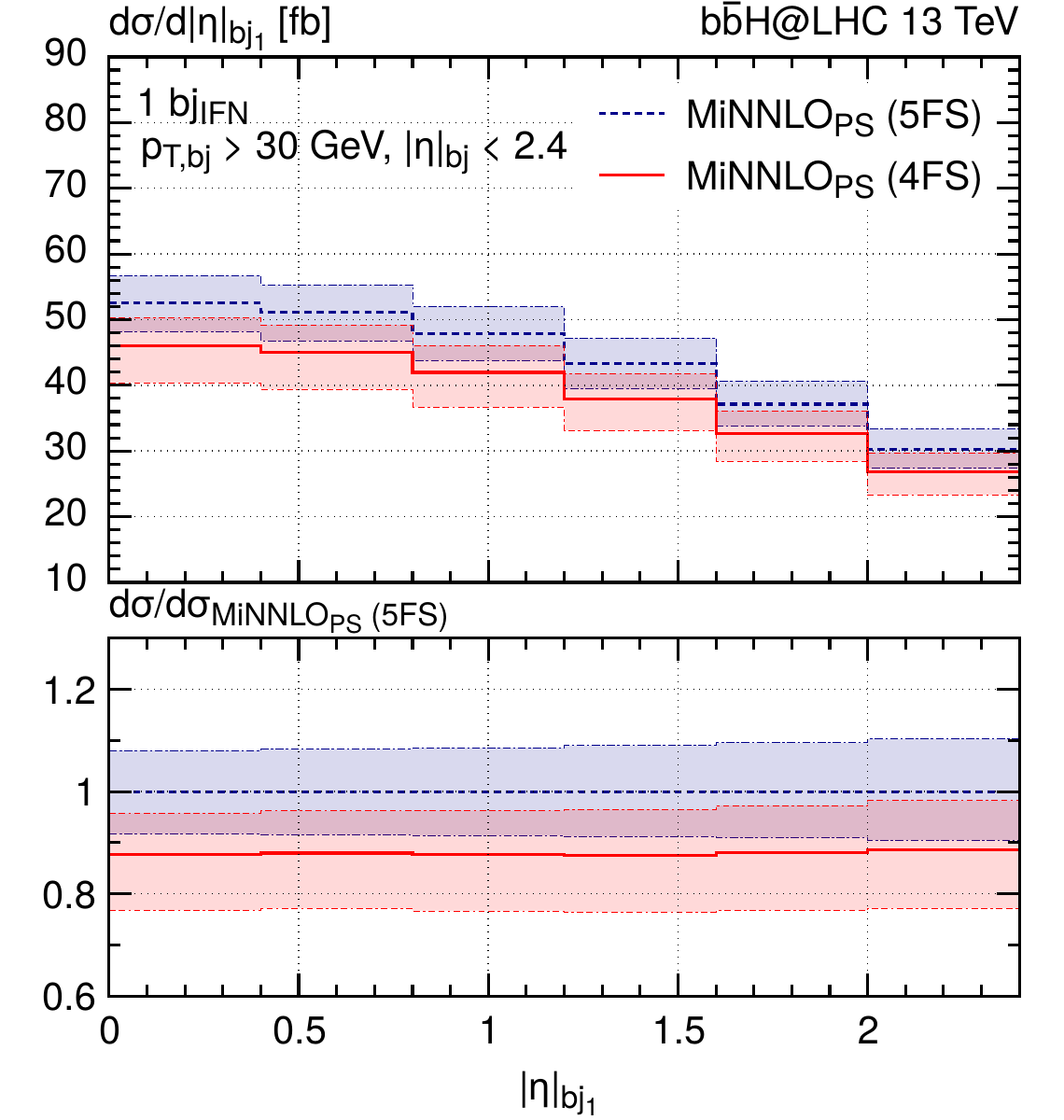}
         \includegraphics[width=.49\textwidth]{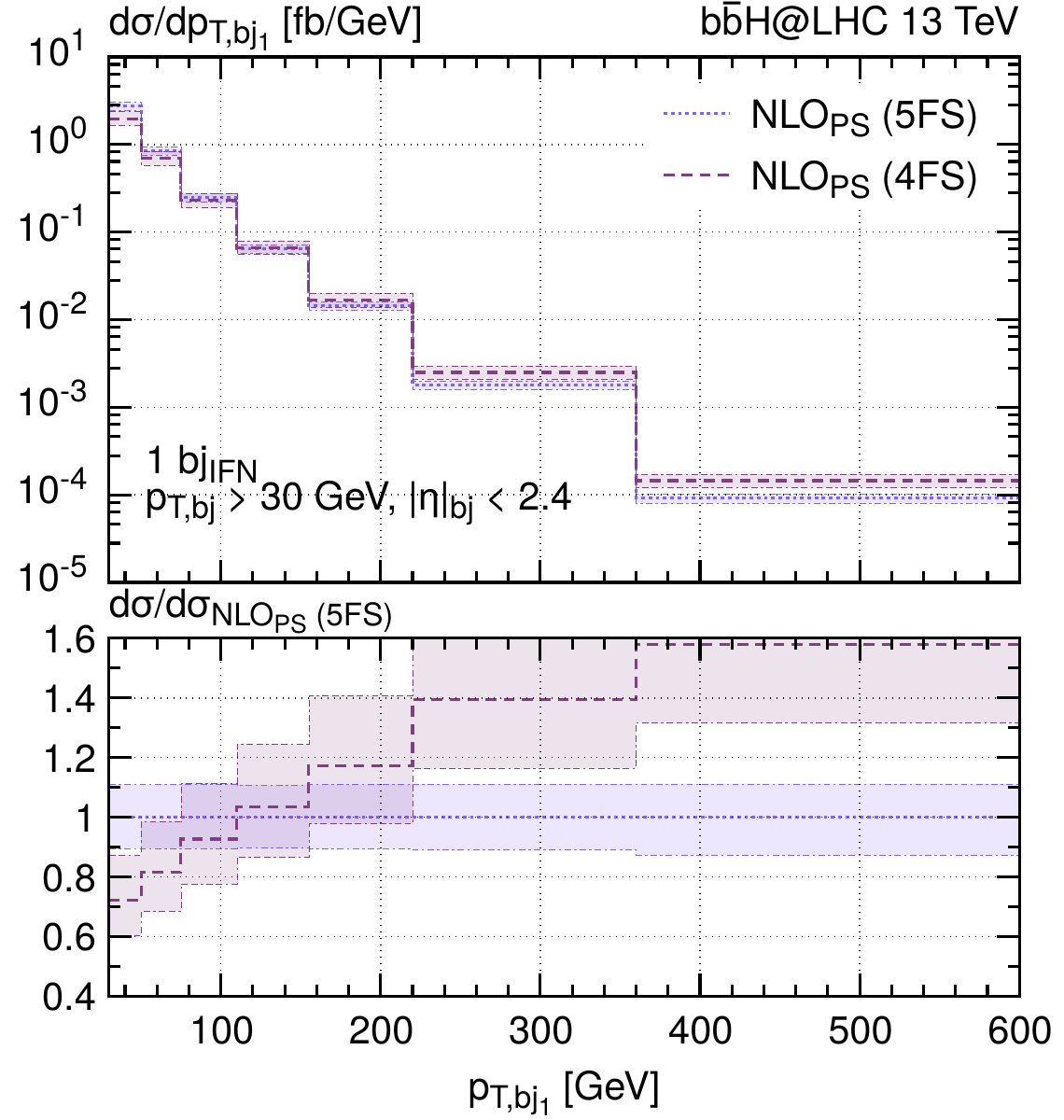}
     \includegraphics[width=.49\textwidth]{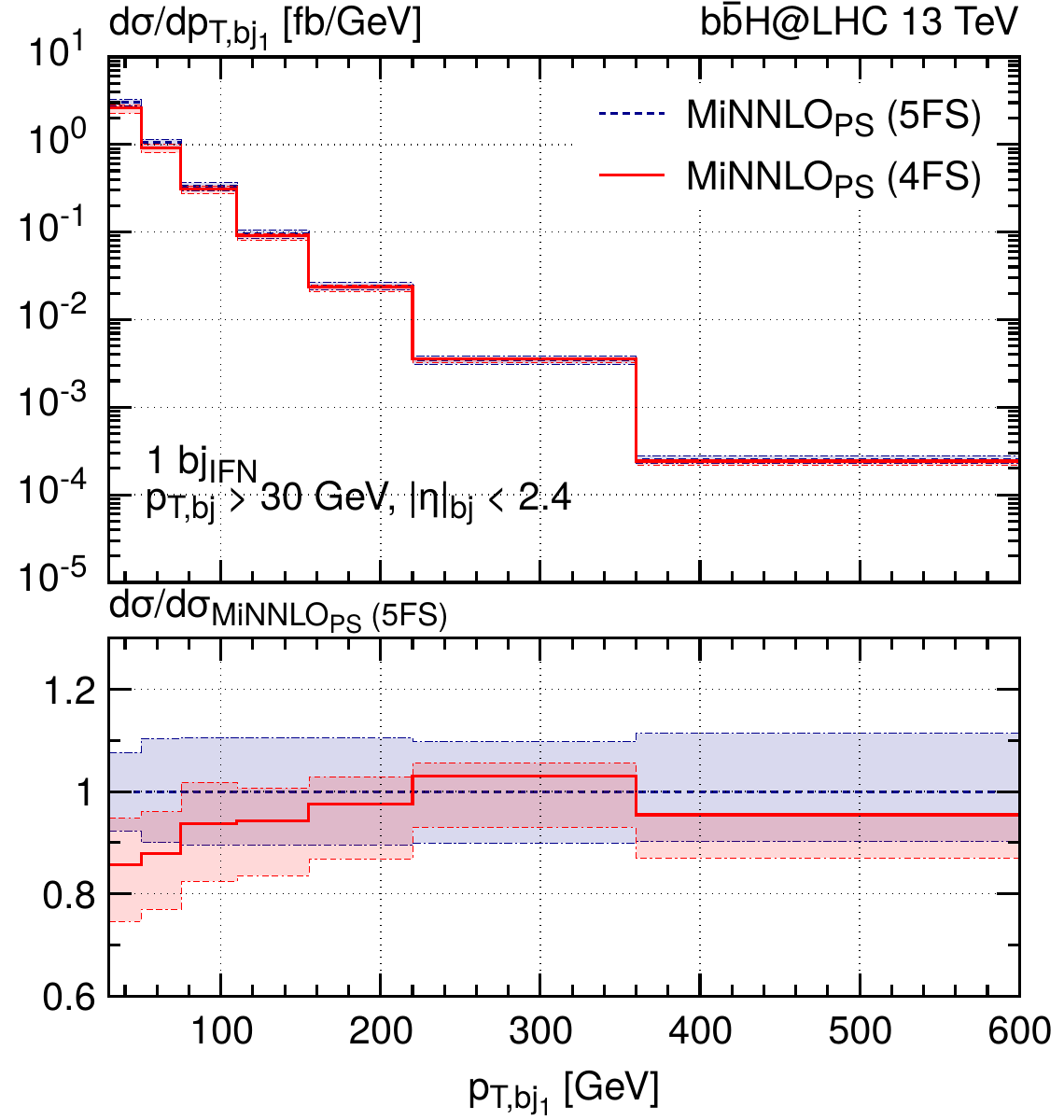}
 \caption{Comparison of 5FS and 4FS results at NLO+PS (left) and \minnlo{} (right) for observables with at least one $b$-jet defined using the \texttt{IFN} definition.}
 \label{fig:5FSvs4FSb1}
 \end{center}
 \end{figure*}
 \afterpage{\clearpage}
 
We begin the flavour-scheme comparison at the differential level by examining the distributions in Higgs transverse momentum and rapidity. Building upon the results presented in \citere{Biello:2024vdh}, which included comparisons with analytic predictions, we now incorporate our new 4FS 
NNLO+PS predictions in that comparison. In the left plot of \fig{fig:ptzoomNNLL}, the \minnlo{} predictions for the Higgs transverse-momentum 
spectrum are compared with the NNLO+NNLL predictions from \citere{Harlander:2014hya}, obtained using the standard setup and the NNPDF 4.0 sets consistent with the respective flavour scheme. By and large, we observe that the three predictions are consistent within their respective uncertainties, although they exhibit some differences in their central values, especially at small transverse momenta. 
However, we expect 
that this region is now well described with the newly developed 4FS \minnlo{} calculation, which can be considered to be
superior to the 5FS ones at small transverse momentum. This is because it includes power corrections in the bottom mass that become
crucial around $\pth \sim m_b$. Also, the uncertainties of the 4FS \minnlo{} predictions appear to be more robust, since the 5FS 
\minnlo{} scale bands are smaller than the more accurate NNLO+NNLL prediction at small transverse momenta.

In the right plot of \fig{fig:ptzoomNNLL}, the Higgs rapidity distribution from the two \minnlo{} generators are in full agreement with the NNLO fixed-order results from \citere{Mondini:2021nck}, which employs the NNLO set of the CT14 PDFs~\cite{Dulat:2015mca} \footnote{We thank the authors of \citere{Mondini:2021nck} for providing us with the relevant results.}. Note that for this particular comparison, we used the CT14 PDF sets in our \minnlo{} generators, both in the 5FS and 4FS. The agreement is particularly
good in the central rapidity region. From this point onward, we revert to the default setting, using the NNPDF4.0 set for all subsequent simulations.

We continue with the flavor-scheme comparison at NLO+PS and NNLO+PS. The first row of \fig{fig:5FSvs4FSH0} shows the Higgs 
transverse-momentum distribution at NLO+PS (left) and NNLO+PS (right). 
At NLO+PS, a significant discrepancy is observed in the small $\pth$ region  between the 4FS and 5FS predictions, while the \minnlo{} generators substantially improve the agreement between the two schemes in that region and achieve excellent consistency for 
$\ptarg{H}>75$ GeV. In this range, the NNLO corrections in the 4FS are flat and well-contained within the scale uncertainties.

 \begin{figure*}[t]
 \begin{center}  
   \includegraphics[width=.49\textwidth]{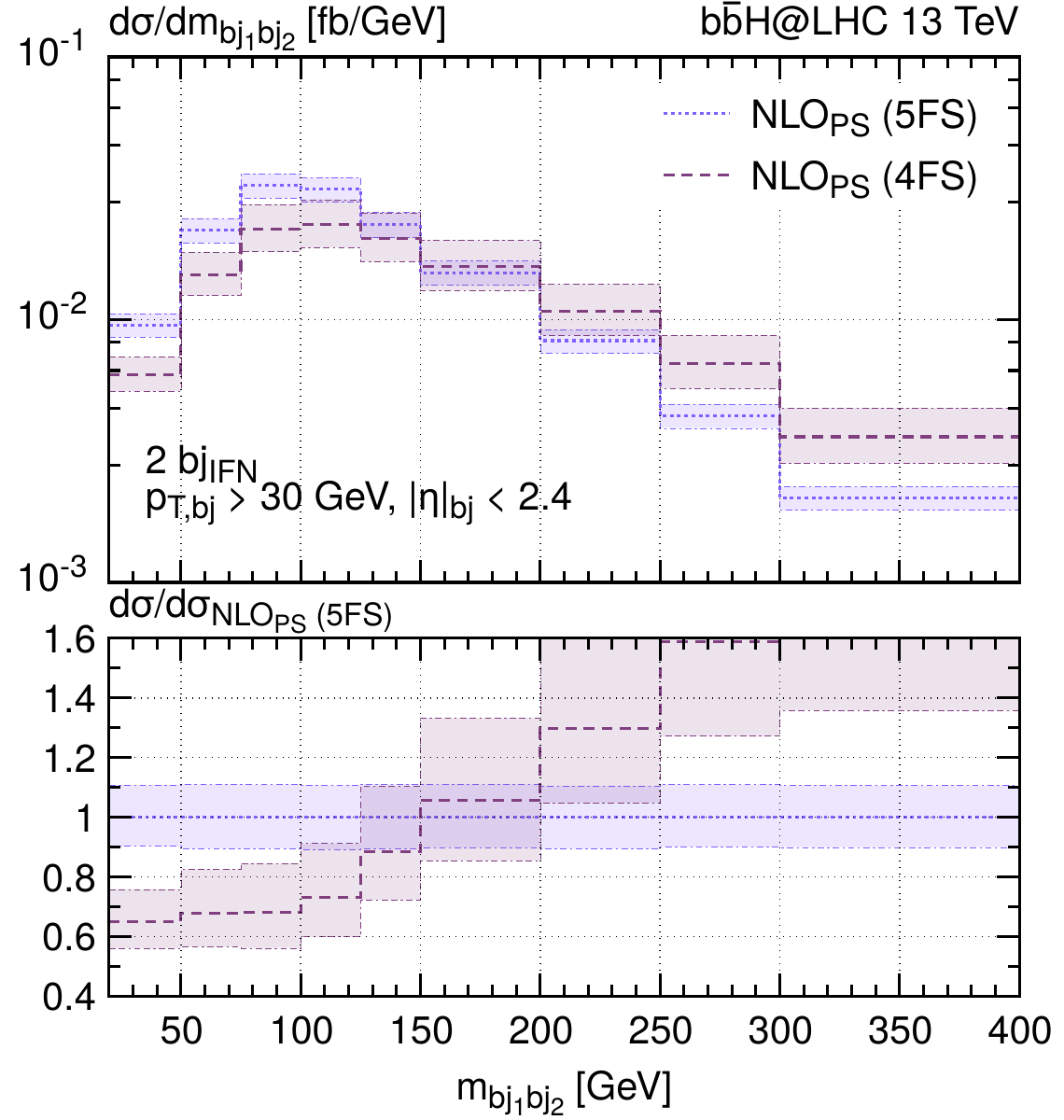}
   \includegraphics[width=.49\textwidth]{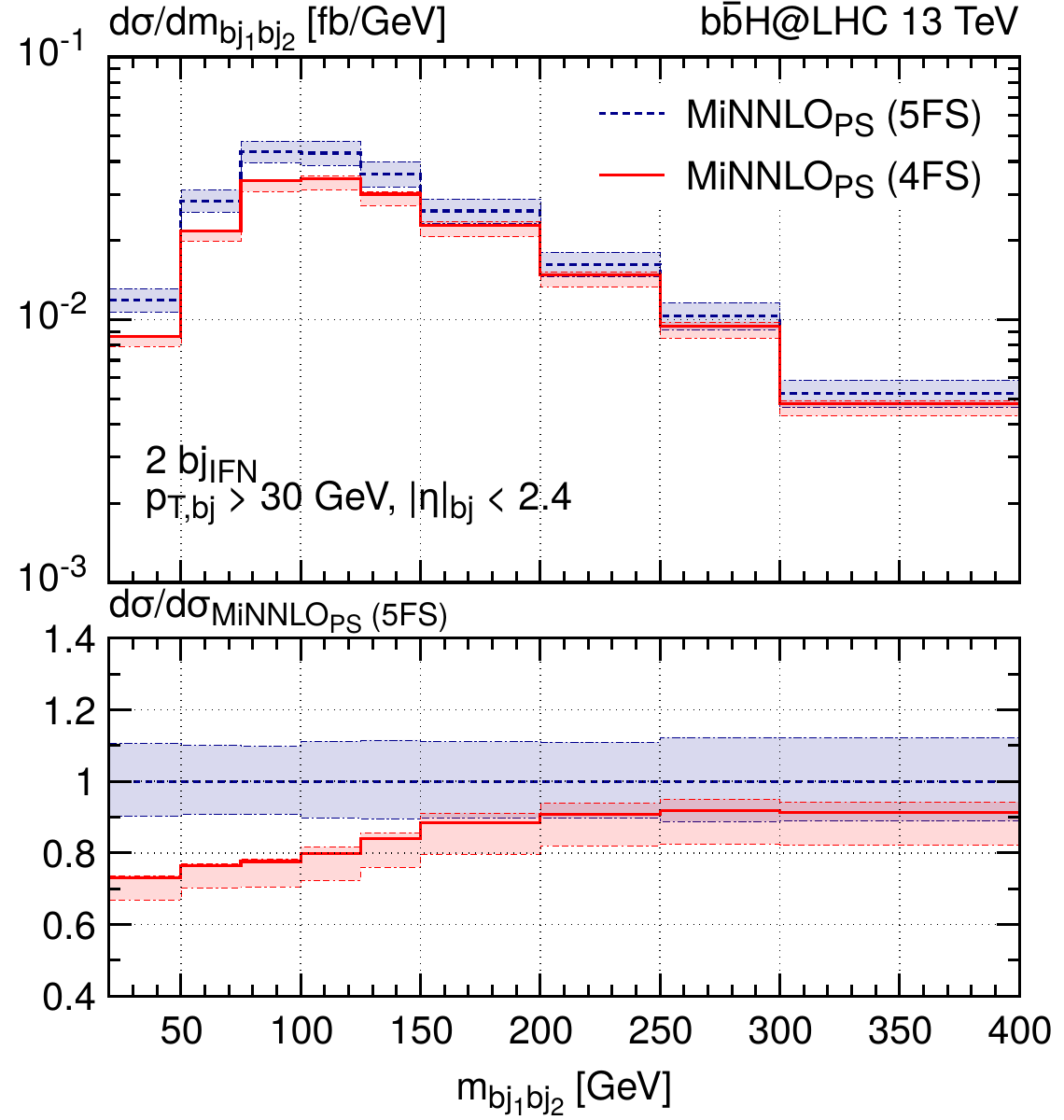}
     \includegraphics[width=.49\textwidth]{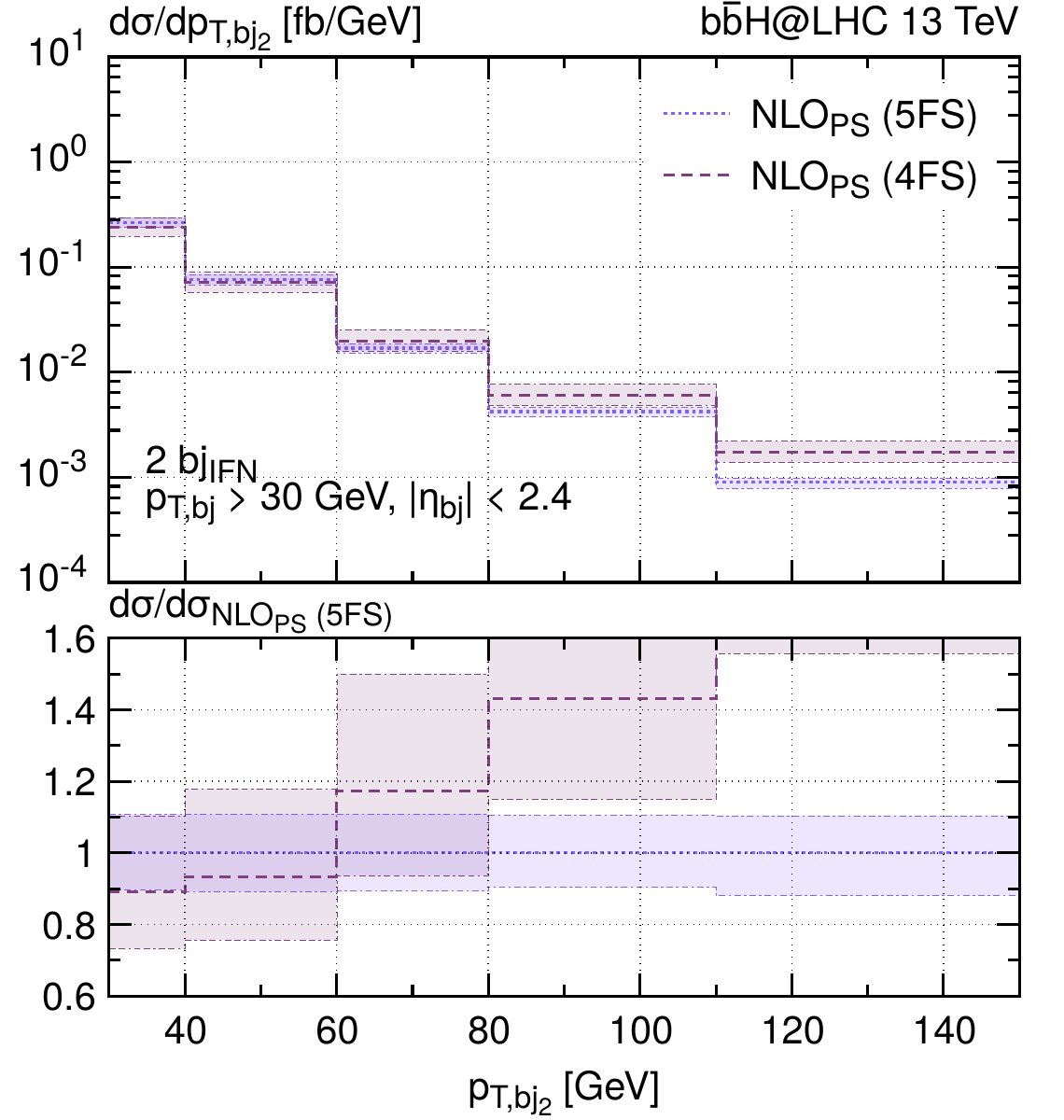}
       \includegraphics[width=.49\textwidth]{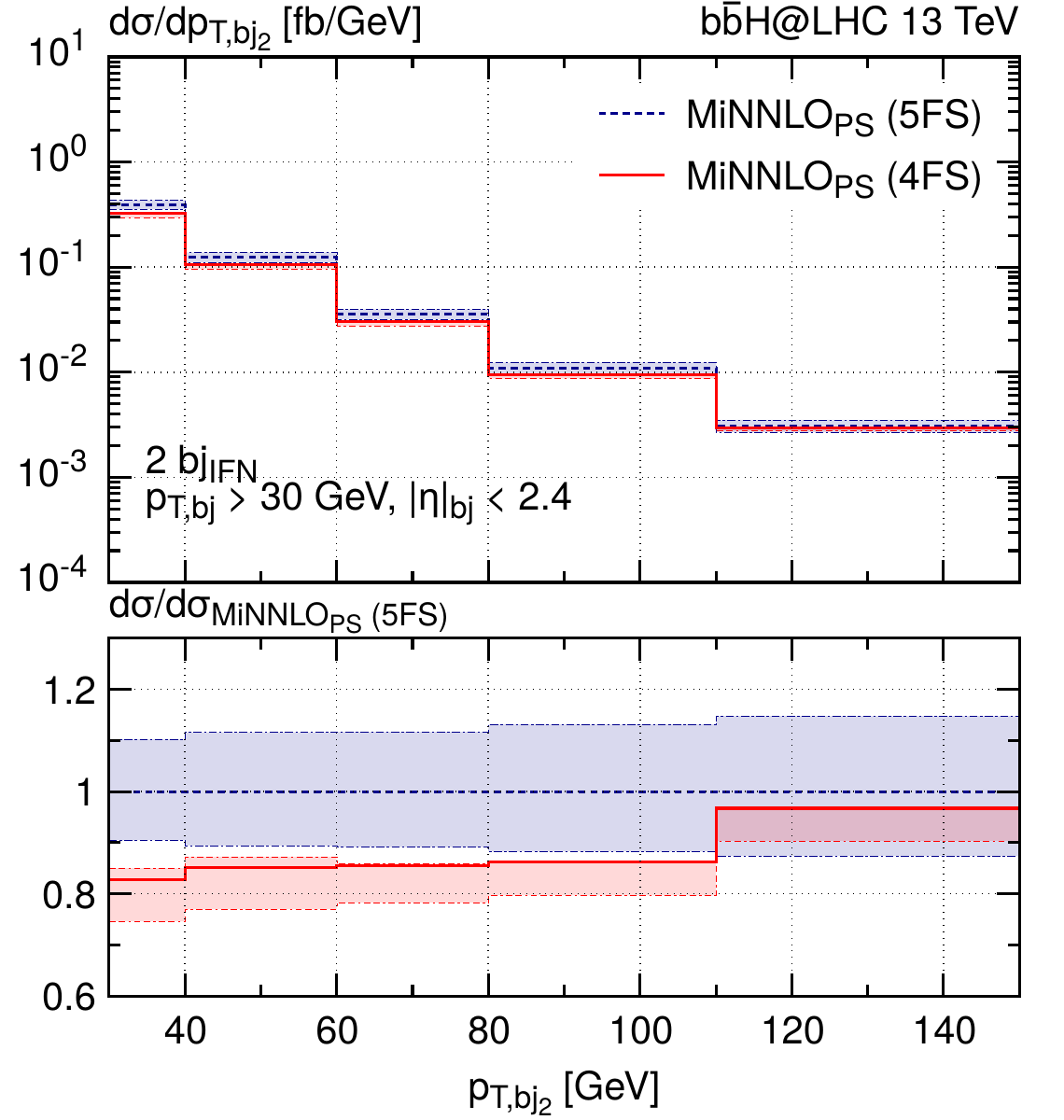}
   \caption{Comparison of 5FS and 4FS predictions at NLO+PS (left) and \minnlo{} (right) for the observables with at least 2 \texttt{IFN} $b$-jets.}
   \label{fig:5FSvs4FSb2b}
\end{center}
\end{figure*}
 
The second row of \fig{fig:5FSvs4FSH0} presents the Higgs rapidity spectrum, where the NLO+PS comparison (left panel) shows 
a substantial discrepancy between the two schemes, which differ by a factor of two or more, well outside the respective uncertainty bands.
By contrast, the NNLO+PS predictions of the two \minnlo{} generators (right panel) are fully consistent within the uncertainty bands. 
These findings strongly indicate that the long-standing tension between the two schemes is fully resolved by the newly computed NNLO QCD
corrections in the 4FS.

Figure\,\ref{fig:5FSvs4FSb1} depicts observables for events tagged with one $b$-jet using the \texttt{IFN} definition. The separation in the $(\eta-\phi)$ plane between the Higgs and the leading $b$-jet reveals a substantial shape discrepancy between the NLO+PS generators (left panel). 
With the inclusion of NNLO corrections in the right panel, the agreement improves significantly.
The shapes become much more similar and the uncertainty bands start overlapping. 
In the second row of \fig{fig:5FSvs4FSb1}, the pseudo-rapidity spectrum of the leading $b$-jet is in reasonable agreement within uncertainties
both at NLO+PS and at NNLO+PS. However, in the latter case, the differences between the two schemes are 
reduced from about 25\% to 10\% and they show very similar shapes.
The last two plots in \fig{fig:5FSvs4FSb1} show the transverse-momentum spectrum of the hardest $b$-jet. 
Again the predictions in 4FS and 5FS, especially in terms of shape, are much closer for the \minnlo{} generators,
with reduced uncertainty bands throughout.

We conclude this section by analysing observables based on events with at least two $b$-jets. 
The first row of \fig{fig:5FSvs4FSb2b} presents the invariant-mass distribution of the two leading $b$-jets. While the shapes at NLO+PS are vastly
different for the two schemes, the agreement is clearly improved upon the inclusion of NNLO corrections, particularly in the high-mass region.
However, notable discrepancies up to 30\% persist below 150 GeV. Despite the clear improvement at NNLO+PS
the difference between the two schemes is not covered by the respective scale uncertainties in that region. 
We emphasise that for these observables the 5FS prediction achieves effectively only LO+PS accuracy, 
while the 4FS provides an NNLO+PS description. This can be seen from the larger uncertainty bands in the 5FS.
The plots in the lower panel of the same figure display the transverse-momentum spectrum of the subleading 
$b$-jet, where we observe a very similar behaviour as for the previous observable.

Regarding the scale variation, we observe that the 4FS NNLO+PS results yield a significantly more conservative uncertainty than the 5FS ones 
for inclusive predictions. This behaviour, shown in \fig{fig:5FSvs4FSH0}, is consistent with the trend seen in other processes where 
bottom quarks appear in the final state of the hard scattering. When requiring at least one $b$-jet, the two \minnlo{} predictions 
exhibit similar scale-variation bands, as can be seen in \fig{fig:5FSvs4FSb1}. 
In contrast, \fig{fig:5FSvs4FSb2b} displays a narrower 
uncertainty band for the 4FS prediction compared to the 5FS one, reflecting the higher accuracy of the massive scheme, 
while the massless scheme is only LO+PS accurate.

\section[Background for $HH$ searches in the $2b2\gamma$ channel]{Background for \boldmath{$HH$} searches in the \boldmath{$2b2\gamma$} channel}
\label{sec:HH}
In this section, we examine a potential application of our novel \minnlo{} generator in the 4FS to model the \bbH{} background to the
di-Higgs ($HH$) signal. For an illustrative \(HH\) search, we consider the \(2b2\gamma\) decay mode, in which one Higgs boson decays into \(b\)-quarks and the other into photons. Ref.\,\cite{Manzoni:2023qaf} investigates the impact of the \bbH{} background on $HH$ searches, considering both $y_b^2$ and $y_t^2$ contributions at NLO+PS within the \textsc{MG5\_aMC@NLO} framework \cite{Alwall:2014hca}. Specifically, they analysed \bbH{} production with the decay $H \to \gamma\gamma$ and compared both the total and various fiducial rates at NLO against the $HH$ signal \cite{Heinrich:2017kxx}. The findings of \citere{Manzoni:2023qaf} reveal that NLO QCD corrections are substantial and essential for obtaining an accurate prediction of the \bbH{} background. Corrections are notably larger for the \ytsq{} contributions, enhancing the leading-order (LO) prediction by approximately $+150\%$, while the impact on the \ybsq{} terms is around $+50\%$. The residual uncertainties, arising from variations in renormalisation and factorisation scale, are about $+50\%$ and $-30\%$ for the combined \ybsq{} and \ytsq{} contributions to the \bbH{} cross section. The study in \citere{Manzoni:2023qaf} also suggests that current limits on $\sigma^{HH}_{\rm SM}$ will improve by a few percent in the $2b2\gamma$ final state
when modelling the \bbH{} background at NLO+PS, and that HL-LHC constraints on $\sigma^{HH}_{\rm SM}$ and the $HH$ discovery significance could improve by 5\%. We note that the estimated improvements are even larger in the $2b2\tau$ channel~\cite{Manzoni:2023qaf}.

By the end of HL-LHC, Higgs pair production is anticipated to be observed at nearly five standard deviations. As a result, accurate modelling of the \bbH{} background in appropriate fiducial phase space regions is critical for improving sensitivity to the $HH$ signal. Given the importance of precise \bbH{} background modelling for $HH$ searches, we aim to improve the accuracy of this background in the $y_b^2$ channel. Specifically, we provide novel NNLO predictions for \bbH{} production proportional to $y_b^2$, including the decay $H \to \gamma\gamma$, using our new \minnlo{} generator in the 4FS. We employ \PYTHIA{8} \cite{Bierlich:2022pfr} to model the Higgs boson decay to two photons in the narrow-width approximation.
The branching fraction is taken to be \({\rm BR}(H \to \gamma\gamma) = 0.227\%\)~\cite{LHCHiggsCrossSectionWorkingGroup:2016ypw}. 

We use the same setup as described in \sct{sec:setup}, except for the phase-space cuts on jets, where we follow the approach given in \citere{ATLAS:2021ifb}. Specifically, we consider anti-$k_T$ jets~\cite{Cacciari:2008gp} with a radius parameter $R=0.4$ as implemented in {\sc FastJet}~\cite{Cacciari:2011ma}. Bottom-flavored jets (b-jets) are defined according to the \texttt{EXP} definition used previously. In this theoretical study, the criteria for selecting jets are as follows:
\begin{equation}
\label{eq:ptjcuts}
p_T(j) > 25,\GeV \qquad {\rm and} \qquad |\eta(j)| < 2.5\,.
\end{equation}
We define the $HH$ signal region by selecting events that contain exactly two $b$-jets and two photons, with QED showering turned off in the simulations. We also impose a cut on the invariant mass of the $b$-jet pair:
\begin{equation}
    \label{eq:mbjcut}
    80\, \GeV < m(b_1,b_2)< 140\,\GeV\,.
\end{equation}
The photon pair is required to meet the following conditions:\footnote{These selection criteria are similar to those used in the ATLAS $HH$ search presented in \citere{ATLAS:2021ifb}}
\begin{equation}
\label{eq:photoncuts}
    105\,\GeV < m(\gamma_1, \gamma_2) < 160\,\GeV, \quad |\eta(\gamma_i)|< 2.37, \quad
    \frac{p_T(\gamma_1)}{m(\gamma_1, \gamma_2)} > 0.35, \quad \frac{p_T(\gamma_2)}{m(\gamma_1, \gamma_2)} > 0.25\,.
\end{equation}
Since the Higgs boson is produced on-shell in our generator and \PYTHIA{8} handles the decay without including QED shower effects, the narrow-width approximation trivially satisfies the condition $m(\gamma_1, \gamma_2) - m_H = \mathcal{O}(\Gamma_H)$.

In addition to the aforementioned cuts, hereafter referred to as \texttt{fiducial cuts}, we introduce the following quantities for the invariant mass of the diphoton plus b-tagged jets system:
\begin{equation}
m_{2b2\gamma} = m(b_1,b_2,\gamma_1,\gamma_2)\,,
\end{equation}
and
\begin{equation}
\label{eq:mbbggs}
    \mbbggs = m_{2b2\gamma} - m(b_1,b_2) - m(\gamma_1,\gamma_2) + 2m_H \,.
\end{equation}
We then consider three different event categories by imposing the following cuts on $\mbbggs$:
\begin{equation}
\label{eq:mbbggscuts}
    \mbbggs < \infty, \qquad \mbbggs < 500\,\GeV, \qquad \mbbggs < 350\,\GeV \,.
\end{equation}
The first condition represents the \texttt{fiducial cuts}, while the other two impose progressively stricter requirements on $\mbbggs$.

\subsection{Inclusive predictions}
\begin{table}[ht!]
  \vspace*{0.3ex}
  \begin{center}
	   \renewcommand{\arraystretch}{1.2}
\begin{tabular}{|c|c|c|c|}
  \hline 
	Fiducial region  & 4FS NLO+PS  & 4FS \minnlo{} & Ratio to NLO+PS  \\
  \hline \hline
   \begin{tabular}{@{}c@{}} No cut \end{tabular}
	   & $805_{-16\%}^{+20\%}$ & $1056_{-14\%}^{+16\%}$ & 1.312\\
  \hline
   \begin{tabular}{@{}c@{}} Fid. cuts \end{tabular}
	   &$3.56_{-18\%}^{+24\%}$ & $5.10_{-9.5\%}^{+2.1\%}$ & 1.432 \\
   \hline
  \begin{tabular}{@{}c@{}} Fid. cuts \\+ $\mbbggs<500\,\GeV$  \end{tabular}
	  & $3.50_{-18\%}^{+24\%}$ & $5.04_{-9.6\%}^{+2.4\%}$ & 1.440\\
   \hline
    \begin{tabular}{@{}c@{}} Fid. cuts \\+ $\mbbggs<350\,\GeV$  \end{tabular}
	    & $3.01_{-18\%}^{+25\%}$ & $4.38_{-9.7\%}^{+2.8\%}$ & 1.455 \\
   \hline
\end{tabular}
  \end{center}
  \vspace{-1em}
  \caption{
Cross sections in ab for the $y_b^2$ contribution to $pp \to b \bar b H$ with $H\to \gamma\gamma$ decay. \label{tab:XS-fid}}
\end{table}

We present the total and fiducial rates for \bbH{} production in the 4FS, including the decay $H \to \gamma\gamma$, at NLO+PS and for \minnlo{} 
in \tab{tab:XS-fid}. The fiducial cuts, which correspond to the $HH$ signal region, are defined according to the definitions 
in \neqn{eq:ptjcuts}--\eqref{eq:photoncuts}, and with the additional constraints on $\mbbggs$ given in \eqn{eq:mbbggscuts}.
For the total inclusive cross section, the NNLO corrections yield a $+30\%$ increase with respect to the NLO result, with scale uncertainties of about $+16\%$ and $-14\%$, which is in line with the inclusive results presented in \tab{tab:XS}. In the fiducial categories, the NNLO corrections contribute about $+40\%$ with respect to the NLO prediction, while scale uncertainties are significantly reduced to $+2\%$ and $-9\%$. 
The \texttt{fiducial cuts} significantly reduce the \bbH{} background, lowering the cross section by roughly a factor of 100 compared to the 
total inclusive cross section. 
Imposing additional constraints on $\mbbggs$ further lowers the cross section, with a reduction of approximately $1\%$ for $\mbbggs<500\,$GeV and $14\%$ for $\mbbggs<350\,$ GeV, respectively.
However, in all categories the $y_b^2$ contribution to the \bbH{} cross section remains at the same order as the $HH$ signal cross section,
making its accurate modelling so crucial. Given the large corrections in the $y_b^2$ contribution to the \bbH{}  background, our analysis
clearly shows the relevance of including NNLO terms in this context.

\subsection{Differential distributions}
\begin{figure*}[h!]
 \begin{center}
 \includegraphics[width=.495\textwidth]{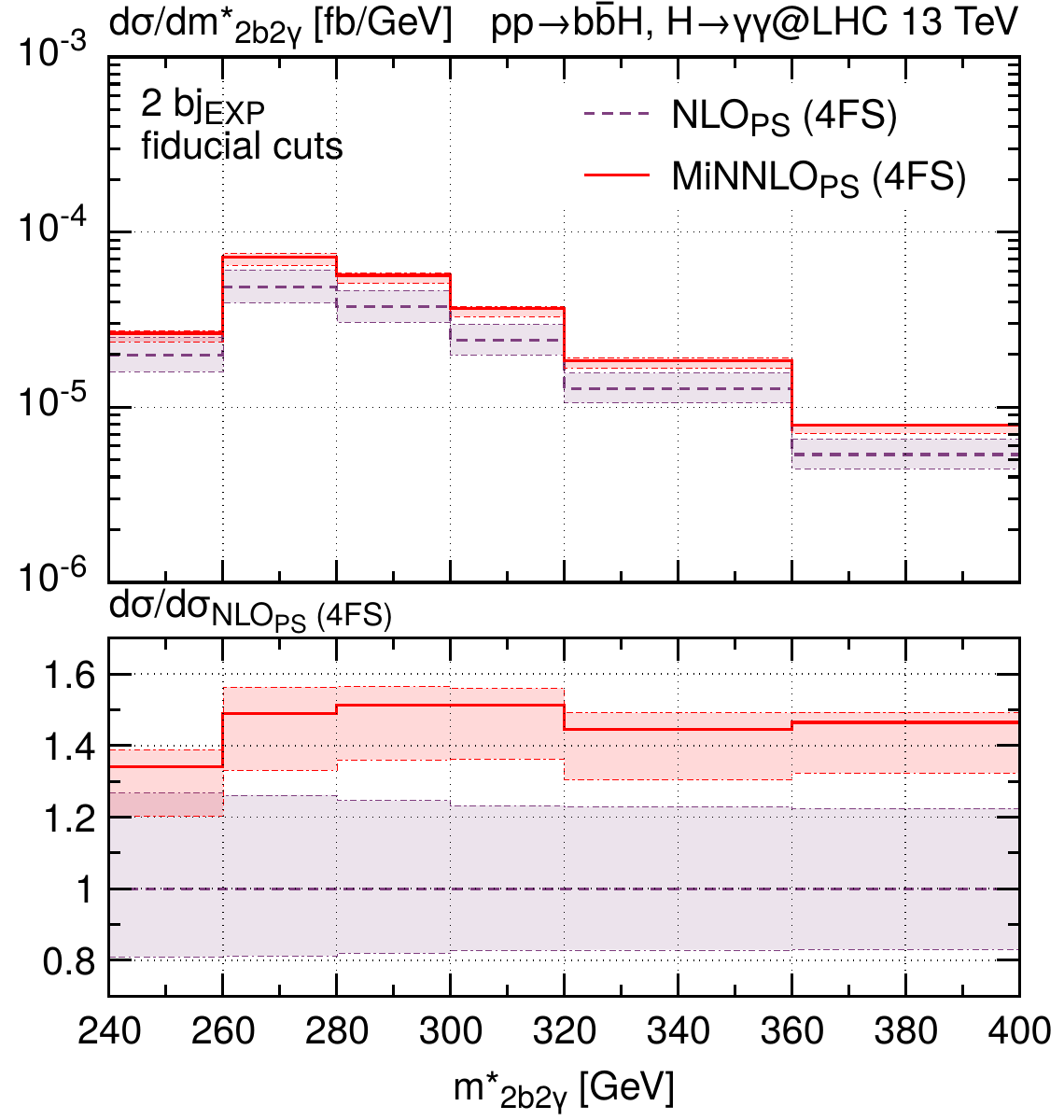}
	 \caption{Comparison of the $\mbbggs{}$ distribution at NLO+PS and \minnlo{} including fiducial cuts.}
 \label{fig:NLOvsMiNNLOoffshell1}
 \end{center}
 \end{figure*}

Finally, we study differential distributions, which provide deeper insights into the kinematics of the \bbH{} process
in the $HH$ signal region. These results could be valuable for understanding which phase-space regions might help reduce the \bbH{} background in $HH$ measurements or, more broadly, for assessing how different cuts affect the \bbH{} process. We start with the $\mbbggs{}$ distribution in \fig{fig:NLOvsMiNNLOoffshell1} within the phase space defined by the \texttt{fiducial cuts}. Both the NLO+PS and \minnlo{} spectra show a suppression as we go to higher values of $\mbbggs{}$.
Comparing the NLO and NNLO predictions, we observe that the NNLO corrections are positive and substantial, with an enhancement of about $40\%$--$50\%$ over the considered $\mbbggs{}$ range. There is only a slight shift in the distribution shape at low $\mbbggs{}$. Scale uncertainties are significantly at NNLO+PS, with no overlap between the NLO and \minnlo{} uncertainty bands, highlighting the importance of the newly calculated NNLO corrections for this process.

\begin{figure*}[t]
 \begin{center}
 \includegraphics[width=.495\textwidth]{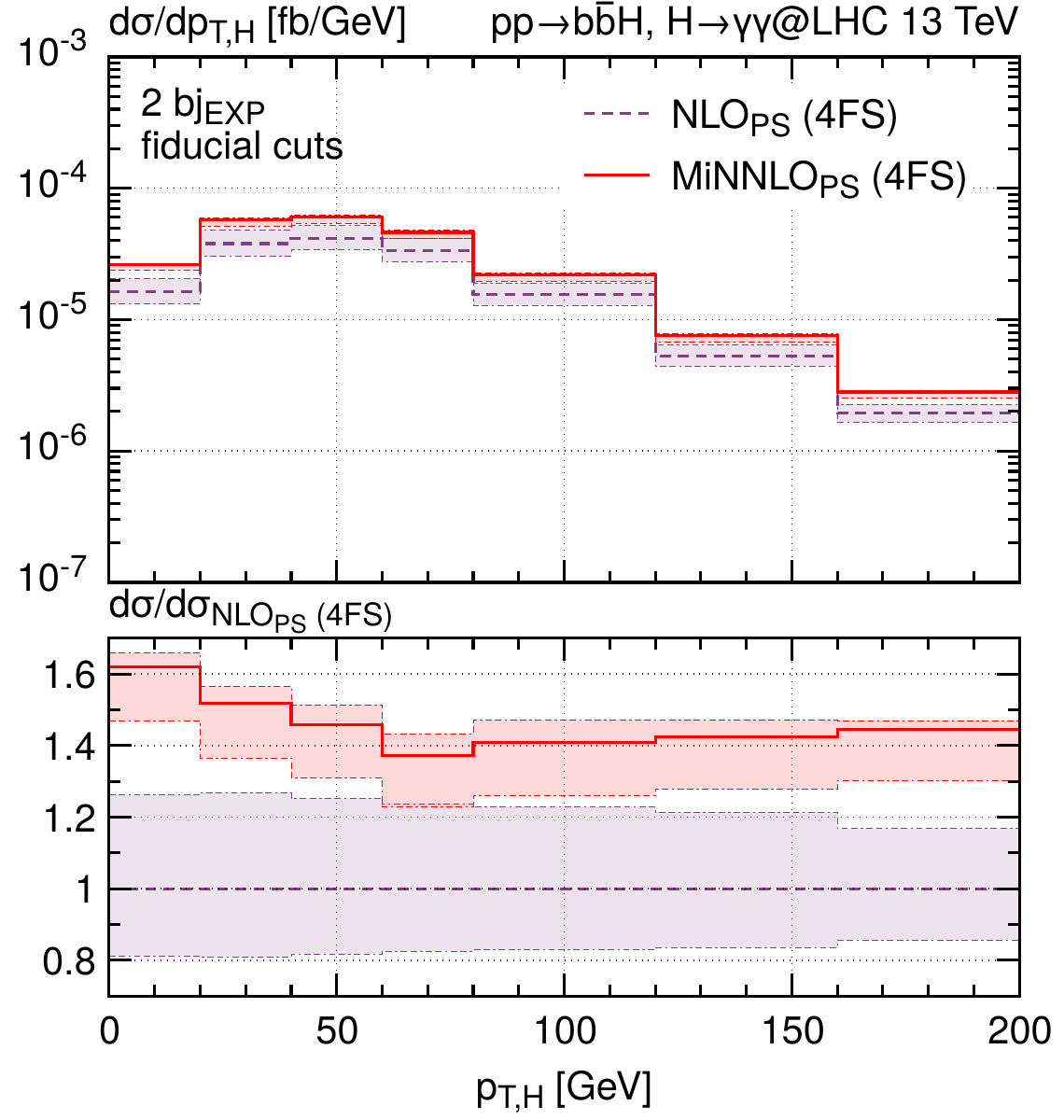}
 \includegraphics[width=.495\textwidth]{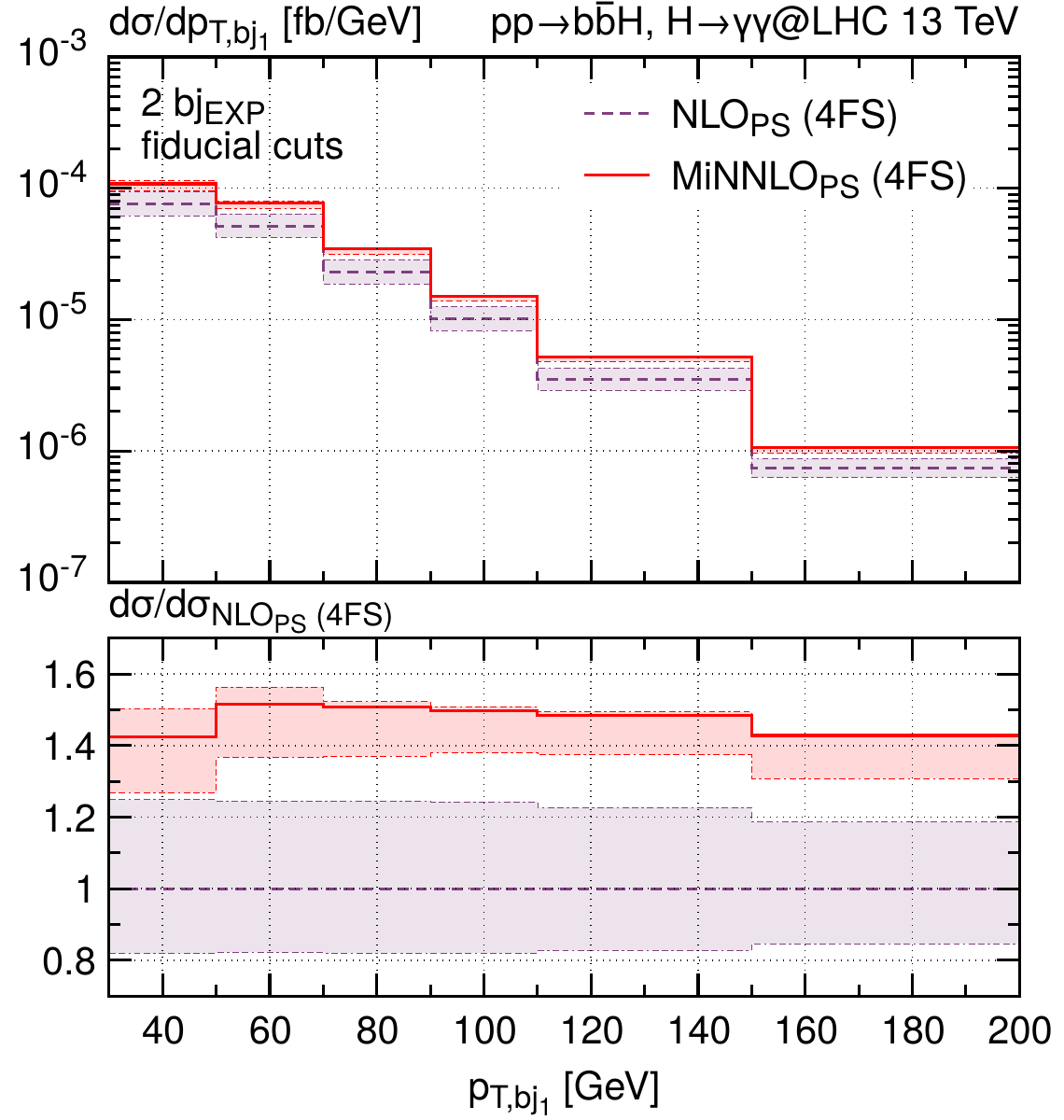}
	 \caption{Comparison of the transverse-momentum spectrum of the Higgs boson reconstructed from the two photons (left) and of the leading $b$-jet (right) at NLO+PS and \minnlo{} including fiducial cuts.}
 \label{fig:NLOvsMiNNLOoffshell2}
 \end{center}
 \end{figure*}
 
Next, we analyse the transverse-momentum distributions for the Higgs boson, reconstructed from the two photons, shown in the left plot of \fig{fig:NLOvsMiNNLOoffshell2}, and for the leading $b$-jet in the right plot of \fig{fig:NLOvsMiNNLOoffshell2}, with fiducial baseline selection cuts applied. The NNLO corrections show an enhancement of $40\%$--$50\%$ across the spectrum except for large $\ptarg{bj_{1}}$. Additionally, there is minimal overlap between the NLO and \minnlo{} uncertainty bands, emphasizing the critical role of including NNLO corrections. The scale uncertainties are significantly reduced for both observables at  NNLO+PS, especially for the $\ptarg{bj_{1}}$ distribution, where the reduction in scale uncertainty is most pronounced.

\section{Conclusions}
\label{sec:con}

In this work, we have presented a novel NNLO+PS event generator for the production of a Higgs boson in association with a bottom-quark pair 
induced by the bottom-Yukawa coupling. The calculation has been performed in the flavour scheme with four massless quark flavours and massive
bottom and top quarks. Our results represent the first computation of NNLO QCD corrections to this process. Moreover, this is only 
the second process of such complexity, namely with a heavy-quark pair and an extra colour singlet particle in the final state, for which NNLO QCD corrections are matched with parton showers.

The present computation has been rendered feasible by two major developments. Firstly, we incorporated in the \minnlo{} method all the essential components 
for computing processes that involve a heavy-quark pair in association with a colour singlet in the final state as well as an overall scale-dependent Yukawa coupling. Secondly,
the major bottleneck on the amplitude side is posed by the complexity of the double-virtual contribution, which we resolved by taking the limit of a small bottom mass, i.e. 
neglecting only power-suppressed terms in the bottom-quark mass while keeping all other contributions in the NNLO+PS calculation exact. 
We stress that this is an excellent approximation for $b\bar bH$ production at LHC energies, since power-suppressed terms 
in the bottom-quark mass are expected to be sufficiently small. Indeed, we have confirmed the high quality of this approximation explicitly at NLO+PS.

We have employed our new $b\bar b H$ generator in an extensive phenomenological analysis, highlighting the importance of NNLO corrections and parton-shower matching 
for this process. The NNLO corrections in the four-flavour scheme are quite sizeable and consistently enhance the $b\bar{b}H$ cross section by about 30\%, which are not fully captured by the 
scale variations of the NLO+PS predictions. Differentially these effects can be significantly enhanced, reaching NNLO corrections up to 100\% in some distributions.

We have compared our NNLO+PS predictions in the four-flavour scheme with massive bottom quarks against the previous NNLO+PS calculation in the five-flavour scheme from \citere{Biello:2024vdh}. Fixed-order results in the five-flavour scheme require an IR-safe algorithm for the identification of the bottom-flavoured jets, and a certain sensitivity
could also be present in calculations matched to parton showers. To study this, we have considered both the standard experimental approach to tag bottom-quark jets and 
the \textit{Interleaved Flavour Neutralisation} definition of bottom-quark jets \cite{Caola:2023wpj}. However, we found (both for the five- and the four-flavour scheme calculations)
that the impact of such an IR-safe algorithm is minor in practice. Only the removal of $g\to b\bar b$ splittings, i.e. identifying a jet including a $b\bar b$ pair not as a bottom-quark jet, 
has a noticeable impact on the predictions, simply because these respective topologies are removed.

A major result of our work is that the NNLO corrections in the four-flavour scheme, computed here for the first time, solve the long-standing discrepancy
between lower-order predictions in the two schemes. Both \minnlo{} generators describe Higgs observables and predictions with at least one identified bottom-flavour jet consistently 
within their respective uncertainties. Notably, the four-flavour scheme calculation delivers more precise predictions for events with identified bottom-quark jets, especially in configurations with at least one of them, where the 5FS NNLO+PS prediction is less accurate.

We have also presented an application of our NNLO+PS generator to estimate the $b\bar b H$ background in di-Higgs searches for the signature 
where one Higgs boson decays into bottom quarks and the other into photons. For typical event categories of the $HH$ signal in the $b\bar b\gamma\gamma$ 
final state we have found that the $b\bar bH$ background proportional to the bottom-quark Yukawa coupling increases by 30\%--50\% upon inclusion of 
NNLO QCD corrections. This provides an important example of the phenomenological impact of the developed $b\bar b H$ generator on future $HH$ measurements.

We reckon that our new $b\bar b H$ \minnlo{} generator will have important implications on various aspects of LHC phenomenology and that it will be a very
useful tool for the experimental collaborations at the LHC. We plan to make the associated code publicly available within the \POWHEGBOXRES{} framework shortly,
and the numerical results, events and code are available upon request.

With the completion of the two NNLO+PS generators in the four- and five-flavour schemes, it will be interesting to consider their consistent
combination, since each scheme describes different contributions in bottom-quark mass more accurately. The 5FS and 4FS MiNNLOPS generators presented in this work can indeed be used to obtain combined 4FS-5FS results. In order to perform the combination, we need to address the overlapping contributions between these two schemes, which requires modifying the current codes to ensure a consistent matching of the 4FS and 5FS. Thus, developing a unified approach, along the
lines of \citeres{Gauld:2021zmq,Guzzi:2024can} for instance,  that combines the two schemes
at NNLO+PS level, which has never been done before, could achieve unprecedented precision in the description of associated $b\bar{b}H$ production and
also provide a pathway to apply such approach to similar processes in the future. 
Additionally, our calculation provides the basis to develop \minnlo{} generators for processes with lighter quark flavours in the final state, such as 
Higgs production in association with charm quarks in the final state. Not least, the gluon-fusion component to $b\bar b H$ production proportional
to the top-Yukawa coupling could be considered with similar techniques.
All these developments are left for future work.

\section{Acknowledgments}
We are grateful to Chiara Savoini for performing cross-checks on the numerical implementation of the double-virtual correction. We acknowledge the \OpenLoops{} authors, particularly Federico Buccioni, for providing a suitable set of amplitudes. We also thank Vasily Sotnikov and Simone Zoia for discussions during the course of this work. All simulations presented here were carried out using the Max Planck Computing and Data Facility (MPCDF) in Garching.

\paragraph{Note added.} In the last stages of our work, the subleading-colour contributions to the double virtual correction for the \bbH{} process in the massless scheme were released in \citere{Badger:2024awe}. These contributions have now been incorporated into the \minnlo{} 4FS generator, and the results are presented in Appendix~\ref{app:FC}.

\appendix
\section{Ingredients for the two-loop approximation}
In this appendix, we provide the momentum mapping and the massification factors that are used for the massification procedure of the two-loop contribution described in \sct{sec:massification}.

\subsection{Momentum mapping}\label{sec:massif_mapping}
The phase-space points $\{p_i\}$ in \POWHEG{} assumes massive bottom quarks,
\begin{align}
  c(p_1) \bar c(p_2) \rightarrow b(p_3)\bar b(p_4) H(p_5)\,,
\end{align}
with $p_3^2=p_4^2=m_b^2$ and $c=q,\bar q,g$. However, the approximation requires the evaluation of the amplitudes with massless bottom quarks by using a suitable set of momenta,
\begin{align}
  c(\tilde p_1) \bar c(\tilde p_2) \rightarrow b(\tilde p_3)\bar	b(\tilde p_4) H(\tilde p_5)
\end{align}
where $\tilde p_3^2=\tilde p_4^2=0$. We define a mapping $\{p_i\}\rightarrow \{\tilde p_i\}$ that introduces only power-corrections $\mathcal{O}(m_b/\mu_h)$ in the redefinition of the momenta and preserve the kinematics of the Higgs state, i.e. $\tilde p_5= p_5$. In the quark-induced channel, we can easily preserve also the momenta of the incoming partons. These conditions imply that $p_3+p_4 = \tilde p_3+\tilde p_4$. Therefore also the invariant mass of the bottom-quark pair is preserved,
\begin{align}
  m_{b\bar b}=(p_3+p_4)^2=(\tilde p_3+\tilde p_4)^2\,,
\end{align}
as an invariant of the mapping. Therefore, by introducing the light-cone factors,
\begin{align}
  \rho_{\pm}=\frac{1\pm \rho}{2\rho},\, \rho=\sqrt{1-\frac{4 m_b^2}{m_{b\bar b}^2}}\,,
\end{align}
we define the massless momenta
\begin{align}
  \tilde p_3 = \rho_+ p_3- \rho_- p_4\,,\\
  \tilde p_4 = \rho_+ p_4- \rho_- p_3\,, 
\end{align}
which tend to $p_3$ and $p_4$, respectively, in the limit $m_b\to0$. If we were to apply the previous mapping to subprocesses with initial state gluons, an initial-final collinear divergence could appear, as also observed in \cite{Mazzitelli:2024ura,Devoto:2024nhl}. Therefore, in the gluon channel, we have adopted a different modification of the bottom-quark momenta,
\begin{align}
  \tilde p_j = p_j + \left( \sqrt{1-\frac{m_b^2 n_j^2}{\left(p_j \cdot n_j\right)^2}} -1 \right) \frac{p_j \cdot n_j}{n_j^2} n_j \,\hspace{1cm} \text{with } j\in\{3,4\}\,,
\end{align}
where we have introduced two vectors, transverse to both the initial-state momenta,
\begin{align}
  n_j=p_j - p_1 \frac{p_2\cdot p_j}{p_1\cdot p_2}-p_2 \frac{p_1\cdot p_j}{p_1\cdot p_2}\,.
\end{align}
In order to restore the momentum conservation, we have used an initial-state recoil, such that
\begin{align}
  \tilde p_1 + \tilde p_2 = p_1 + p_2 - \left( \tilde p_3 + \tilde p_4 - p_3 - p_4 \right)\,.
\end{align}
We stress that the Higgs kinematics is also preserved in the gluon channel with the previous choices.
\subsection{Massification factors} \label{sec:massif_ingredients}
\newcommand{\lmb}{~\ell_b}
\newcommand{\Nl}{n_l}
\newcommand{\Nh}{n_h}
The coefficients of the factor $\mathcal{\bar F}$ up to $\mathcal{O}(\alpha_s^2)$ in \eqn{eq:Fbar} are presented in \citere{Mazzitelli:2024ura} with the needed corrections from heavy-quark loops. For completeness, we report the first-order coefficient as follows,
\begin{align}
  \bar{\mathcal{F}}_\ccbar^{(1)} &= 2 \CF \lmb^{2} + \CF \lmb + \CF \left( 2 + \frac{\pi^{2}}{12}  \right) ~+~ \Nh \bar{\mathcal{F}}^{(1)}_{\Nh,c\bar c}\, ,
 \end{align}
 with
  \begin{align}
  \bar{\mathcal{F}}^{(1)}_{\Nh,q\bar{q}} = 0\, , \quad \bar{\mathcal{F}}^{(1)}_{\Nh,gg}= -\frac{2}{3} \lmb\, , 
  \end{align}
  while the second-order term reads as
  \begin{align}
  \nonumber
  \bar{\mathcal{F}}_\ccbar^{(2)} &=   2 \CF^{2} \lmb^{4} + \lmb^{3} \left( \frac{22}{9} \CA \CF + 2 \CF^{2} - \frac{4}{9} \CF \Nl \right) + \\ \nonumber
  \lmb^{2} &\left(\CA \CF \left( \frac{167}{18} - \frac{\pi^{2}}{3}\right) + \CF^{2} \left( \frac{9}{2} + \frac{\pi^{2}}{6}  \right) -\frac{13}{9} \CF \Nl \right) + \\ \nonumber 
  \lmb &\left( \CA \CF \left( \frac{1165}{108} + \frac{14}{9} \pi^{2} - 15 \zeta_3 \right) + \CF^{2} \left( \frac{11}{4} - \frac{11}{12} \pi^{2} + 12 \zeta_3 \right) - \CF \Nl \left( \frac{77}{54} + \frac{2}{9} \pi^{2} \right)\right) +  \\
  \nonumber &  \CF^{2} \left( \frac{241}{32} - \frac{163}{1440} \pi^{4} + \pi^{2} \left( \frac{13}{12} - 2 \log (2) \right) - \frac{3}{2} \zeta_3 \right) +\\
  \nonumber & \CA \CF \left( \frac{12877}{2592} - \frac{47}{720} \pi^{4} + \pi^{2} \left( \frac{323}{432} + \log (2) \right) + \frac{89}{36} \zeta_3 \right) -  \\ 
   & \CF\, \Nl \left( \frac{1541}{1296} + \frac{37}{216} \pi^{2} + \frac{13}{18} \zeta_3 \right) \;+\; \Nh\, \bar{\mathcal{F}}^{(2)}_{\Nh,c \bar{c}}\, , 
\end{align}
with
\begin{align}
  &\bar{\mathcal{F}}^{(2)}_{\Nh,q\bar{q}} = - \frac{20}{9} \lmb^{3} \CF  + \frac{32}{9} \lmb^{2} \CF  + \lmb \CF \left( -\frac{157}{27} - \frac{7}{18} \pi^{2} \right) + \CF \left( \frac{1933}{324} - \frac{13}{108} \pi^{2} - \frac{7}{9} \zeta_3 \right),  \\ \nonumber
  &\bar{\mathcal{F}}^{(2)}_{\Nh,gg} =  \lmb^{3} \left( -\frac{4}{9} \CA - \frac{28}{9} \CF \right) + \lmb^{2} \left( \frac{10}{9} \CA + \frac{7}{9} \CF \right) + \\
  &\lmb \left( \CF \left( -\frac{319}{54} - \frac{5}{18} \pi^{2} \right) + \CA \left( -\frac{92}{27} + \frac{\pi^{2}}{18}  \right) \right)  + \nonumber \\
  &\,\CA \left( \frac{179}{108} - \frac{5}{216} \pi^{2} - \frac{7}{18} \zeta_3 \right) + \CF \left( \frac{2677}{1296} - \frac{41}{216} \pi^{2} - \frac{\zeta_3}{18}  \right)  ~+~ \frac{4}{9} \Nh\lmb^2\,.
\end{align}
Here, the number $n_l=4$ represents the light flavours in the 4FS calculation, while $n_h=1$ is the number of quarks running in the loops that we want to promote to a massive state. The $n_h$ part is the novel contribution from the massification factors of the generalised massification approach in \eqn{eq:BecherMelnikov} compared to the procedure encoded in \eqn{eq:MitovMoch}. The logarithmic contributions are expressed as $\lmb = -\log\left( m_b/\mu_R \right)$ for compactness.\\

\noindent The non-trivial coefficient of the soft function in \eqn{eq:Sdef} is
\begin{align}\label{eq:massification_coeffs_end}
  \mathcal{S}^{(2)} &=  \Nh \left( - \frac{2}{3} \lmb^{2} + \frac{10}{9} \lmb -\frac{14}{27} \right)\, .
\end{align}

\section{Inclusion of subleading-colour terms in the double virtual contribution}
\label{app:FC}
The \minnlo{} generator in the massive scheme has been interfaced with the library described in~\citere{Badger:2024awe} to compute subleading-color contributions in the two-loop amplitude with massless bottom quarks, which is used to determine the non-logarithmic part of the massive double-virtual contribution. With the inclusion of the missing subleading terms, the scale dependence of the two-loop massified amplitude in the minimal subtraction scheme has been verified and found to be consistent with the prediction from the Renormalisation Group Flow in the small bottom-mass limit. This agreement serves as a nontrivial check, as both the massless two-loop amplitude and several contribution in the massification procedure depend on the infrared-subtraction scale. To mitigate numerical instabilities, a rescue system similar to that employed for the leading-colour library has been implemented in the \POWHEG{} interface. On average, evaluating the full-colour amplitude takes approximately 20 times longer than computing the planar component. In order to verify the stability of the full-colour result, we have evaluated the amplitude with both the momenta and the rescaled variables as described in~\eqref{eq:rescmomenta}. The novel simulations are performed using the setup described in section~\ref{sec:setup} with the default choice of the Higgs mass as the central renormalisation scale for the Yukawa coupling. The cross section results, with and without the subleading-color contributions, are compared in~\tab{tab:FCxsec}. 

\begin{table}[h!]
  \vspace*{0.3ex}
  \begin{center}
	             \renewcommand{\arraystretch}{1.4}
\begin{tabular}{|c|c|c|c|}
  \hline
    Fiducial region & Precision & $\sigma_{\rm integrated}$ [fb]   & \begin{tabular}{@{}c@{}} Ratio to \vspace{-2mm}\\  \minnlo{} wo/SLC\end{tabular} \\
  \hline \hline
      \multirow{2}{*}{$pp\rightarrow H$} & \minnlo{} wo/SLC & $466.(0)^{+16\%}_{-14\%}$ & 1.000 \\
      & \minnlo{} w/SLC & $451.(3)^{+15\%}_{-14\%}$ & 0.968 \\
      \hline
       \multirow{2}{*}{$pp\rightarrow H+\leq1\,b$ jets} & \minnlo{} wo/SLC & $92.(8)^{+9.8\%}_{-12\%}$ & 1.000 \\
      & \minnlo{} w/SLC & $88.(6)^{+8.2\%}_{-11\%}$ & 0.955 \\
      \hline
       \multirow{2}{*}{$pp\rightarrow H+\leq2\,b$ jets} & \minnlo{} wo/SLC & $6.5(4)^{+1.8\%}_{-9.2\%}$ & 1.000 \\
      & \minnlo{} w/SLC & $6.5(1)^{+3.5\%}_{-9.8\%}$ & 0.995 \\
      \hline
       \multirow{2}{*}{$pp\rightarrow H+ 0\,b$ jets} & \minnlo{} wo/SLC & $373.(2)^{+22\%}_{-20\%}$ & 1.000 \\
      & \minnlo{} w/SLC & $362.(7)^{+21\%}_{-19\%}$ & 0.972 \\
      \hline
\end{tabular}
  \end{center}
  \vspace{-1em}
  \caption{
	  Cross section rates for $b\bar bH$ production in the 4FS with and without the subleading-colour (SLC) terms in the estimation of the two-loop contribution. The fiducial region is defined according to the \texttt{EXP} definition of $b$-jets.}
   \label{tab:FCxsec}
\end{table}

\begin{figure*}[h!]
  \begin{center}
    \includegraphics[width=.49\textwidth, page=3]{defplots/figure-appB.pdf}
    \includegraphics[width=.49\textwidth, page=4]{defplots/figure-appB.pdf}
    \includegraphics[width=.49\textwidth, page=1]{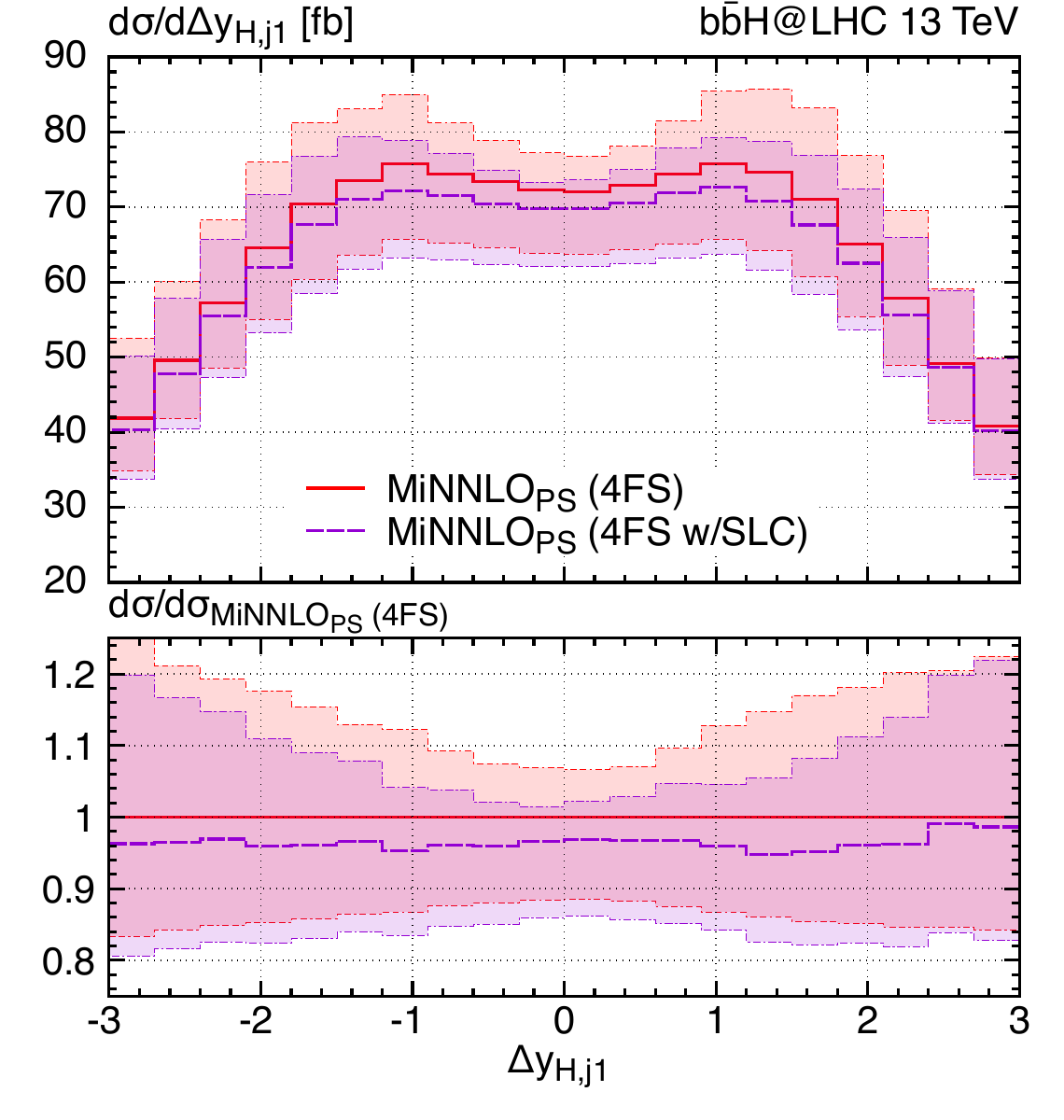}
    \includegraphics[width=.49\textwidth, page=2]{defplots/figure-appB.pdf}
    \caption{Comparison of NNLO+PS prediction with (violet, dashed) and without (red, solid) the inclusion of the subleading-colour contributions from the two-loop amplitude with massless bottom quarks.}
    \label{fig:FCplots}
  \end{center}
\end{figure*}

Although the inclusion of subleading effects is numerically challenging, we have obtained some differential predictions, shown in figure~\ref{fig:FCplots}. The first row describes the transverse-momentum and the rapidity spectra of the Higgs boson. In the second row of figure~\ref{fig:FCplots}, we present the rapidity difference between the Higgs and the hardest jet as a representative example of a jet-related observable. The last plot shows the Higgs transverse-momentum spectrum when at least a $b$-jet is required. The phenomenological impact of the subleading-colour contributions is approximately $\lesssim 5\%$, primarily acting as a negative and flat correction. The inclusion of the full-colour two-loop amplitude leads to a slight reduction in the theoretical uncertainty estimated through the 7-point scale variation. Due to the numerical efficiency of the planar component, we continue to use the leading-colour approximation in the \minnlo{} predictions for differential studies in more exclusive regions.



\bibliography{MiNNLO_bbH4FS}
\bibliographystyle{JHEP}

\end{document}